\documentclass[twocolumn]{aastex63}

\shorttitle{}
\shortauthors{Law-Smith et al.}

\graphicspath{{./}{figures/}}

\begin{document}

\title{Successful Common Envelope Ejection and Binary Neutron Star Formation in 3D Hydrodynamics}

\correspondingauthor{Jamie A.P. Law-Smith}
\email{jamie.law-smith@cfa.harvard.edu}

\author[0000-0001-8825-4790]{Jamie A.P. Law-Smith}
\affiliation{Center for Astrophysics $|$ Harvard \& Smithsonian, Cambridge, MA 02138, USA}
\affiliation{Department of Astronomy and Astrophysics, University of California, Santa Cruz, CA 95064, USA}
\affiliation{Niels Bohr Institute, University of Copenhagen, Blegdamsvej 17, 2100 Copenhagen, Denmark}

\author[0000-0001-5256-3620]{Rosa Wallace Everson}
\altaffiliation{NSF Graduate Research Fellow}
\affiliation{Department of Astronomy and Astrophysics, University of California, Santa Cruz, CA 95064, USA}
\affiliation{Niels Bohr Institute, University of Copenhagen, Blegdamsvej 17, 2100 Copenhagen, Denmark}

\author[0000-0003-2558-3102]{Enrico Ramirez-Ruiz}
\affiliation{Department of Astronomy and Astrophysics, University of California, Santa Cruz, CA 95064, USA}
\affiliation{Niels Bohr Institute, University of Copenhagen, Blegdamsvej 17, 2100 Copenhagen, Denmark}

\author[0000-0001-9336-2825]{Selma E. de Mink}
\affiliation{Center for Astrophysics $|$ Harvard \& Smithsonian, Cambridge, MA 02138, USA}
\affiliation{Max-Planck-Institut für Astrophysik, Karl-Schwarzschild-Straße 1, 85740 Garching bei München, Germany}
\affiliation{Anton Pannekoek Institute for Astronomy, University of Amsterdam, Science Park 904, 1098XH Amsterdam, The Netherlands}

\author[0000-0001-5484-4987]{Lieke A.C. van Son}
\affiliation{Center for Astrophysics $|$ Harvard \& Smithsonian, Cambridge, MA 02138, USA}
\affiliation{Anton Pannekoek Institute for Astronomy, University of Amsterdam, Science Park 904, 1098XH Amsterdam, The Netherlands}

\author[0000-0002-6960-6911]{Ylva Götberg}\thanks{Hubble Fellow}
\affiliation{The observatories of the Carnegie institution for science, 813 Santa Barbara Street, Pasadena, CA 91101, USA}

\author[0000-0003-2880-9090]{Stefan Zellmann}
\affiliation{Institute of Computer Science, University of Cologne, Weyertal 121, 50931 Cologne, Germany}

\author[0000-0003-1817-3586]{Alejandro Vigna-Gómez}
\affiliation{Niels Bohr Institute, University of Copenhagen, Blegdamsvej 17, 2100 Copenhagen, Denmark}

\author[0000-0002-6718-9472]{Mathieu Renzo}
\affiliation{Center for Computational Astrophysics, Flatiron Institute, New York, NY 10010, USA}
\affiliation{Department of Physics, Columbia University, New York, NY 10027, USA}

\author[0000-0003-2872-5153]{Samantha C. Wu}
\affiliation{California Institute of Technology, 1200 East California Boulevard, MC 249-17, Pasadena, CA 91125, USA}

\author[0000-0003-1735-8263]{Sophie L. Schrøder}
\affiliation{Niels Bohr Institute, University of Copenhagen, Blegdamsvej 17, 2100 Copenhagen, Denmark}

\author[0000-0002-2445-5275]{Ryan J. Foley}
\affiliation{Department of Astronomy and Astrophysics, University of California, Santa Cruz, CA 95064, USA}

\author[0000-0002-3472-2453]{Tenley Hutchinson-Smith}
\affiliation{Department of Astronomy and Astrophysics, University of California, Santa Cruz, CA 95064, USA}
\affiliation{Niels Bohr Institute, University of Copenhagen, Blegdamsvej 17, 2100 Copenhagen, Denmark}

\begin{abstract}
A binary neutron star merger has been observed in a multi-messenger detection of gravitational wave (GW) and electromagnetic (EM) radiation. Binary neutron stars that merge within a Hubble time, as well as many other compact binaries, are expected to form via common envelope evolution. Yet five decades of research on common envelope evolution have not yet resulted in a satisfactory understanding of the multi-spatial multi-timescale evolution for the systems that lead to compact binaries. In this paper, we report on the first successful simulations of common envelope ejection leading to binary neutron star formation in 3D hydrodynamics. We simulate the dynamical inspiral phase of the interaction between a 12$M_\odot$ red supergiant and a 1.4$M_\odot$ neutron star for different initial separations and initial conditions. For all of our simulations, we find complete envelope ejection and final orbital separations of $a_{\rm f} \approx 1.3$--$5.1 R_\odot$ depending on the simulation and criterion, leading to binary neutron stars that can merge within a Hubble time. We find $\alpha_{\rm CE}$-equivalent efficiencies of $\approx 0.1$--$2.7$ depending on the simulation and criterion, but this may be specific for these extended progenitors. We fully resolve the core of the star to $\lesssim 0.005 R_\odot$ and our 3D hydrodynamics simulations are informed by an adjusted 1D analytic energy formalism and a 2D kinematics study in order to overcome the prohibitive computational cost of simulating these systems. The framework we develop in this paper can be used to simulate a wide variety of interactions between stars, from stellar mergers to common envelope episodes leading to GW sources.
\end{abstract}


\section{Introduction} \label{sec:intro}
The majority of astrophysical sources detected by the advanced Laser Interferometer Gravitational-Wave Observatory (LIGO) and Virgo
observatory have involved stellar mass binary black hole (BBH) mergers, with the two most notable exceptions being the (likely) binary neutron star (BNS) mergers GW170817 and GW190425 \citep{2017ApJ...848L..13A,2020ApJ...892L...3A}.
While dynamical encounters may play a role in the origin of BBHs, they are not an effective pathway for the assembly of binary neutron star mergers \citep[e.g.,][]{2020ApJ...888L..10Y}, which are thought to form almost exclusively in interacting binaries \citep{1973NInfo..27...70T,1993MNRAS.260..675T,2016ApJ...819..108B,2017ApJ...846..170T}.

Massive stars are the progenitors of neutron stars and black holes, and the majority of massive (i.e., type B and O) stars are in close enough binaries such that interaction is inevitable as the stars evolve \citep{2012Sci...337..444S, 2017ApJS..230...15M}.
A BNS is expected to form from the cores of well-evolved stars, and thus have much lower orbital energy and angular momentum than the original binary progenitor. For BNSs that merge in a Hubble time, after one of the progenitor stars evolves through the red giant phase and overflows its Roche lobe, the original binary is believed to significantly shrink during a phase of unstable mass transfer, which leads to a spiral-in of the binary and ejection of the envelope---this is collectively commonly referred to as common envelope (CE) evolution \citep[e.g.,][]{1975MitAG..36R..93R, 1976IAUS...73...75P,1993PASP..105.1373I,2013A&ARv..21...59I}. If this process leads to a deposition of orbital energy that is sufficient to eject the envelope of the giant, the predicted properties of the resulting compact binary could match the observed properties of the BNS population. Past attempts to model this process have failed because they cannot reproduce these observed properties. A more complete (and in particular, multidimensional) theoretical description is required in order to provide an accurate description of the evolution of a NS embedded in a common envelope. This work focuses on the decades-long pursuit of this elusive phenomenon.

A critical juncture in the life of a binary occurs just after mass transfer commences in the system. The system
either coalesces or may survive to become an interacting binary. 
This is the case of the recently discovered M Supergiant High Mass X-Ray Binary (HMXB) 4U 1954+31 \citep{2020ApJ...904..143H}, which contains a late-type supergiant of mass $9^{+6}_{-2} M_\odot$; it is the only known binary system of its type. 
It is difficult and rare to observe a system in this state, as the system evolves rapidly, yet this discovery may be the first observation of a system similar to the progenitor studied in this work. 
If mass transfer becomes unstable in this system it could lead to a CE episode. Two outcomes are then possible: (1) one star has a clear core/envelope separation and the other star is engulfed into its envelope, or (2) both stars have a core/envelope separation and the envelopes of the two stars overfill their Roche lobes \citep[see e.g.,][]{2020PASA...37...38V}.
Usually, the term CE is used to describe a situation in which the envelope is not co-rotating with the binary and is not necessarily in hydrostatic equilibrium.
The state of the primary at onset of CE evolution is determined by the initial separation and the orbital evolution of the binary \citep[see e.g.,][]{2021A&A...645A..54K}---generally, it will begin when the radius of the primary overflows its Roche lobe.
The outcome of the CE phase can be either a stellar merger or the formation of a close binary. 
If the binary remains bound and on a tight orbit after the second NS has been created, the system will merge due to the dissipation of gravitational waves (GWs).
The merger timescale depends on the final orbital separation and energy of the binary; if these are small enough such that the binary merges within a Hubble time, the stellar remnants---either black holes, neutron stars, or white dwarfs---will merge and produce GW and possibly electromagnetic (EM) radiation. 

In particular, BNS mergers expel metallic, radioactive debris (the light from which is referred to as a kilonova) that can be seen by telescopes \citep[e.g.,][]{2017Natur.551...80K}. In August 2017, for the first time, we detected both GWs and EM radiation \citep[e.g.,][]{2017Sci...358.1556C,2017ApJ...848L..12A,2017ApJ...848L..14G} coming from the same astrophysical event. This landmark discovery, which has opened up new lines of research into several areas in astrophysics and physics, makes the study of interacting binaries and common envelope in particular, even more essential in our attempts to discern the assembly history of these probes of extreme physics.  
Yet, their formation process remains an open question. 

In this work we present the first 3D hydrodynamics simulations of successful CE ejection that can lead to a BNS system. Simulations of this kind have not been performed so far due to the prohibitive computational cost---the relevant dynamic ranges of density and physical distance are $\gtrsim$10$^6$ (e.g., the global problem must resolve a $R \approx 10^6$ cm neutron star within the envelope of a $R \approx 10^{13}$ cm giant star, whose density varies from $\rho \approx 10^6$ g/cm$^3$ to $\rho \approx 10^{-9}$ g/cm$^3$ within the relevant regions).
Most 3D hydrodynamics simulations of CE evolution have been at relatively equal mass ratios and for relatively low stellar masses ($M_\star \lesssim 3 M_\odot$) 
\citep[e.g.,][]{2001ApJ...550..357Z, 2008ApJ...672L..41R, 2012ApJ...746...74R, 2012ApJ...744...52P, 2016MNRAS.460.3992N, 2016ApJ...816L...9O, 2017MNRAS.464.4028I, 2018MNRAS.477.2349I, 2019MNRAS.486.5809P, 2020A&A...642A..97K, 2020A&A...644A..60S, 2020MNRAS.495.4028C}, and there has been an early attempt and characterization of the difficulties faced by simulating a massive star binary by \citet{2019IAUS..346..449R}.
Higher mass ratios involving NSs have been studied in 1D \citep[e.g.,][]{2015ApJ...803...41M, 2019ApJ...883L..45F}. 

In contrast to other contemporary studies, the initial conditions of our 3D hydrodynamics simulations are informed by an adjusted 1D analytic energy formalism and a 2D kinematics study.
We start the 3D hydrodynamics simulation once the secondary has ejected $< 0.1$\% of the star's binding energy. This corresponds to a relatively small radius compared to the full radius of the star for the extended progenitors we study, which have very tenuous envelopes: we trim the star to $10 R_\odot$ (or $20 R_\odot$ for a convergence test).
In contrast to other contemporary work in which the core is often replaced with a point mass, we fully resolve the gas in the core to $\lesssim 0.005 R_\odot$. 

This paper is organized as follows. \S\ref{sec:methods} describes our methods, including the 1D analysis, 2D kinematics, and 3D hydrodynamics, \S\ref{sec:results} describes our results, \S\ref{sec:discussion} compares to other work and discusses caveats and future work, and \S\ref{sec:conclusion} concludes.

\section{Methods} \label{sec:methods}

We simulate the CE evolution of an initially 12$M_\odot$ red supergiant primary (donor) and a 1.4$M_\odot$ point mass secondary (NS) in 3D hydrodynamics, for different initial separations and initial conditions. We build the primary with a 1D stellar evolution code (\texttt{MESA}). We use an adjusted 1D energy formalism to predict the likely CE ejection regime, and we use a 2D kinematics study to inform the initial conditions of the 3D hydrodynamics simulations. We import the stellar model to the 3D hydrodynamics simulation (\texttt{FLASH}), in which we excise the outermost layers of the star with negligible binding energy and start the secondary relatively close to the core of the primary where the CE ejection is predicted to take place.

\subsection{\texttt{MESA} model}

We use the 1D stellar evolution code \texttt{MESA} v8118 \citep{2011ApJS..192....3P, 2013ApJS..208....4P, 2015ApJS..220...15P} to construct the primary.
We use an inlist from \citet{2018A&A...615A..78G}, which is publicly available on Zenodo.\footnote{\texttt{MESA} inlists, v8118: \url{https://zenodo.org/record/2595656}.}
We construct a 12$M_\odot$ solar-metallicity \citep[X=0.7154, Y=0.270, Z=0.0142;][]{2009ARA&A..47..481A} single-star primary as this is a typical mass to form a NS \citep{2003ApJ...591..288H}.
In \S\ref{sec:results} we show the evolutionary history of this model and in \S\ref{sec:MESA_profiles} and \S\ref{sec:1D_energy_plots} we show mass, density, composition, and binding energy profiles for the models we simulate in 3D hydrodynamics.

See Section 2.1 of \citet{2018A&A...615A..78G} for details on the \texttt{MESA} setup. Additional uncertainties in the \texttt{MESA} modeling are discussed in Section~\ref{sec:discussion}.
In brief, our setup is to use the \texttt{mesa\_49.net} nuclear network of 49 isotopes, account for overshooting following \citet{2011A&A...530A.115B}, and account for mass loss using the wind schemes of \citet{1988A&AS...72..259D} and \citet{2001A&A...369..574V}.

\subsection{Energy formalism}\label{subsec:1D_energy}

We perform CE energy formalism \citep[$\alpha$ formalism;][]{1988ApJ...329..764L, 1976IAUS...73...35V, 1984ApJ...277..355W, 1990ApJ...358..189D, 1993PASP..105.1373I} calculations on the profiles to predict the radius ranges in which CE ejection is possible and to inform the initial conditions of the 3D hydrodynamics simulations.
As in \citet{2020ApJ...901...44W}, we calculate the gravitational binding energy and orbital energy loss profiles (see \S\ref{sec:1D_energy_plots}) for the \texttt{MESA} model at all ages throughout its giant branch evolution.
These profiles help determine the predicted ejection ranges (see \S\ref{sec:results}).
See \citet{2020ApJ...901...44W} and \citet{2020ApJ...899...77E} for further details of these calculations.

Local 3D hydrodynamical simulations of CE have shown that during dynamical inspiral, the energy deposition from the secondary's plunge extends inward from the secondary's location, heating and unbinding deeper envelope material \citep[see, e.g.,][]{2017ApJ...838...56M, 2020ApJ...897..130D}. To incorporate the effects of this energy deposition on the ejection radius range, we also apply an adjusted $\alpha$ formalism \citep{Everson_in_prep} that requires the orbital energy loss to overcome the binding energy at radii deeper than that given by the orbital separation, corresponding to $r-R_a$ (where $R_a$ is the Bondi accretion radius) or $r-R_{\rm Roche}$ (where $R_{\rm Roche}$ is the Roche radius); see below.
The ejection ranges shown in Figure~\ref{fig:MESA} were calculated using this adjusted $\alpha$ formalism as well as using work from \citet{2020ApJ...899...77E}.

All $\alpha$ formalism calculations in, e.g., Figure~\ref{fig:MESA_profiles}, are based on, e.g., \citet{2013A&ARv..21...59I} and \citet{2016A&A...596A..58K}. The change in orbital energy is defined as \citep[see e.g., Eq. 2 of][]{2016A&A...596A..58K}:
\begin{equation}\label{eq:Delta_E_orb}
\Delta E_{\rm orb} = -\frac{GM(r<a_{\rm f}) M_{\rm secondary}}{2a_{\rm f}}+\frac{GM_{\rm donor}M_{\rm secondary}}{2a_{\rm i}},
\end{equation}
where $a_{\rm i}$ is the initial orbital separation, $a_{\rm f}$ is the final orbital separation, and $M(r<a_{\rm f})$ is the mass interior to the final orbital separation (this is $\approx M_{\rm core}$).
The gravitational binding energy is defined as
\begin{equation}\label{eq:E_bind}
E_{\rm bind} = -\int \frac{G m}{r} dm.
\end{equation}
For the $r-R_a$ adjusted formalism we use the accretion radius as defined in, e.g.,  \citet{1944MNRAS.104..273B} and \citet{1939PCPS...35..405H}:
\begin{equation}
R_a = \frac{2 G M_{\rm secondary}}{v_\infty^2}
\end{equation}
and for the $r-R_{\rm Roche}$ adjusted formalism we use the Roche radius (the radius equivalent to the volume of the Roche lobe) as in the approximation of \citet{1983ApJ...268..368E}.

We adapted a 2D integrator used to study the kinematics of CE inspiral with drag \citep{2017ApJ...838...56M} with the results from a 3D study of drag coefficients in CE evolution with density gradients \citep{2020ApJ...897..130D} to determine an initial velocity vector for the radius at which we begin our 3D hydrodynamics simulations. We also compare to results using a circular initial velocity vector.

\subsection{\texttt{FLASH} setup}

The outline of our 3D hydrodynamics setup is the following: (1) excise (cut out) the tenuous outer layers of the primary (donor) star, (2) initialize the primary on the grid, (3) relax the point particle secondary (neutron star) onto the grid, (4) initialize the point particle's velocity vector based on the 2D kinematics results, and (5) simulate the system in 3D hydrodynamics until the orbital separation stalls.

In this paper, we focus on evolutionary stages which we expect that will lead to a CE ejection \textit{a priori} and then use 3D hydrodynamics to simulate the crucial dynamical inspiral phase of the CE evolution.

Numerical diffusion prohibits us from evolving the system for $\gtrsim 30$ orbits, since for many orbits, numerically-driven drag results in the companion inspiraling toward the core of the donor. 
See \S\ref{sec:discussion} for a detailed discussion of this.
Thus, we consider only evolution in the 3D hydrodynamics on a timescale much shorter than the thermal timescale, to prevent including artificially merging or ejected cases.

We use a custom setup of the 3D adaptive-mesh refinement (AMR) hydrodynamics code \texttt{FLASH} \citep{2000ApJS..131..273F}, version 4.3\footnote{The updates in later versions do not affect our setup.}.
Our \texttt{FLASH} setup is based on that of \citet{2020ApJ...901...44W}, which was based on that of \citet{2019ApJ...882L..25L} and \citet{2020ApJ...905..141L}, which was in turn based on that of \citet{2009ApJ...705..844G} and \citet{2013ApJ...767...25G}. 
See these references for more details on the numerics. A brief summary including salient features and changes to the setup is below.

We use a Helmholtz equation of state with an extended Helmholtz table\footnote{As of time of writing available at \url{http://cococubed.asu.edu/code_pages/eos.shtml}.} spanning $10^{-12} \leq \rho~{\rm [g~cm^{-3}]} \leq 10^{15}$ and $10^3 \leq T~{\rm [K]} \leq 10^{13}$.
The Helmholtz equation of state assumes full ionization \citep{2000ApJS..126..501T} and thus does not include recombination energy in the internal energy.
We track the same chemical abundances in the 3D hydrodynamics as in the \texttt{MESA} nuclear network for the star, for all elements above a mass fraction of $10^{-5}$ (this value is somewhat arbitrary but does not affect the results); this is 22 elements ranging from hydrogen ($^1$H) to iron ($^{56}$Fe). While including an arbitrary number of the elements tracked in \texttt{MESA} is possible in our setup, including all of the elements would unnecessarily increase the memory load of the 3D hydrodynamics.

We excise the outer envelope of the primary donor star that constitutes $<0.1$\% of the total binding energy (see \S\ref{subsec:1D_energy}, \S\ref{sec:results}, and \S\ref{sec:1D_energy_plots} for further discussion) and is easily ejected, trimming the star to $R = 10 R_\odot$ (or $R = 20 R_\odot$ for a convergence test). 
Our box size is $\Delta X_{\rm max} = 40 R_\odot$ on a side.
This technique was also employed in \citet{2020ApJ...901...44W}.
We refine such that $\Delta X_{\rm min} \lesssim 0.005 R_\odot$ within a factor of 100 of the maximum density, then derefine in the AMR with decreasing density, for $N \approx 272$ cells across the diameter of the star for the nominal simulations presented in this paper.
We verified the hydrostatic equilibrium of our initial conditions for several dynamical timescales of the star (and 100s of dynamical timescales of the core).
Hydrostatic equilibrium following the relaxation scheme in our setup has also been tested in e.g., \citet{2020ApJ...905..141L}.
We initialize the secondary point mass (NS) at $a_{\rm i}=8 R_\odot$, within the envelope of the $10 R_\odot$ trimmed star.
After initializing the star on the grid, we gradually introduce the point mass secondary inside the envelope of the primary by gradually increasing its velocity to its initial velocity vector (see also \S\ref{subsec:1D_energy}).
This technique is also used in \citet{2017ApJ...838...56M} and \citet{2020ApJ...901...44W}.

More realistic initial conditions would start at the point of Roche lobe overflow to take into account the transfer of energy and angular momentum from the orbit to the envelope, but this is computationally prohibitive for a $R_\star \approx 1000 R_\odot$ primary with a density range of 15 orders of magnitude (from $\rho \approx 10^6$ to $\rho \approx 10^{-9}$ g/cm$^3$).
However, we argue that the initial conditions used in this work are similar to the configuration if we had begun the simulation at this earlier stage and evolved it to the time we start our simulation.
This is justified in \S\ref{sec:1D_energy_plots} and using the methods of \S\ref{subsec:1D_energy}.

We use two initial velocity vectors: (1) circular and (2) informed by a 2D kinematics study using the stellar density profile. The 2D kinematics velocity vectors are derived from orbits that are more eccentric than a circular orbit.
However, we find that the initial velocity vector does not have a significant effect on the final outcome of the simulation, with both velocity vectors leading to qualitatively similar results. This weak dependence on the initial velocity vector is due to the fact that the point mass relatively quickly encounters drag and spirals inward dynamically, as was also found in \citet{2020ApJ...901...44W}.

We also perform a numerical convergence study (see \S\ref{sec:numerical_convergence} for details). First, we run two simulations with the same initial conditions but one with 2.5X lower linear resolution than the other, and find similar results in the orbital evolution and final orbital separations, verifying that our nominal resolution of $\Delta X_{\rm min} \lesssim 0.005 R_\odot$ is adequately converged.
Second, we verify that the NS clears the material interior to its orbit and exterior to the Roche radius of the core, which is necessary to successfully stall at a final orbital separation.
As part of this analysis, in order to isolate the effect of the NS vs. mass leakage due to numerical effects, we run an additional simulation without the NS.
Third, we run an additional simulation where the NS is initialized at $a_{\rm i}=16 R_\odot$ and the star is trimmed to $R=20R_\odot$, twice the values in our nominal study, and verify that the NS reaches the same radius at which we start our nominal simulations (i.e., it does not stall exterior to our initial conditions if we start our simulation at this larger radius).

\section{Results} \label{sec:results}

\subsection{1D modeling}

\begin{figure*}[htp!]
\epsscale{1.15}
\plotone{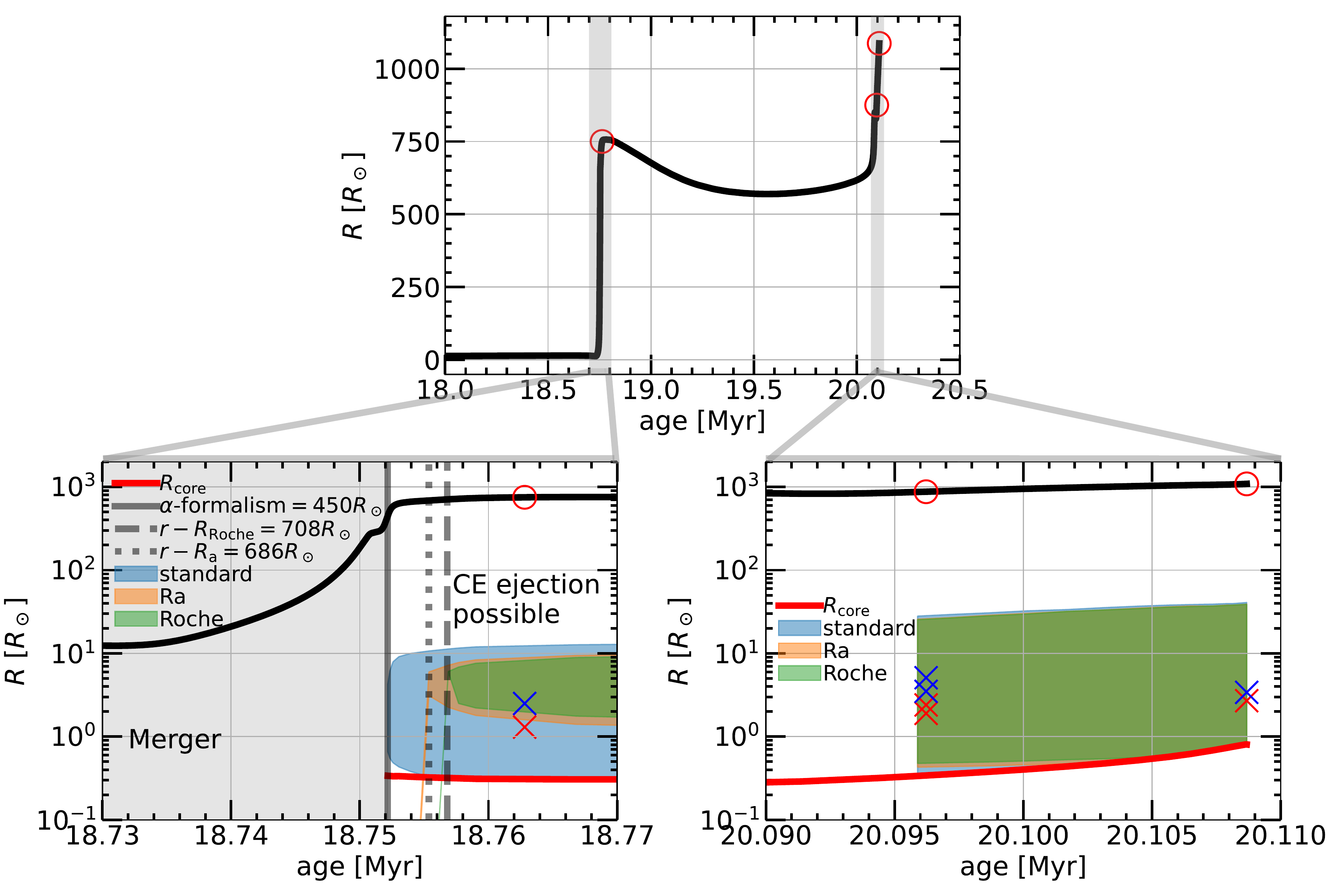}
\caption{
\texttt{MESA} evolutionary history for the 12$M_\odot$ primary (donor) star.
Top: radius vs. time.
Red circles indicate the models we simulate in 3D hydrodynamics.
Bottom left: focus on the first rise (expansion).
Vertical lines indicate the earliest ages where CE ejection is possible and shaded regions indicate the radius ranges where CE ejection is possible according to our adjusted 1D energy formalism.
Red line indicates the radius of the helium core.
Blue and red `X's indicate the final orbital separations from our 3D hydrodynamics simulations (see text).
Bottom right: focus on the second rise.
See \S\ref{sec:MESA_profiles} for further details on the \texttt{MESA} model.
\label{fig:MESA}
}
\end{figure*}

The top panel of Figure~\ref{fig:MESA} shows radius vs. time for the initially $12 M_\odot$ donor, evolved as a single star using the setup of \citet{2018A&A...615A..78G}. 
The red circles indicate the three different models we simulate in 3D hydrodynamics: near the first peak ($R_\star=750 R_\odot$, $M_\star=11.8 M_\odot$), on the second rise ($R_\star=900 R_\odot$, $M_\star = 9.9 M_\odot$), and at the second peak ($R_\star=1080 R_\odot$, $M_\star = 9.8 M_\odot$).
The first peak corresponds to RLOF (Roche-lobe overflow) during late hydrogen-shell burning \citep[case B; e.g.,][]{1967ZA.....65..251K} and the second peak to RLOF after core-helium burning \citep[case C; e.g.,][]{1970A&A.....7..150L}.
In all three cases, the donor has a deep convective envelope and the mass transfer is dynamically unstable.

In the bottom panels, we zoom in on the first and second rises (expansions).
The radius of the helium core is shown in red (defined by the \texttt{he\_core\_mass} attribute in \texttt{MESA}, using \texttt{he\_core\_boundary\_h1\_fraction $\geq$ 0.01} and \texttt{min\_boundary\_fraction $\geq$ 0.1}).
It is $R_{\rm core}=0.31 R_\odot$ for the first peak, $R_{\rm core}=0.36 R_\odot$ for the second rise, and $R_{\rm core}=0.8 R_\odot$ for the second peak.
For a given stellar age, the predicted radius ranges where CE ejection is possible as predicted by the three 1D energy formalisms (standard $\alpha$ formalism, $r-R_a$ adjusted formalism, and $r-R_{\rm Roche}$ adjusted formalism) are shown in shaded blue, orange, and green respectively (see \S\ref{subsec:1D_energy}).
We start the \texttt{FLASH} simulations just within these ranges (see \S\ref{sec:methods}).
Blue `X's indicate the final orbital separations when the envelope outside the current orbit of the secondary is ejected (see Figure~\ref{fig:E_bind_vs_time}) and red `X's indicate the final orbital separations when the entire envelope outside the helium core is ejected (see Figure~\ref{fig:f_ej}) in our 3D hydrodynamics simulations.

The bottom left panel focuses on the first rise. The earliest ages at which CE ejection is possible from the 1D energy formalisms are indicated by the vertical lines.
The bottom right panel focuses on the second rise. Here the different energy formalisms predict a similar range of radii for possible CE ejection, and in the 3D hydrodynamics we eject the envelope within these ranges.

We calculate the minimum radius on the second rise in which Roche-lobe overflow is possible, accounting for orbital widening of the binary as a result of mass loss by fast stellar winds during its prior evolution (see \S\ref{sec:forbidden_radii} for discussion and details on this).
We find that after the first peak (at $R_\star = 757 R_\odot$), for radii less than $R_\star = 900 R_\odot$ on the second rise, RLOF will not occur.
Thus, we simulate three models in 3D hydrodynamics that are chosen to span the range of stellar structures in which dynamical CE ejection is possible for a $12 M_\odot$ primary: near the first peak ($R_\star=750 R_\odot$), on the second rise ($R_\star=900 R_\odot$), and at the second peak ($R_\star=1080 R_\odot$). 
We note that the $750 R_\odot$ and $1080 R_\odot$ models may appear fine-tuned in isolation, but they are chosen so that our suite of 3D hydrodynamics simulations in this paper span the parameter space of stellar structures that will lead to dynamical CE ejection.

\subsection{3D hydrodynamics}

\begin{figure*}[htp!]
\epsscale{1.12}
\plotone{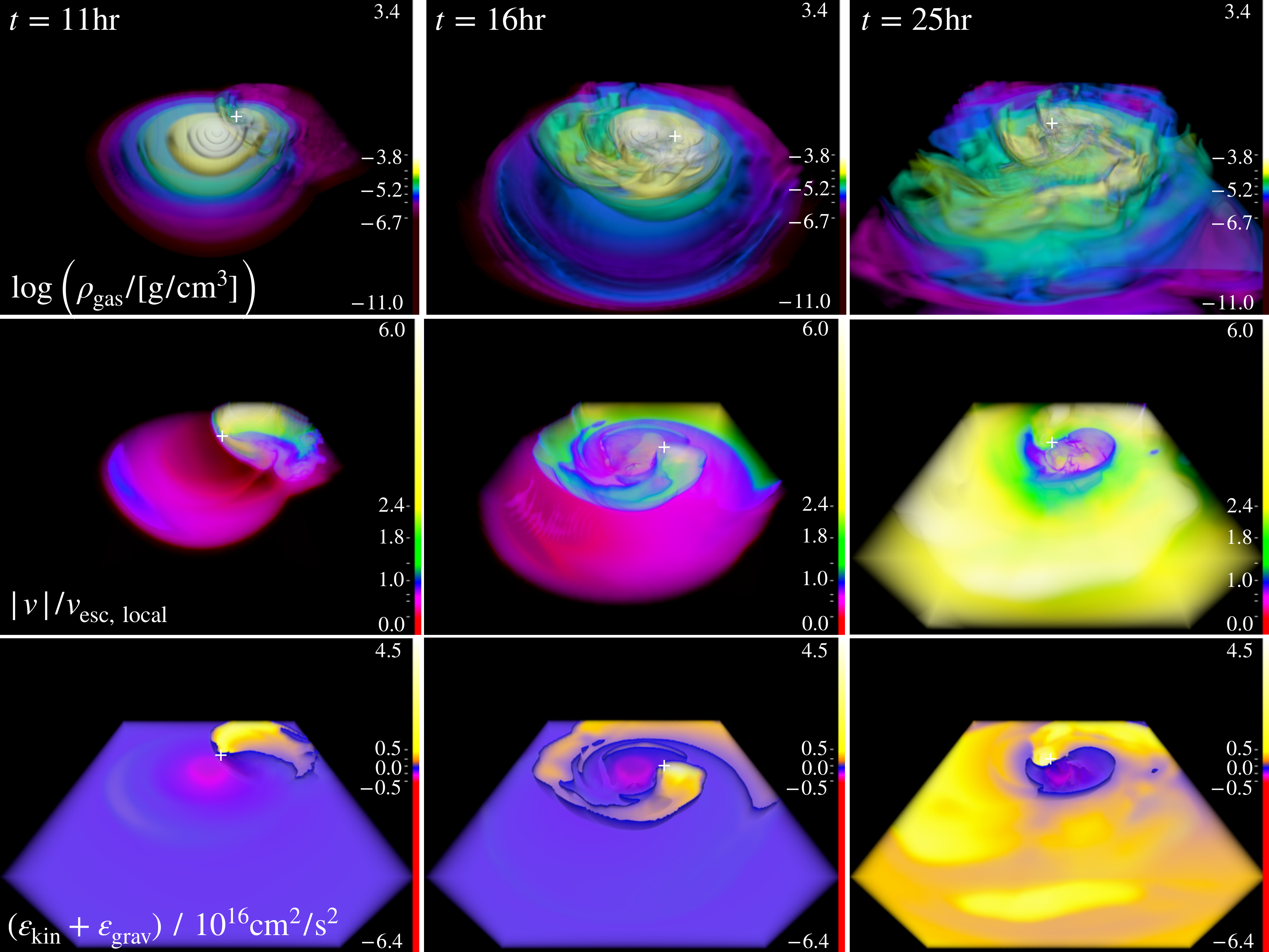}
\caption{
3D renderings of three fields (density, velocity, and energy) at three times: early in the evolution (11 hr), at an intermediate time (16 hr), and at a moderately late time (25 hr) when the envelope has just been ejected.
We show the $900 R_\odot, v_i=v_{\rm circ}$ simulation; results are qualitatively similar for all our other simulations.
1st row: logarithm of gas density. Shells corresponds to different density isosurfaces.
2nd row: ratio of velocity magnitude to local escape velocity, $|v|/v_{\rm esc, local}$. 
Blue isosurface is at $|v|/v_{\rm esc, local}=1$, pink-red is $<1$, green-yellow is $>1$.
3rd row: sum of specific kinetic and potential energy. 
Blue isosurface at $\varepsilon=0$, pink-purple corresponds to bound ($\varepsilon<0$) and yellow corresponds to unbound ($\varepsilon>0$).
White `+' indicates position of secondary.
Videos available at \url{https://youtube.com/channel/UCShahcfGrj5dOZTTrOEqSOA}.
\label{fig:9panel}
}
\end{figure*}

Figure~\ref{fig:9panel} shows 3D volume renderings of three fields (density, velocity, and energy) at three times: early in the evolution (11 hr), at an intermediate time (16 hr), and at a relatively late time (25 hr) when the envelope has just been ejected.
We show renderings for the $900 R_\odot, v_i=v_{\rm circ}$ (circular initial velocity) simulation. Results are qualitatively similar for all of the other simulations.
The volume renderings are of the bottom half of the orbital plane ($z<0$, with $J_{\rm orb} \parallel z$), with a color map and transfer function chosen to highlight the dynamic range and structure of the field being studied.
See \S\ref{sec:detailed_time_evolution} for the detailed time evolution of these three fields and a zoom-in on the core.

\begin{figure*}[htp!]
\epsscale{0.57}
\plotone{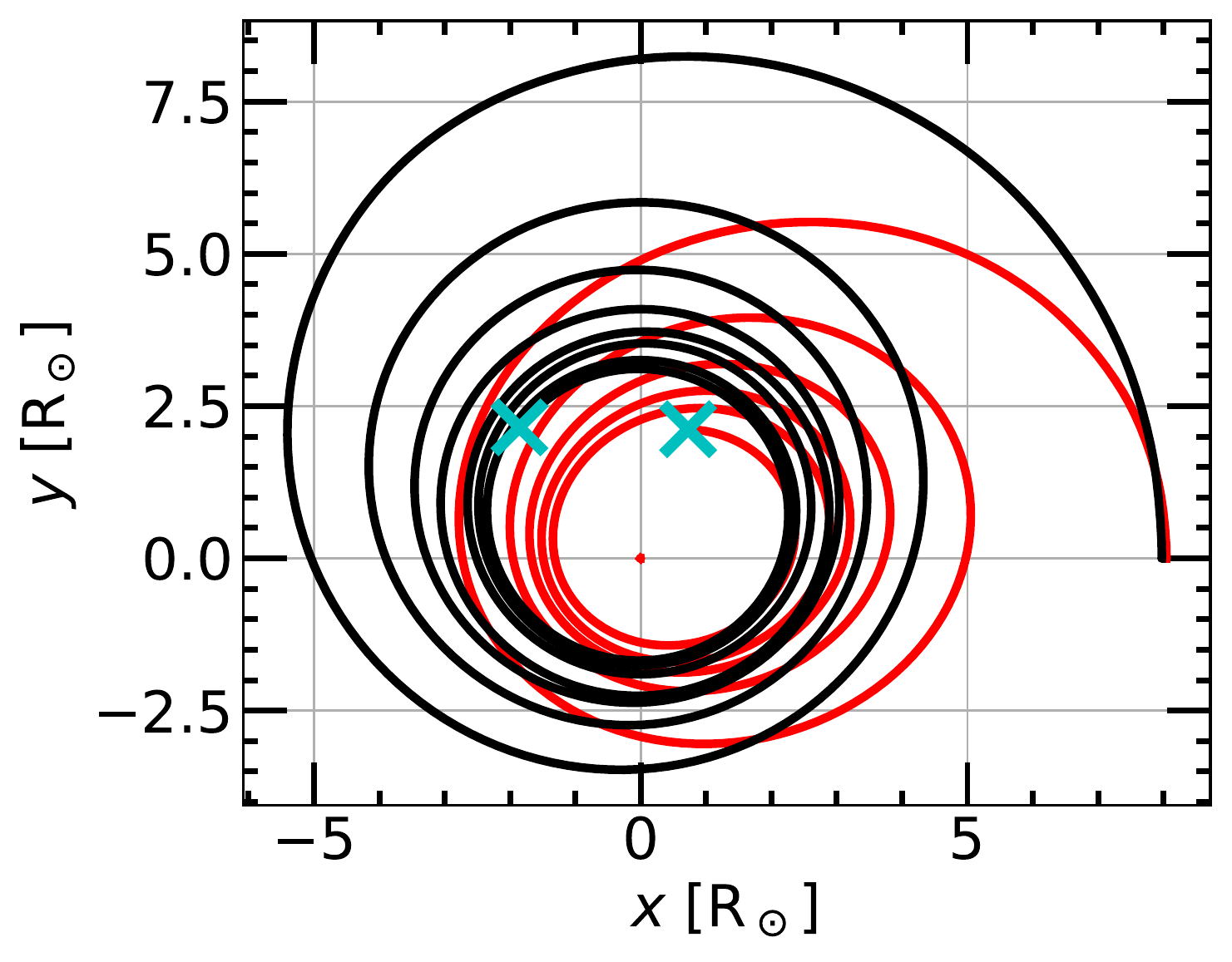}
\plotone{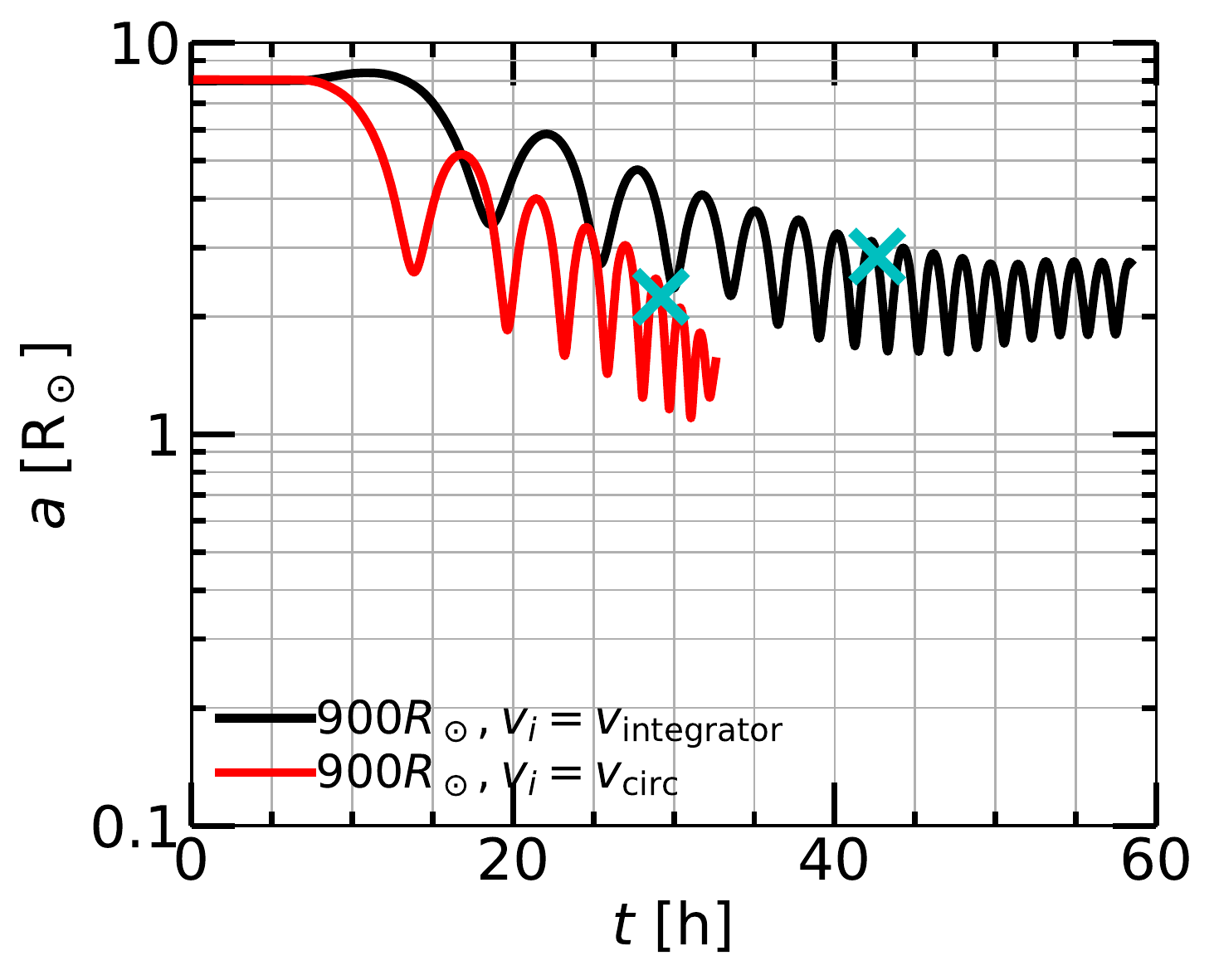}
\caption{
Trajectory and orbital separation for two $900 R_\odot$ simulations with different initial velocities, $v_i=v_{\rm integrator}$ (black) and $v_i=v_{\rm circ}$ (red).
Cyan `X's mark the time at which the envelope is completely ejected (see Figure~\ref{fig:f_ej}).
Left panel: trajectory. Black/red line is secondary (NS), red dot is center of mass of primary (donor star). In this panel we do not plot the orbital evolution past the point of complete envelope ejection to avoid many overlapping lines.
Right panel: separation $a(t)$ between center of mass of primary and position of point mass secondary vs. time.
\label{fig:trajectory_and_a_vs_t}
}
\end{figure*}

The 1st row of Figure~\ref{fig:9panel} shows the logarithm of gas density.
In the first panel, one can see the density shells that are progressively disturbed as the secondary sweeps through the primary's envelope.
At late times, the structure is quite disturbed and resembles a differentially rotating disk, though at even later times, the secondary stalls at its final orbital separation (see Figure~\ref{fig:trajectory_and_a_vs_t}).

The 2nd row of Figure~\ref{fig:9panel} shows the ratio of absolute magnitude of velocity to the local escape velocity for each cell, $|v|/v_{\rm esc, local}$. 
Pink corresponds to gas that is bound to the system (values $<1$) and green corresponds to gas that is not bound to the system (values $>1$). The blue isosurface is at $|v|/v_{\rm esc, local}=1$. 
At late times (after a few orbits of the secondary), nearly all of the envelope is at $|v|>v_{\rm esc, local}$ and is gravitationally unbound from the star. 
Some of the envelope material is shocked to $|v|\gtrsim 6 v_{\rm esc, local}$ on the leading edge of a spherically expanding shell.
One can see the envelope being shocked and swept preferentially outwards as the secondary orbits the center of mass of the primary.
As the secondary moves through the envelope of the primary, it acts as a local diffusive source term, giving surrounding material roughly outward velocities.
We also analyzed the velocity vectors of each grid cell in the envelope, and found that they are nearly all pointed outwards from the core as a result of the secondary's repeated passages.

The 3rd row of Figure~\ref{fig:9panel} shows specific energy, $\varepsilon = \varepsilon_{\rm kin} + \varepsilon_{\rm grav}$ (the sum of specific kinetic and potential energy; internal energy is not included).
Pink-purple corresponds to bound ($\varepsilon<0$) and yellow corresponds to unbound ($\varepsilon>0$). There is a blue isosurface at $\varepsilon=0$.
At early times, the binding energy of most cells is negative. At late times, nearly all of the material in the box (except for the surviving core) has positive energy.
The core and secondary have separate Roche lobes, and the equipotential surface of $\varepsilon=0$ (blue isosurface) is confined to a small region around the core. The size of this region decreases with time and number of orbits until the secondary stalls at its final orbital separation.
This qualitatively shows envelope ejection.

In the bottom left panel of Figure~\ref{fig:9panel} one can see a crescent-shaped sliver of material on the left hand side of the panel that becomes unbound.
This is due to the change in the mass distribution interior to the radius of this sliver (initially at $r\approx 10 R_\odot$) caused by the secondary sweeping out mass on the right hand side. The gravitational potential due to the enclosed mass changes and this sliver of material becomes unbound due to gravitational effects (acting nearly instantaneously) as opposed to hydrodynamical effects (acting on the dynamical time).

\begin{figure*}[htp!]
\epsscale{1.17}
\plotone{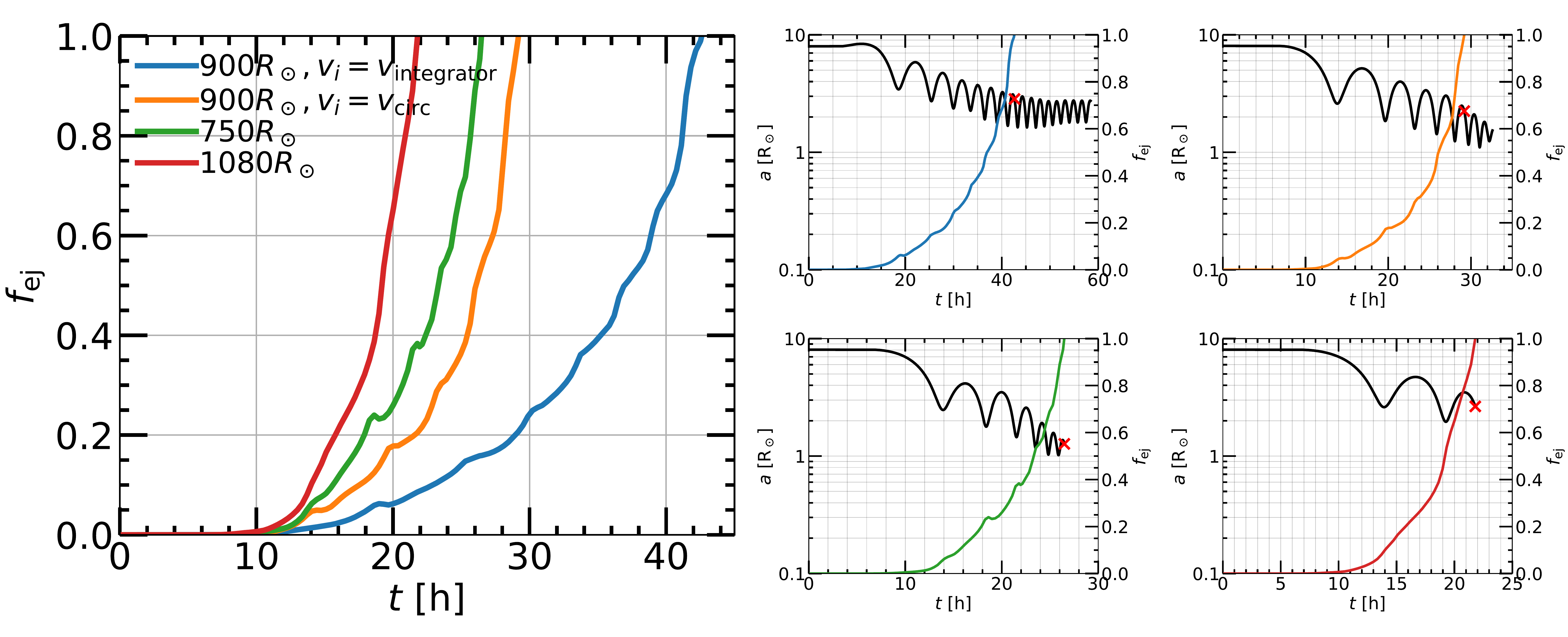}
\caption{
Fraction of envelope mass ejected $f_{\rm ej}(t)$ vs. time for the four main simulations (see \S\ref{sec:numerical_convergence} for additional numerical convergence simulations).
Left panel: all the $f_{\rm ej}(t)$ on one plot.
Right panels: orbital separation $a(t)$ vs. time (black lines) for each simulation with $f_{\rm ej}(t)$ (colored lines; colors match left panel) overlaid.
Red `X's mark the time at which the envelope outside the helium core is completely ejected.
\label{fig:f_ej}
}
\end{figure*}

We now discuss the orbital parameters of the two objects and in particular the position of the secondary as it orbits the center of mass of the primary donor star.
The left panel of Figure~\ref{fig:trajectory_and_a_vs_t} shows the trajectory of the secondary as a function of time in two of our simulation box coordinates ($x$ and $y$; because our simulation is symmetric along the $z$-axis, there is little evolution of the center of masses in $z$). 
We show the evolution for two $900 R_\odot$ simulations with different initial velocities, $v_i=v_{\rm integrator}$ (initial velocity vector informed by the 2D kinematics study) and $v_i=v_{\rm circ}$ (circular initial velocity vector).
Cyan `X's mark the time at which the envelope is completely ejected (see Figure~\ref{fig:f_ej}).

The right panel of Figure~\ref{fig:trajectory_and_a_vs_t} shows the orbital separation $a(t)$ between the center of mass of the primary and the position of the point mass secondary vs. time for the same two simulations. 
The orbital evolution of our other simulations is shown in the right panels of Figure~\ref{fig:f_ej} and in \S\ref{sec:numerical_convergence}.
Oscillations are observed in the orbital separation vs. time, reflecting an eccentric orbit, and in both cases the secondary is still on an eccentric orbit when the envelope is completely ejected.
The initial velocity vector informed by the 2D kinematics study (see Section~\ref{sec:methods}) occurs near the pericenter of an eccentrically inspiraling orbit and it is thus higher energy (larger velocity) than the circular initial velocity.
This leads to a relatively larger orbital separation, including a larger final separation.

\begin{deluxetable*}{l | c c c c c c}
\tablecaption{
Summary of simulations and results. 
$R_\star$ is the radius of the red giant primary, $v_{\rm i}$ is the initial velocity vector of the NS secondary, $\Delta X_{\rm max}$ is the simulation box size on one side ($\Delta X_{\rm max}=40 R_\odot$ unless otherwise indicated), $a_{\rm i}$ is the initial orbital separation of the NS secondary ($a_{\rm i}=8 R_\odot$ unless otherwise indicated), and $\dagger$ indicates numerical convergence simulations, discussed in \S\ref{sec:numerical_convergence}.
$a_{\rm f}^\ast$ is the final orbital separation determined by requiring ejection of all of the envelope material outside the helium core, $a_{\rm f}^{\ast\ast}$ is the final orbital separation determined by requiring ejection of the envelope material outside the secondary, $f_{\rm ej}^\ast$ is the fraction of envelope mass outside the helium core ejected, $f_{\rm ej}^{\ast\ast}$ is the fraction of envelope mass outside the secondary ejected, $\alpha_{\rm CE}^\ast$ is the $\alpha_{\rm CE}$-equivalent efficiency using $a_{\rm f}^\ast$, and $\alpha_{\rm CE}^{\ast\ast}$ is the $\alpha_{\rm CE}$-equivalent efficiency using $a_{\rm f}^{\ast\ast}$.
\label{tab:table1}}
\tablehead{
\colhead{Simulation} & \colhead{$a_{\rm f}^\ast~[R_\odot]$} & \colhead{$a_{\rm f}^{\ast\ast}~[R_\odot]$} & \colhead{$f_{\rm ej}^\ast$} & \colhead{$f_{\rm ej}^{\ast\ast}$} & \colhead{$\alpha_{\rm CE}^\ast$} & \colhead{$\alpha_{\rm CE}^{\ast\ast}$} 
}
\startdata
$R_\star=750 R_\odot, v_{\rm i}=v_{\rm circ}$ & $\approx 1.3$ & $\approx 2.5$ & 1.0 & 1.0 & $\approx 2.65$ & $\approx 0.38$ \\
$R_\star=900 R_\odot, v_{\rm i}=v_{\rm integrator}$ & $\approx 2.4$ & $\approx 3.5$ & 1.0 & 1.0 & $\approx 1.23$ & $\approx 0.15$ \\
($R_\star=900 R_\odot, v_{\rm i}=v_{\rm int.}, \Delta X_{\rm max}=100 R_\odot$) $\dagger$ & $\approx 2.8$ & $\approx 5.0$ & 1.0 & 1.0 & $\approx 1.43$ & $\approx 0.20$ \\
$R_\star=900 R_\odot, v_{\rm i}=v_{\rm circ}$ & $\approx 1.9$ & $\approx 5.1$ & 1.0 & 1.0 & $\approx 0.97$ & $\approx 0.21$ \\
($R_\star=900 R_\odot, v_{\rm i}=v_{\rm circ}, a_{\rm i}=16 R_\odot$) $\dagger$ & $\approx 3.2$ & $\approx 3.6$ & 1.0 & 1.0 & $\approx 1.65$ & $\approx 0.15$ \\
$R_\star=1080 R_\odot, v_{\rm i}=v_{\rm circ}$ & $\approx 2.7$ & $\approx 3.4$ & 1.0 & 1.0 & $\approx 0.37$ & $\approx 0.11$
\enddata
\end{deluxetable*}

Table~\ref{tab:table1} lists the final orbital separations for all of the simulations studied in this work.
The final separations determined by requiring all of the material outside the helium core to be ejected (see below) range from $a_{\rm f}^\ast \approx 1.3$--$2.8 R_\odot$.
Because there is non-negligible eccentricity at the time of complete envelope ejection, rather than quoting the exact orbital separation at which $f_{\rm ej}^\ast=1$ (i.e., the exact location of the `X's), we quote the average of the local maximum and minimum immediately preceding and following this time.
The final separations determined by considering only the envelope material outside the current orbit of the secondary (see \S\ref{sec:old_fig4}) range from $a_{\rm f}^{\ast\ast} \approx 2.5$--$5.1 R_\odot$.
After supernova kicks \citep[calculated with 5000 randomly oriented kicks and a kick magnitude drawn from a Maxwellian distribution with 1D RMS $\sigma=265$ km/s;][]{2005MNRAS.360..974H}, we calculate that a significant fraction of these systems will form binary neutron stars that merge within a Hubble time. 
See \S\ref{sec:merger_time_distribution} for the numbers and details of this calculation.

We estimate the $\alpha_{\rm CE}$-equivalent efficiency
\begin{equation}\label{eq:alpha_CE}
\alpha = \frac{E_{\rm bind, env}}{\Delta E_{\rm orb}},
\end{equation}
where $\Delta E_{\rm orb}$ is defined in Eq.~\ref{eq:Delta_E_orb} (with $M_{\rm secondary}=1.4 M_\odot$, $M_{\rm donor}=M_\star$, $a_{\rm f}=a_{\rm f}^\ast$, and $a_{\rm i}=R_\star$\footnote{The result depends very weakly on $a_{\rm i}$ for the large $R_\star$ models we study. Eq.~\ref{eq:alpha_CE} can be approximated by $\alpha \approx 2 a_{\rm f} E_{\rm bind} / G M_{\rm core}M_{\rm secondary}$, but we do not use this approximation.}) and $E_{\rm bind, env}$ is defined in Eq.~\ref{eq:E_bind}. 
In the calculation of $\alpha_{\rm CE}^\ast$, we take $E_{\rm bind, env}$ at the radius of the helium core, not at the radius of the final orbital separation; i.e., we consider the entire envelope mass. We do this because in the $a_{\rm f}^\ast$ criterion (see below) the entire envelope is ejected. 
So, to be clear, we are taking $\Delta E_{\rm orb}$ at some larger radius and $E_{\rm bind, env}$ at the core radius.
As an example, for the ($R_\star=900 R_\odot, v_{\rm i}=v_{\rm integrator}$) run, for $\Delta E_{\rm orb}$ we take $M(r<a_{\rm f}^\ast)=4.6 M_\odot$, $M_{\rm secondary}=1.4 M_\odot$, $M_{\rm donor}=9.9 M_\odot$, $a_{\rm f}^\ast=2.4 R_\odot$, $a_{\rm i}=900 R_\odot$, and for $E_{\rm bind, env}$ we take a binding energy at $R_{\rm core}=0.36 R_\odot$ of $E_{\rm bind, env} \approx 6.2\times 10^{48}$ erg (see e.g. Figure~\ref{fig:Rosa_profiles})\footnote{We note that this is the binding energy calculated from the MESA model. The actual binding energy at the radius of the helium core after the numerical relaxation scheme in FLASH is a few percent lower, as the helium core has expanded slightly during relaxation. Accounting for this would lower the $\alpha_{\rm CE}^\ast$ values quoted in Table~\ref{tab:table1} by a few percent.}. 
This gives an $\alpha_{\rm CE}$-equivalent efficiency of $\alpha_{\rm CE}^\ast \approx 1.23$.
Note that if we had taken $E_{\rm bind, env}$ at the Roche radius of the core, $R_{\rm Roche, core}\approx 1.2 R_\odot$ for $a_{\rm f}^\ast=2.4 R_\odot$ (this is mixing and matching criteria; see below), this would have given $\alpha_{\rm CE} \approx 0.17$.
Values for the other simulations are listed in Table~\ref{tab:table1}, and the range of values using this criterion is $\alpha_{\rm CE}^\ast \approx 0.4$--$2.7$.
Thus, the $\alpha_{\rm CE}$ ranges from $<1$ to $>1$ using this criterion, depending on the progenitor and initial conditions.
We also calculate the $\alpha_{\rm CE}^{\ast\ast}$ obtained using $a_{\rm f}^{\ast\ast}$, i.e., using the criterion that the total energy of the envelope outside the secondary is positive (unbound) (see \S\ref{sec:old_fig4}).
Here, we take $E_{\rm bind, env}$ at the radius of the final orbital separation, not at the radius of the helium core; i.e., we only consider the envelope mass outside the secondary. We do this because in the $a_{\rm f}^{\ast\ast}$ criterion it is only the envelope outside the secondary that is ejected. 
These values are also listed in Table~\ref{tab:table1}, and the range of values using this criterion is $\alpha_{\rm CE}^{\ast\ast} \approx 0.1$--$0.4$.
Thus, the $\alpha_{\rm CE}$ is small using this criterion.
We note that $\alpha_{\rm CE}^\ast$ uses perhaps the strictest conceivable criterion and $\alpha_{\rm CE}^{\ast\ast}$ uses a weaker criterion. 
One can imagine other criteria. For example, we could require that the secondary eject the material outside the (time-dependent) Roche radius of the core in order to determine the final separation and binding energy.
This criterion would result in $\alpha_{\rm CE}$ values in between $\alpha_{\rm CE}^\ast$ and $\alpha_{\rm CE}^{\ast\ast}$.
We also note that the range of values quoted above may be specific for these extended progenitors.

We calculate the fraction of envelope mass unbound (ejected) as a function of time as
\begin{equation}
    f_{\rm ej}(t) = \frac{M_{\rm unbound}(t) + M_{\rm exited~box}(t)}{M_{\rm env,~initial}}.
\end{equation}
$M_{\rm unbound}$ is the sum of the masses of all the grid cells that have positive energy.
We only consider kinetic energy and gravitational potential energy in determining whether a grid cell is unbound, $\varepsilon_{\rm kin} + \varepsilon_{\rm grav} > 0$.\footnote{This corresponds to the ``kinetic energy criterion'' of \citet{2022A&A...660L...8O}.}
$M_{\rm exited~box}$ is the amount of mass that has exited the simulation box.
We assume that material that has exited the simulation box is unbound. 
This is a reasonable assumption, as an examination of the energy of material that is exiting the box at late times reveals that $\approx 100$\% of it is already unbound.
Finally, 
\begin{equation}
M_{\rm env,~initial} = M_{\rm total~in~box,~initial} - M_{\rm helium~core,~initial}.
\end{equation}
Thus, we consider all material that is outside the helium core to be envelope material.
We note that this is a strict condition for envelope ejection\footnote{This is as compared to, e.g., \citet{2022MNRAS.512.5462L} or \citet{2021arXiv211112112M}, who only consider material outside $R_{\rm cut}=20 R_\odot$. However, these authors simulate the entire envelope exterior to this point and to do excise it exterior to $10 R_\odot$ as we do in this work, so their total envelope masses are much higher than ours.}; the radius of the helium core is only $R_{\rm core}=0.31 R_\odot$, $0.36 R_\odot$, and $0.8 R_\odot$ for the $750 R_\odot, 900 R_\odot$, and $1080 R_\odot$ models respectively.
We also consider a condition for envelope ejection that only considers the material outside the current orbit of the secondary in \S\ref{sec:old_fig4}.

Figure~\ref{fig:f_ej} shows $f_{\rm ej}(t)$ vs. time for the four main 3D hydrodynamics simulations in this paper (see \S\ref{sec:numerical_convergence} for additional numerical convergence simulations).
The left panel shows $f_{\rm ej}(t)$ for all of the simulations and the right panels show the orbital separation $a(t)$ vs. time for each simulation with $f_{\rm ej}(t)$ overlaid.
The fraction of unbound material increases with time and results in complete envelope ejection for all of the models.
The envelope of the $1080 R_\odot$ model is ejected earliest, followed by the $750 R_\odot$ model, the $900 R_\odot$ model with circular initial velocity vector, and finally the $900 R_\odot$ model with initial velocity vector informed by the 2D kinematics study.
The time of complete envelope ejection is a function of the initial binding energy profile of the envelope (see e.g. Figure~\ref{fig:Rosa_profiles}) and the orbital separation as a function of time.
The $1080 R_\odot$ model's initial envelope is the least bound and so it makes sense that it is the easiest to eject. 
The $750 R_\odot$ model's envelope is more bound than the $900 R_\odot$ model's, yet it is ejected slightly earlier. This may be because the secondary inspirals more deeply at an earlier time in the $750 R_\odot$ simulation, due to the higher density gas it encounters, allowing the secondary to deposit more orbital energy at earlier times.
The ($900 R_\odot, v_{\rm i}=v_{\rm integrator}$) simulation takes considerably longer to eject the envelope that the $v_{\rm i}=v_{\rm circ}$ simulation. This may also be due to the effect of orbital separation, as the secondary remains at larger orbital separations for $v_{\rm i}=v_{\rm integrator}$.

The envelope is successfully ejected for all of our simulated models, which span the range of stellar age and radii in which dynamical CE ejection is predicted to be possible for an initially $12 M_\odot$ primary.
We note that we do not include internal or recombination energy in the calculation of the envelope energy (which some contemporary studies do, and which is a positive quantity that helps with envelope ejection; however (see discussion in \S\ref{sec:old_fig4}), these energies are relatively small compared to the envelope energy for our simulations), only kinetic and gravitational potential energy.
We note that $f_{\rm ej}$ increases to $>1$ at late times. This is due to the NS ejecting material that has leaked from the highly concentrated core; this is studied in detail in \S\ref{sec:numerical_convergence}, which also contains several numerical convergence tests.
We additionally verify that the secondary clears the material interior to its orbit and exterior to the Roche radius of the core, which is necessary to successfully stall at a final orbital separation (see Figure~\ref{fig:mass_enclosed} for details).

\begin{figure*}[htp!]
\epsscale{1.12}
\plotone{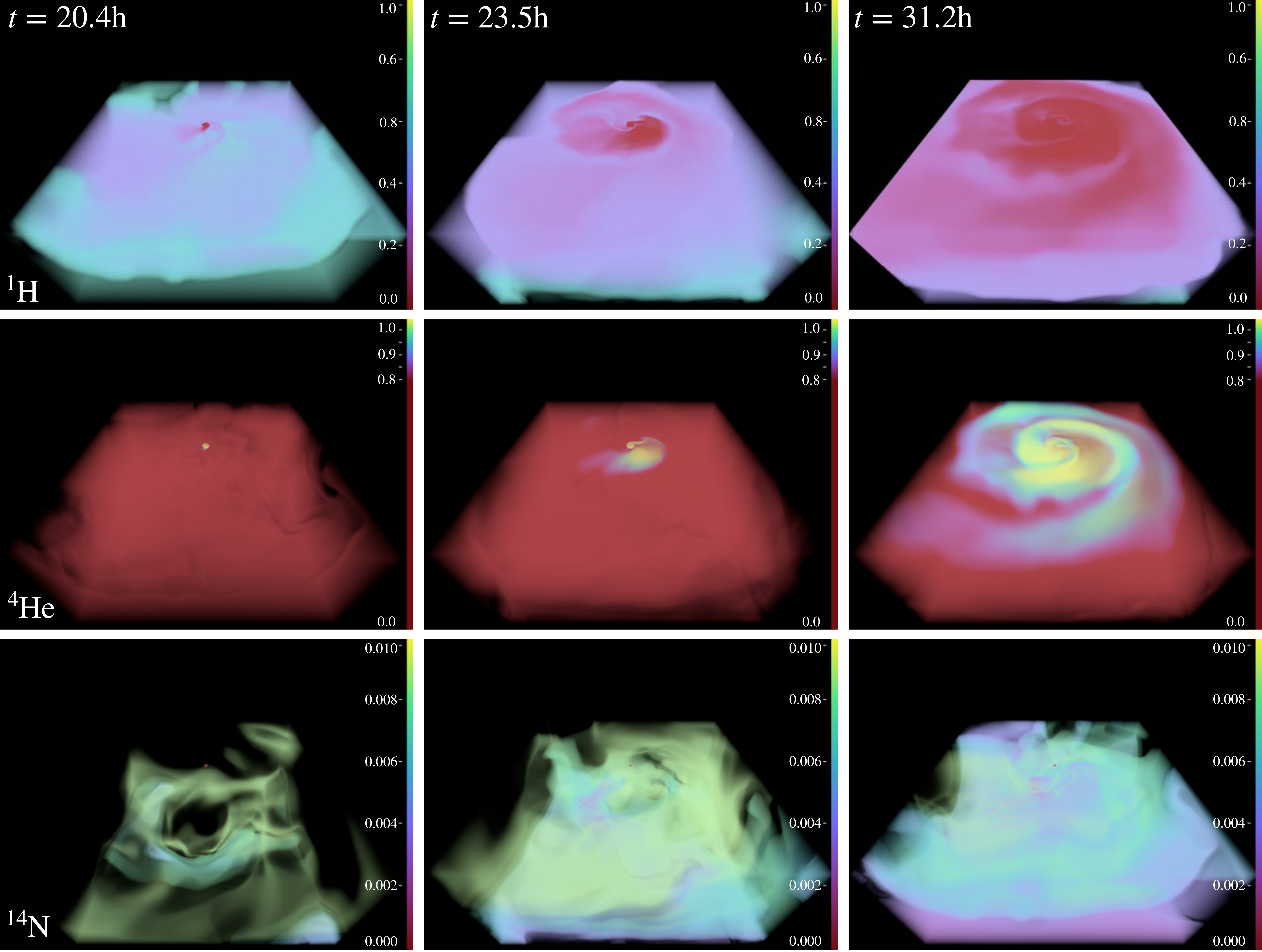}
\caption{
3D renderings of mass fraction of hydrogen, helium, and nitrogen as a function of time, for the ($900 R_\odot, v_i=v_{\rm circ}$) run. 
Results for other simulations are qualitatively similar.
The colormap is the same for each element but note that the scale changes for each element in order to highlight the structure.
For hydrogen and helium, mass fractions range from 0 (dark purple) to 1 (light yellow).
For nitrogen, mass fractions range from 0 (dark purple) to 0.01 (light yellow).
\label{fig:composition_3d}
}
\end{figure*}

We now briefly discuss the chemical abundance of the ejecta.
Figure~\ref{fig:composition_3d} shows 3D renderings of the mass fraction of hydrogen, helium, and nitrogen, at three times for the ($900 R_\odot, v_i=v_{\rm circ}$) run. 
Hydrogen, helium, and nitrogen mix with the outer debris; results for other runs are qualitatively similar. 
See \S\ref{sec:MESA_profiles} for 1D composition profiles of these elements at the beginning of the simulation.
All composition data is available upon request. 
The most notable result is that in our simulations the hydrogen envelope is completely ejected at late times; this implies that hydrogen will not be visible in the spectrum of the surviving stripped star as the surface hydrogen mass fraction is comparable to the core's hydrogen abundance, roughly $\lesssim 10^{-3}$.

\subsection{Recombination Transient}
The expanding hydrogen bubble is observable as a hydrogen recombination transient (or a luminous red nova) \citet{2013Sci...339..433I}.
We use Eqns. (A1), (A2), and (A3) of \citet{2017ApJ...835..282M}, based on \citet{2013Sci...339..433I}'s application of the analytic  theory of recombination transients \citep[e.g.,][]{1993ApJ...414..712P,2009ApJ...703.2205K,2010ApJ...714..155K} to estimate the luminosity, timescale, and total energy of this recombination transient (see \S\ref{sec:hydrogen_recombination_transient} for details of the calculation).
Using $R_{\rm init} \approx 2 R_\odot$ (approximate stalling orbital separation of the secondary across our models), $\Delta M \approx 5 M_\odot$ (the entire mass of the envelope), $v_{\rm ej}\approx 18~{\rm km/s}$ (the velocity at $10 R_\odot$ at the end of our simulation), $\kappa \approx 0.32~{\rm cm^2~g}$, and $T_{\rm rec} \approx 4500~{\rm K}$, we find
$L_{\rm p} \approx 10^{37}~{\rm erg~s}^{-1}$,
$t_{\rm p} \approx 274~{\rm d}$, and
$E_{\rm rad,p} \approx 2\times 10^{44}~\rm{erg}$.
The mass of the stripped star is $M_\star \approx  4.5 M_\odot$, radius $R_\star \approx 1 R_\odot$.
We note a stripped star and neutron star are also interesting as a ``living'' gravitational wave source potentially observable with LISA (\citet{2020ApJ...904...56G}; for lower mass systems see also \citet{2004MNRAS.349..181N, 2008AstL...34..620Y, 2020A&A...634A.126W}).
If the system is tight enough, the future evolution is determined by the radiation of GWs and no longer the evolution of the stripped star.
See \S\ref{sec:discussion} for discussion on extensions to our framework to study the remnant in more detail and for longer timescales.

Roughly 10\% of the brightest luminous red novae (LRN) transients, which have been previously associated with stellar mergers and common-envelope ejections, are predicted to occur at some point in binary neutron star forming systems \citep{2018MNRAS.481.4009V,2020MNRAS.492.3229H,2020PASA...37...38V}. LRN have come to be associated with stellar mergers through detailed study of a few landmark events. M31 RV was one of the first LRN to be identified, in 1988, but the light curve of the transient is only captured during the decline \citep[e.g.,][]{1990ApJ...353L..35M}.  The galactic transient V1309 Sco proved essential in establishing the nature of these events as stellar mergers \citep{2010A&A...516A.108M, 2013MNRAS.431L..33N}.
Noteworthy transients arising from relatively massive stars include M31LRN 2015 with a progenitor of $M_\star\approx3$--$5.5M_\odot$ \citep{2017ApJ...835..282M} and M101 OT2015-1 with a progenitor of $M_\star\approx18 M_\odot$ \citep{2017ApJ...834..107B}.

\section{Discussion} \label{sec:discussion}

Here we briefly compare to other work, discuss uncertainties in the 1D stellar modeling, resolution-dependent effects in the 3D hydrodynamics, and comment on future work.

\subsection{Comparison to other work}

Since the first posting of this paper, \citet{2022MNRAS.512.5462L} and \citet{2021arXiv211112112M} presented 3D hydrodynamics simulations of the CE interaction between a massive red supergiant primary and a compact object secondary. 
Both studies simulate the entire red supergiant envelope, but replace the material interior to $R_{\rm cut}=18.5 R_\odot$ or $R_{\rm cut}=20 R_\odot$ (respectively) by a point mass and a small amount of gas matched via a polytropic profile to the original envelope.
This is in contrast to the present work, where we resolve the interior gas and excise the envelope exterior to $10 R_\odot$ (or $20 R_\odot$ for a convergence test).

\citet{2022MNRAS.512.5462L} study a $12 M_\odot$ primary and a $1.26 M_\odot$ NS and $3 M_\odot$ BH in the SPH code \texttt{PHANTOM}. 
The red supergiant model does not include wind mass loss, has a radius of $R_\star = 619 R_\odot$, and has a helium core mass of $2.8 M_\odot$ \citep[compare to our helium core masses of $\gtrsim 4 M_\odot$, motivated by the higher overshoot determined from observations of massive stars by][]{2011A&A...530A.115B}.
For the NS, they do not report a final separation as the NS reaches $a_{\rm f} < 22.8 R_\odot$, at which point the softened regions of the two point masses overlap. Similarly, they report the fraction of envelope ejected as $f > 0.03$--$0.1$ depending on the EOS and the ejection criteria. For the BH, they report a final separation of $a_{\rm f} = 33$--$44 R_\odot$ and a fraction of envelope ejected of $f = 0.16$--$1$ depending on the EOS, ejection criteria, and resolution.
They find that successively including radiation energy and recombination energy leads to higher fractions of envelope ejected and larger final separations than an ideal gas EOS.
They additionally find that helium recombination is more important than hydrogen recombination, as the latter takes place in material that is already unbound.

\citet{2021arXiv211112112M} study an initially $10 M_\odot$ primary and a $1.4 M_\odot$ NS and $5 M_\odot$ BH in the moving-mesh code \texttt{AREPO}.
The red supergiant model used as input to the 3D hydrodynamics is $9.4 M_\odot$, has a radius of $R_\star=395 R_\odot$, and has a helium core mass of $2.74 M_\odot$.
They report a final separation for the NS of $a_{\rm f}=15 R_\odot$, which is still decreasing slightly at $0.3 R_\odot$/yr (but the decay rate itself is also decreasing), and nearly complete envelope ejection with an associated $\alpha_{\rm CE}=0.61$.
This orbital separation will not lead to a merger via gravitational waves in a Hubble time; thus they conclude that favorable SN kicks are required for mergers.
Their $\alpha_{\rm CE}$-equivalent efficiency is similar to what we find, particularly for our smallest radius primary (which is still larger than their primary).
However, in contrast to this work, they find that the dynamical plunge-in terminates at an orbital separation larger than we use for our initial conditions.\footnote{Note that we do perform an additional convergence simulation with $a_{\rm i}=16 R_\odot$, but to effectively test their finding of $a_{\rm f}=15 R_\odot$ in our setup we would need to use, e.g., $a_{\rm i}=32 R_\odot$ or larger (in addition to addressing differences in the red supergiant model and the numerical methods).}
Because the final separation is within $R_{\rm cut}$, there is a concern that envelope material that should contribute to the drag force and further shrink the orbit is missing in this region. 
However, they run a numerical convergence test with $R_{\rm cut}=10 R_\odot$ and $R_{\rm cut}=40 R_\odot$ and find that the average final separation is converged for $R_{\rm cut}=20 R_\odot$.
It should be noted that since the gravitational softening length is adaptively reduced when the point masses approach each other, their approximate treatment of gravity within $R_{\rm cut}$ is not responsible for the large final separation.

In 1D, \citet{2019ApJ...883L..45F} also study BNS formation through the CE phase. These authors study a different (though comparable) 
\texttt{MESA} model to ours (see Figure~\ref{fig:MESA_profiles}) and thus a direct comparison is not possible.
We note our 1D formalism predicts that the model studied by \citet{2019ApJ...883L..45F} is in a boundary region where the outcome of CE evolution is unclear. 
\citet{2019ApJ...883L..45F} find a final orbital separation of $a_{\rm f} \approx 3.3$--$5.7 R_\odot$, and a high $\alpha_{\rm CE}$-equivalent efficiency of $\approx 5$, though we note that this study finds envelope ejection in the self-regulated regime and $\alpha_{\rm CE}$ is calculated after a mass-transfer phase which occurs after the envelope is ejected.
In contrast, we study CE ejection in the dynamical regime, and find that for all of the models we simulate in 3D hydrodynamics, the envelope is ejected in the dynamical regime.
After envelope ejection, in our simulations, we expect a stable mass transfer phase to occur between the surviving core and the NS \citep[as in][]{2019ApJ...883L..45F}, which will further alter the separation before the supernova takes place. 
While it is valuable to model the CE evolution from start to finish, the 1D treatment that is necessary to facilitate this has inherent limitations. For example,  \citet{2019ApJ...883L..45F} assume complete and instantaneous spherically symmetric sharing of orbital energy with the envelope. This is a nonphysical assumption that can only be addressed by 3D hydrodynamics.

There has been five decades of work on the CE phase \citep[see e.g.,][]{2013A&ARv..21...59I}, and there is an extensive literature on CE ejection.
We briefly compare to other work that does not address CE ejection leading to a binary neutron star system (i.e., with lower mass ratios and smaller stellar radii) below.
Besides the stars studied, the main methodological difference with other work is that we excise the outer envelope of the primary star that contains $<0.1\%$ of the total binding energy, trimming the star to $10 R_\odot$ (or $20 R_\odot$ for a convergence test), and the initial conditions of our 3D hydrodynamics simulations are informed using an adjusted 1D energy formalism and a 2D kinematics study (see \S\ref{sec:methods}).
Whereas some contemporary studies replace the core of the star with a point mass (see e.g. above), this technique allows us to fully resolve the gas in the core of the star, allowing for a more realistic treatment of the inspiral and material interior to the secondary's location as it stalls at a final orbital separation.
However, this means that we do not simulate the entire star in an ``ab-initio'' way.

Generally, studies at lower mass ratios have had difficulty ejecting the envelope in the course of the 3D simulation.
The fact that the envelope is successfully ejected in our study, without including internal or recombination energy (which is claimed to be essential to CE ejection in some contemporary work at lower masses; see below),
is likely due to the fact that we study an evolved $12 M_\odot$ red supergiant primary; thus, the secondary encounters a very different density profile during its inspiral that the density profiles in the works listed below.
\citet{1998ApJ...500..909S} find 23-31\% envelope ejection in simulations with 3$M_\odot$ and 5$M_\odot$ AGB primaries.
\citet{2016MNRAS.455.3511S} find 25\% envelope ejection with a 3.5$M_\odot$ AGB primary.
\citet{2020A&A...644A..60S} find $<$20\% envelope ejection when not accounting for recombination energy, and complete envelope ejection when including recombination energy, for a 1$M_\odot$, 174$R_\odot$ early-AGB star with companions of different masses.
\citet{2020MNRAS.495.4028C} find an envelope unbinding rate of 0.1–0.2$M_\odot~{\rm yr^{-1}}$, implying envelope unbinding in $<$10 yr, for a 1.8$M_\odot$, 122$R_\odot$ AGB primary  with 1$M_\odot$ secondary.
\citet{2020A&A...642A..97K} find complete envelope ejection if recombination energy is included for a $1M_\odot$, $173 R_\odot$ tip-of-the-RGB primary with $0.01$--$0.08 M_\odot$ secondaries.
Note that in determining envelope ejection, what consists of the ``envelope'' is different across studies {\citep[for a discussion of this point, see also][]{2001A&A...369..170T}}; e.g., it is material outside $R_{\rm cut}$ for studies using a point mass for the core of the giant star, or material outside the Roche radius of the core in other studies, or all material outside the helium core or alternatively outside the secondary in the two criteria in our study.

\subsection{Uncertainties due to prior evolution}

There are four main disclaimers to our analysis, and indeed to our initial stellar models in general:

First, our model of the $12 M_\odot$ donor was evolved as a single star. However, for the progenitor system of a BNS merger, the typical scenario includes a stable mass transfer phase before the formation of the NS \citep[e.g.,][]{2017ApJ...846..170T}. Therefore, the donor star at the CE phase is the initially less massive star which has possibly accreted mass from the NS progenitor and survived the passage of the supernova shock. While the latter has only a moderate effect on the stellar structure \citep[e.g.,][]{2018ApJ...864..119H}, the phase of stable mass transfer can lead to high rotation \citep[e.g.,][]{1981A&A....99..126H, 2007A&A...465L..29C, 2013ApJ...764..166D}, chemical pollution with He \citep[e.g.,][]{1993ASPC...35..207B}, mixing of fresh hydrogen in the core, and other effects such as the development of a convective layer inside the post-MS star \citep[e.g.,][]{2021ApJ...923..277R}. These effects can influence the stellar radius significantly (e.g., rotation can increase the equatorial radius, He-richness can contribute to keep the star more compact), and most importantly change the density profile just outside the core  (i.e., in the domain of our 3D simulation) with the rejuvenation-inducing mixing. 
A second order effect is the impact on the wind mass loss rate (and thus orbital evolution) of the system \citep[e.g.,][]{2017A&A...603A.118R}. While these require further investigation, our models provide a proof-of-concept of our methods that could be applied to more realistic post-RLOF CE donors.

Second, we do not accurately know the distribution of separations that systems have at the time when star one is a neutron star and the other star is a red supergiant \citep[e.g.,][]{2020MNRAS.498.4705V,2020A&A...638A..39L}. 

Third, in considering the orbital evolution prior to filling the Roche lobe, we use the Jeans approximation for widening as a result of stellar wind mass loss (see \S\ref{sec:forbidden_radii}).
The Jeans approximation may not actually hold for the donor star. The mass loss occurs in the late phases and the systems of interest in this work will be very close to Roche-lobe filling at this stage. We may have wind focusing \citep[e.g.,][]{2007ASPC..372..397M}.  It is possible that the systems shrink instead of widening. In that case, the forbidden region (see \S\ref{sec:forbidden_radii}) might no longer be forbidden.

Fourth, our results depend on how accurate our progenitor models are \citep[e.g.,][]{2016ApJS..227...22F}. These are subject to all of the uncertainties that affect massive star evolution, most notably those related to mass loss \citep[e.g.,][]{2017A&A...603A.118R} and internal mixing \citep[e.g.,][]{2019MNRAS.484.3921D}. These affect the final structure and core mass at the moment of Roche-lobe filling.

\subsection{Numerical resolution}

Our resolution is sufficient to achieve common envelope ejection and stall at a final orbital separation in our simulations. However, there is mass leakage and redistribution from the highly centrally concentrated core ($\rho_c \approx 10^3$--$10^6$ g/cm$^3$) at radii $r \lesssim 0.5 R_\odot$ (see \S\ref{sec:numerical_convergence}). Because it occurs at radii smaller than the position of the secondary, this redistribution of mass should not have an effect on the secondary's orbit (Gauss's law).
The largest numerical effect on the secondary's orbit is the numerical diffusion introduced by the grid (as in any 3D hydrodynamics simulation). This effect decreases with increasing resolution. We discuss this further in \S\ref{sec:numerical_convergence}.

Our \texttt{FLASH} setup uses a cartesian grid, which does not conserve angular momentum $L$ (this happens any time there is rotational motion across a grid cell). This causes the secondary to inspiral more rapidly. 
This is in comparison to explicitly Galilean-invariant codes such as moving-mesh codes. For example, \citet{2016ApJ...816L...9O} quote that $L$ was conserved during their run with an error below 1\%.
Technically, our \texttt{FLASH} setup violates Galilean invariance, as do other conventional grid-based hydrodynamics codes (when altering the background velocity at the same resolution), but as \citet{2010MNRAS.401.2463R} showed, this is a resolution-dependent effect, and $L$ in grid codes approaches perfect conservation at very high resolutions. The non-conservation of $L$ becomes larger with each orbit (the longer the simulation is run). 
Thus, if we have successful CE ejection, which we do, this likely represents a ``lower limit'' of possible CE ejection, because with perfect conservation of $L$ the point mass would orbit more times and have longer to strip and eject the CE.
While our detailed results are resolution-dependent to a certain extent (see \S\ref{sec:numerical_convergence}), the main result of this work---successful CE ejection that can lead to binary neutron star formation for all of the models we study---is robust and will only become stronger at higher resolutions.

\subsection{Future work}

The framework developed in this work can be used to study various binary stellar phenomena. 
First, one can study the large parameter space of systems that can be accurately modeled as a star--point mass interaction, including different mass ratios, primary/donor stars, and metallicities; i.e., one can perform a parameter-space study of CE systems leading to BNSs and BH/NS binaries.
One can also study the long-term evolution by exporting the \texttt{FLASH} simulation back to \texttt{MESA} \citep[this capability was already explored in][]{2020ApJ...901...44W}.

One can accurately calculate and include the effects of accretion onto the neutron star and the associated feedback and energy injection in the envelope. This has not been studied in sufficient detail yet.
\citet{2015ApJ...798L..19M} found that accretion onto the neutron star is suppressed by 1 to 2 orders of magnitude compared to the Hoyle-Lyttleton prediction, and that during the CE phase neutron stars accrete only modest amounts of envelope material, $\lesssim$ 0.1$M_\odot$.
However, \citet{2021ApJ...910L..22H} claim that the energy that accretion liberates via jets can be comparable to the orbital energy.

The astrophysical context provided by a detailed physical understanding of the CE phase allows one to use GW and EM observations of binary neutron star mergers as tools to answer a broader set of questions than the raw GW data alone can answer, for example, on the lives and deaths of stars, the difficult-to-probe physics of the deep interiors of stars, and how nucleosynthesis operates in the Universe. 

In another direction, one can adjust our framework to follow the ejected material in more detail, to inform our understanding of supernovae that interact with material from CE ejections. This may also help to understand some stars in the Galaxy that have interacted with CE material.

In the longer term, one can extend our \texttt{FLASH} setup to initialize two separate \texttt{MESA} stars. This would (in theory) allow one to study the entire parameter space of star-star interactions, leading to both stellar mergers and CE ejections.

\section{Conclusion}\label{sec:conclusion}

The main points of this paper are summarized below. 

\begin{enumerate}

\item We study the dynamical common envelope evolution of an initially 12$M_\sun$ red supergiant star and a 1.4$M_\sun$ neutron star in 3D hydrodynamics.

\item Most earlier studies have focused on low mass stars. This is the first successful 3D hydrodynamics simulation of a high mass progenitor that can result in a binary neutron star that merges within a Hubble time.

\item We excise the outer envelope that contains $<0.1\%$ of the total binding energy, trimming the star to $10 R_\odot$ (or $20 R_\odot$ for a convergence test), our 3D hydrodynamics simulations are informed by an adjusted 1D analytic energy formalism and a 2D kinematics study, and we fully resolve the core of the star to $\lesssim 0.005 R_\odot$, 

\item We study different initial separations where the donor fills its Roche lobe during the first ascent of the giant branch and after the completion of central helium burning.

\item We find complete envelope ejection (without requiring any other energy sources than kinetic and gravitational energy) during the dynamical inspiral for all of the models we study.

\item We find final orbital separations of $a_{\rm f}^\ast \approx 1.3$--$2.8 R_\odot$ (when requiring complete envelope ejection outside the helium core) and $a_{\rm f}^{\ast\ast} \approx 2.5$--$5.1 R_\odot$ (when requiring envelope ejection outside the orbit of the secondary) for the models we study, which span the range of initial separations in which dynamical CE ejection is possible for a $12 M_\odot$ star.
A significant fraction of these systems can form binary neutron stars that merge within a Hubble time.
We find $\alpha_{\rm CE}$-equivalent efficiencies of $\alpha_{\rm CE}^\ast \approx 0.4$--$2.7$ and $\alpha_{\rm CE}^{\ast\ast} \approx 0.1$--$0.4$ for the respective criteria above, but this may be specific for these extended progenitors.

\item The framework developed in this work can be used to study the diversity of common envelope progenitors in 3D hydrodynamics.\\

\end{enumerate}

We thank Jeff Andrews, Andrea Antoni, Robert Fisher, Tassos Fragos, Stephen Justham, Matthias Kruckow, Mike Lau, Dongwook Lee, Morgan Macleod, and Paul Ricker for intellectual contributions.
We thank NVIDIA for helping with visualizations and volume renderings of the simulations.
We acknowledge use of the {\it lux} supercomputer at UCSC, funded by NSF MRI grant AST 1828315, and the HPC facility at the University of Copenhagen, funded by a grant from VILLUM FONDEN (project number 16599). 
The UCSC and NBI team is supported in part by NASA grant NNG17PX03C, NSF grant AST-1911206, AST-1852393, and AST-1615881, the Gordon \& Betty Moore Foundation, the Heising-Simons Foundation, the Danish National Research Foundation (DNRF132), and by a fellowship from the David and Lucile Packard Foundation to R.J.F. 
R.W.E. is supported by the National Science Foundation Graduate Research Fellowship Program (Award \#1339067), the Heising-Simons Foundation, and the Vera Rubin Presidential Chair for Diversity at UCSC. Any opinions, findings, and conclusions or recommendations expressed in this material are those of the authors and do not necessarily reflect the views of the NSF.
S.d.M. and L.v.S. are funded in part by the European Union’s Horizon 2020 research and innovation program from the European Research Council (ERC, Grant agreement No. 715063), and by the Netherlands Organization for Scientific Research (NWO) as part of the Vidi research program BinWaves with project number 639.042.728.
S.C.W. is supported by the National Science Foundation Graduate Research Fellowship Program under Grant No. DGE‐1745301.
Support for this work was provided by NASA through the NASA Hubble Fellowship Program grant \#HST-HF2-51457.001-A awarded by the Space Telescope Science Institute, which is operated by the Association of Universities for Research in Astronomy, Inc., for NASA, under contract NAS5-26555.
\software{\texttt{MESA} \citep{2011ApJS..192....3P, 2013ApJS..208....4P, 2015ApJS..220...15P}, 
\texttt{FLASH} \citep{2000ApJS..131..273F},
\texttt{astropy} \citep{2013A&A...558A..33A,2018AJ....156..123A}, 
\texttt{yt} \citep{2011ApJS..192....9T},
\texttt{matplotlib} \citep{2007CSE.....9...90H},
\texttt{py\_mesa\_reader} \citep{2017zndo....826958W}.
}

\appendix

\section{Detailed time evolution}\label{sec:detailed_time_evolution}

\begin{figure*}[tp!]
\begin{interactive}{animation}{video1-2.mp4}
\plotone{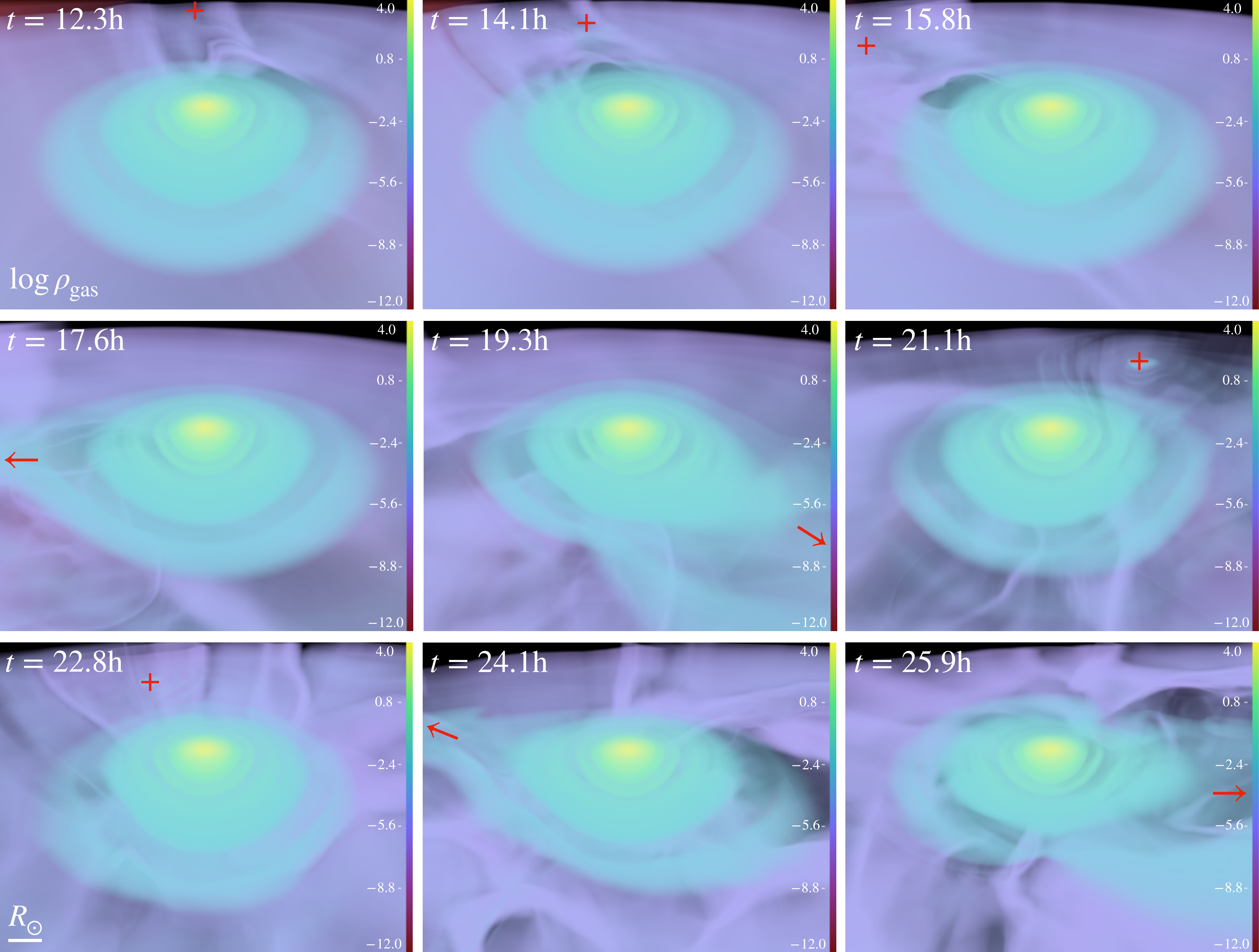}
\end{interactive}
\caption{
Video figure (viewable online). Video shows a 3D rendering of the logarithm of gas density ([g/cm$^3$]) for material near the core of the primary (the domain is of the video here is $x\approx 10 R_\odot$ on a side) during the neutron star's inspiral for the ($900 R_\odot$, $v_i=v_{\rm integrator}$) run. 
Shells correspond to different density isosurfaces; white is highest density, dark purple is lowest density.
Video shows that the neutron star significantly disturbs the density structure of the envelope as it orbits and stalls at a final orbital separation, but that the core of the star remains largely undisturbed.
Position of the neutron star is indicated by the red `+', or `$\rightarrow$' if it is outside the domain.
Videos also available at \url{https://youtube.com/channel/UCShahcfGrj5dOZTTrOEqSOA}.
\label{fig:video1-2}
}
\end{figure*}

Here we show the evolution 
during the neutron star's inspiral for the ($900 R_\odot$, $v_i=v_{\rm integrator}$) run. See Figure~\ref{fig:trajectory_and_a_vs_t} for the trajectory and orbital separation as a function of time (black line).
The animated video Figure~\ref{fig:video1-2} shows a 3D rendering of the material near the core of the primary, from initial inspiral through common envelope ejection and stalling of the neutron star at its final orbital separation. Different shells corresponds to different density isosurfaces.
While the material inside the core of the primary remains relatively undisturbed (as the closest approach of the secondary is $r\approx 2R_\odot$ and the radius of the core is $R_{\rm core} \approx 0.36 R_\odot$), the material outside the core (both interior to and exterior to the orbit of the neutron star) is swept away and cleared with each successive passage of the neutron star.
Red `+' (or `$\rightarrow$' if it is outside the visualization domain) indicates the position of the neutron star.

\begin{figure*}[tp!]
\begin{interactive}{animation}{video2-2.mp4}
\plotone{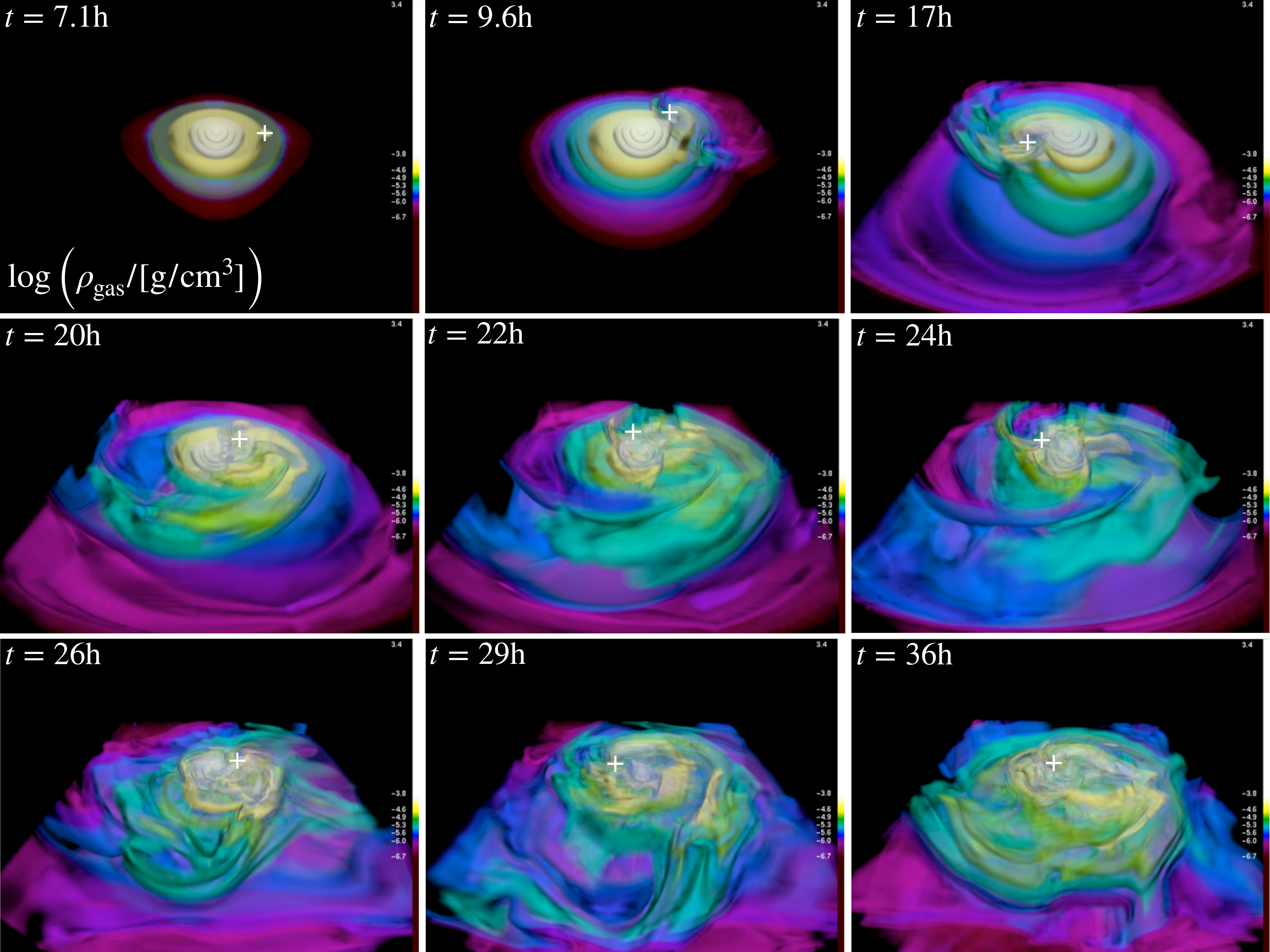}
\end{interactive}
\caption{
Video figure (viewable online). Video shows a 3D rendering of the logarithm of gas density for the full domain (compare to Figure~\ref{fig:video1-2}) during the neutron star's inspiral for the ($900 R_\odot$, $v_i=v_{\rm integrator}$) run. 
Shells correspond to different density isosurfaces; white is highest density, dark purple is lowest density.
Video highlights the severely shocked and disturbed density structure of the outer envelope, which is ejected as the neutron star orbits the giant star.
Position of the neutron star is indicated by the white `+'.
Videos also available at \url{https://youtube.com/channel/UCShahcfGrj5dOZTTrOEqSOA}.
\label{fig:video2-2}
}
\end{figure*}

The animated video Figure~\ref{fig:video2-2} shows a 3D rendering of the material for the entire domain, as opposed to a zoom-in on the material near the core in Figure~\ref{fig:video1-2}.
While in Figure~\ref{fig:video1-2} we saw that the core remained relatively undisturbed and that there was not significantly more material in between the orbit of the neutron star and the core, here the focus is the severely disturbed material in the envelope.
One can see the ``spiral-wave'' feature as the neutron star sweeps out envelope mass with each successive passage in its orbit.
One can also see that some of the higher density material closer to the core is moved outward toward the periphery as the neutron star ejects this material.

\begin{figure*}[tp!]
\begin{interactive}{animation}{video3-2.mp4}
\plotone{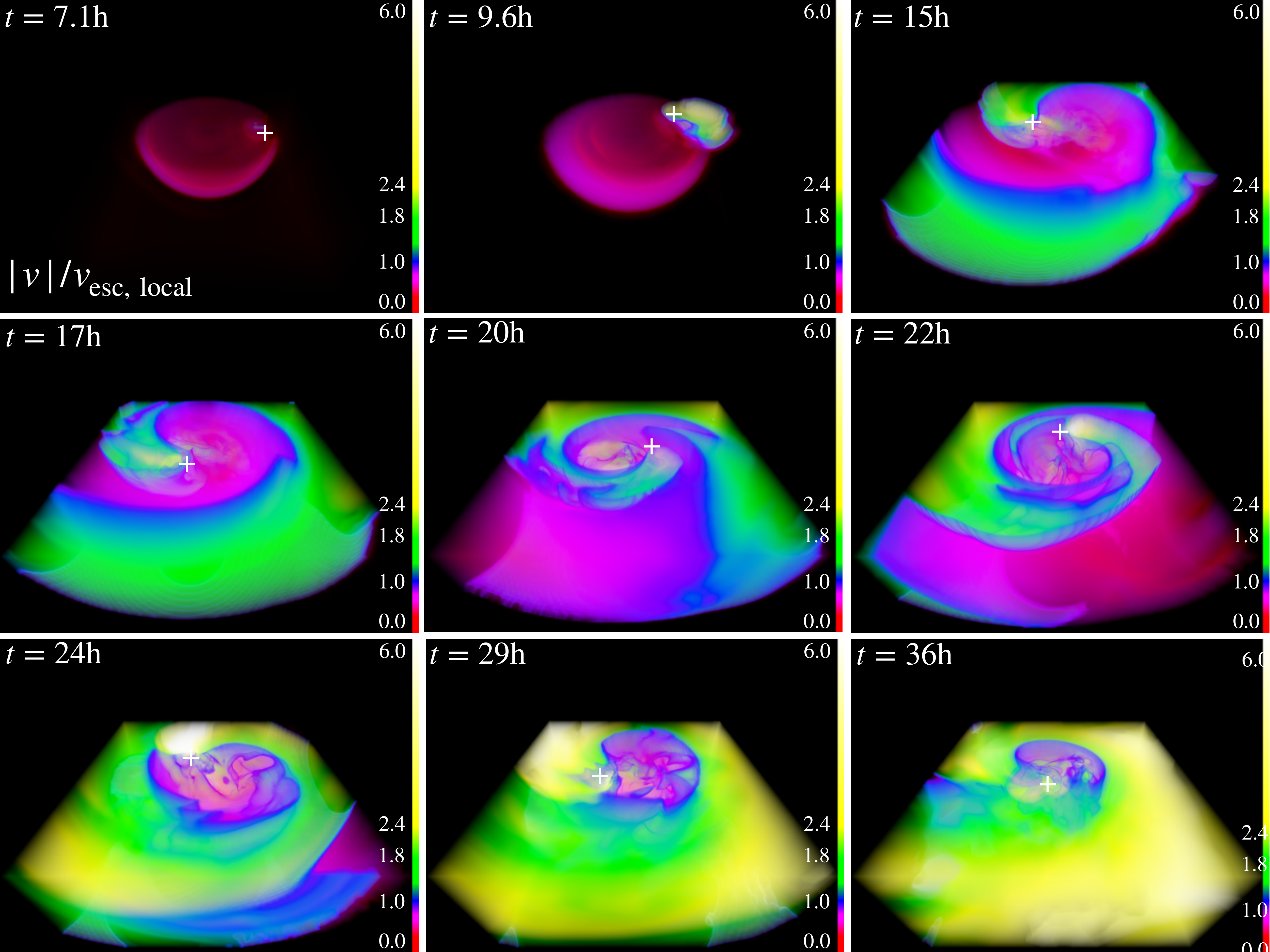}
\end{interactive}
\caption{
Video figure (viewable online). Video shows a 3D rendering of the ratio of velocity magnitude to local escape velocity for the full domain for the ($900 R_\odot$, $v_i=v_{\rm integrator}$) run. 
Blue isosurface is at $|v|/v_{\rm esc, local}=1$, pink-red is $<1$, green-yellow is $>1$.
Position of the neutron star is indicated by the white `+'.
Videos also available at \url{https://youtube.com/channel/UCShahcfGrj5dOZTTrOEqSOA}.
\label{fig:video3-2}
}
\end{figure*}

The animated video Figure~\ref{fig:video3-2} shows a 3D rendering of the ratio of the velocity magnitude to the local escape velocity, $|v|/v_{\rm esc,local}$, as a function of time.
As the neutron star orbits the center of mass of the red supergiant star, it strips off the envelope material outside its orbit, unbinding it and shocking this material to velocities in excess of $6 v_{\rm esc,local}$. These large velocities are an indication of how efficiently the orbital energy of the neutron star is transferred to the energy of the envelope.

\begin{figure*}[tp!]
\begin{interactive}{animation}{video4-2.mp4}
\plotone{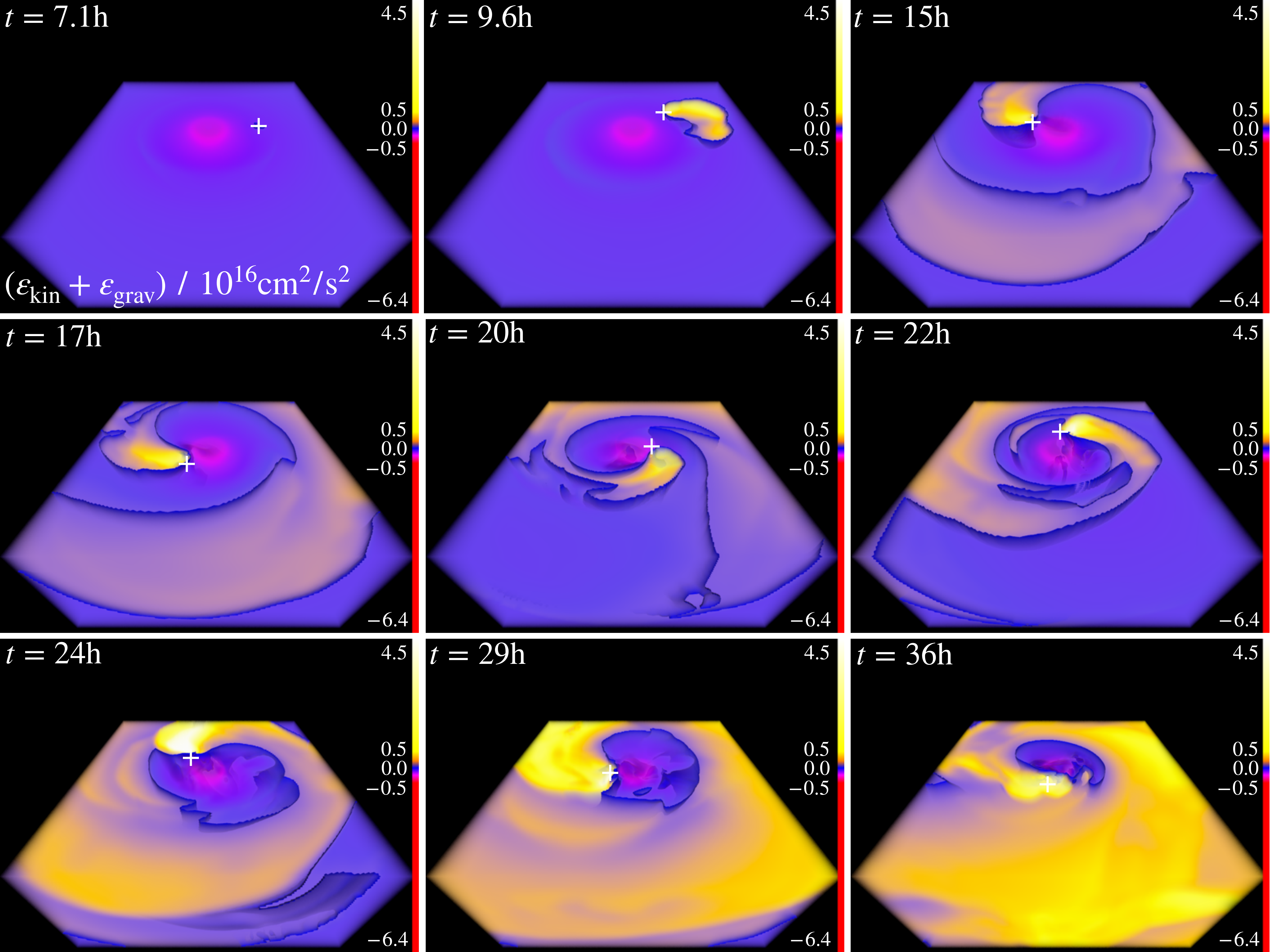}
\end{interactive}
\caption{
Video figure (viewable online). Video shows a 3D rendering of the sum of specific kinetic and potential energy for the full domain for the ($900 R_\odot$, $v_i=v_{\rm integrator}$) run. 
Blue isosurface at $\varepsilon=0$, pink-purple corresponds to bound ($\varepsilon<0$) and yellow corresponds to unbound ($\varepsilon>0$).
Position of the neutron star is indicated by the white `+'.
Videos also available at \url{https://youtube.com/channel/UCShahcfGrj5dOZTTrOEqSOA}.
\label{fig:video4-2}
}
\end{figure*}

The animated video Figure~\ref{fig:video4-2} shows a 3D rendering of the sum of the specific kinetic and potential energy ($\varepsilon = \varepsilon_{\rm kin} + \varepsilon_{\rm grav}$) as a function of time.
There is a blue isosurface at $\varepsilon=0$, pink-purple corresponds to bound material ($\varepsilon<0$), and yellow corresponds to unbound material ($\varepsilon>0$).
As in Figure~\ref{fig:video3-2}, the envelope gains more energy with each orbital passage of the neutron star and becomes progressively more unbound (the colors become a brighter yellow with time).

\section{Hydrogen recombination transient}\label{sec:hydrogen_recombination_transient}

Here we outline the details of our estimate of the properties of the hydrogen recombination transient from the ejected hydrogen envelope (see \S\ref{sec:results}).
We use Eqns. (A1), (A2), and (A3) of \citet{2017ApJ...835..282M}, based on \citet{2013Sci...339..433I}'s application of the analytic theory of recombination transients \citep[e.g.,][]{1993ApJ...414..712P,2009ApJ...703.2205K,2010ApJ...714..155K} to estimate the luminosity, timescale, and total energy of the hydrogen recombination transient predicted by our 3D hydrodynamics simulations:
\begin{equation}
L_{\rm p} \approx 4.2 \times 10^{37}~{\rm erg~s}^{-1}  
 \left(R_{\rm init} \over 10 R_\odot \right)^{2/3}
  \left( \Delta M \over 0.1 M_\odot \right)^{1/3}
   \left(v_{\rm ej} \over 100~{\rm km~s}^{-1} \right)^{5/3}
    \left( \kappa \over 0.32~{\rm cm^2~g}^{-1}\right)^{-1/3} 
    \left(T_{\rm rec} \over 4500~{\rm K} \right)^{4/3},
\end{equation}
\begin{equation}
t_{\rm p} \approx  42~{\rm d} \left(R_{\rm init} \over 10 R_\odot \right)^{1/6}
 \left( \Delta M \over 0.1 M_\odot \right)^{1/3} 
 \left(v_{\rm ej} \over 100~{\rm km~s}^{-1} \right)^{-1/3} 
 \left( \kappa \over 0.32~{\rm cm^2~g}^{-1}\right)^{1/6} 
 \left(T_{\rm rec} \over 4500~{\rm K} \right)^{-2/3},
\end{equation}
\begin{equation}
E_{\rm rad,p}\approx 1.5\times 10^{44} \rm{erg}
\left(R_{\rm init} \over 10 R_\odot \right)^{5/6}
 \left( \Delta M \over 0.1 M_\odot \right)^{2/3} 
 \left(v_{\rm ej} \over 100~{\rm km~s}^{-1} \right)^{4/3} 
 \left( \kappa \over 0.32~{\rm cm^2~g}^{-1}\right)^{-1/6} 
 \left(T_{\rm rec} \over 4500~{\rm K} \right)^{2/3}.
\end{equation}
Using $R_{\rm init} \approx 2 R_\odot$ (approximate final orbital separation of the secondary across our models), $\Delta M \approx 5 M_\odot$ (the entire mass of the envelope), $v_{\rm ej}\approx 18~{\rm km/s}$ (a characteristic velocity at $10 R_\odot$ at the end of our simulation), $\kappa \approx 0.32~{\rm cm^2~g}$, and $T_{\rm rec} \approx 4500~{\rm K}$, we find
$L_{\rm p} \approx 10^{37}~{\rm erg~s}^{-1} \approx 3 \times 10^3 L_\odot$,
$t_{\rm p} \approx 274~{\rm d}$, and
$E_{\rm rad,p} \approx 2\times 10^{44}~\rm{erg}$.

\section{\texttt{MESA} profiles}\label{sec:MESA_profiles}

\begin{figure*}[tp!]
\epsscale{0.56}
\plotone{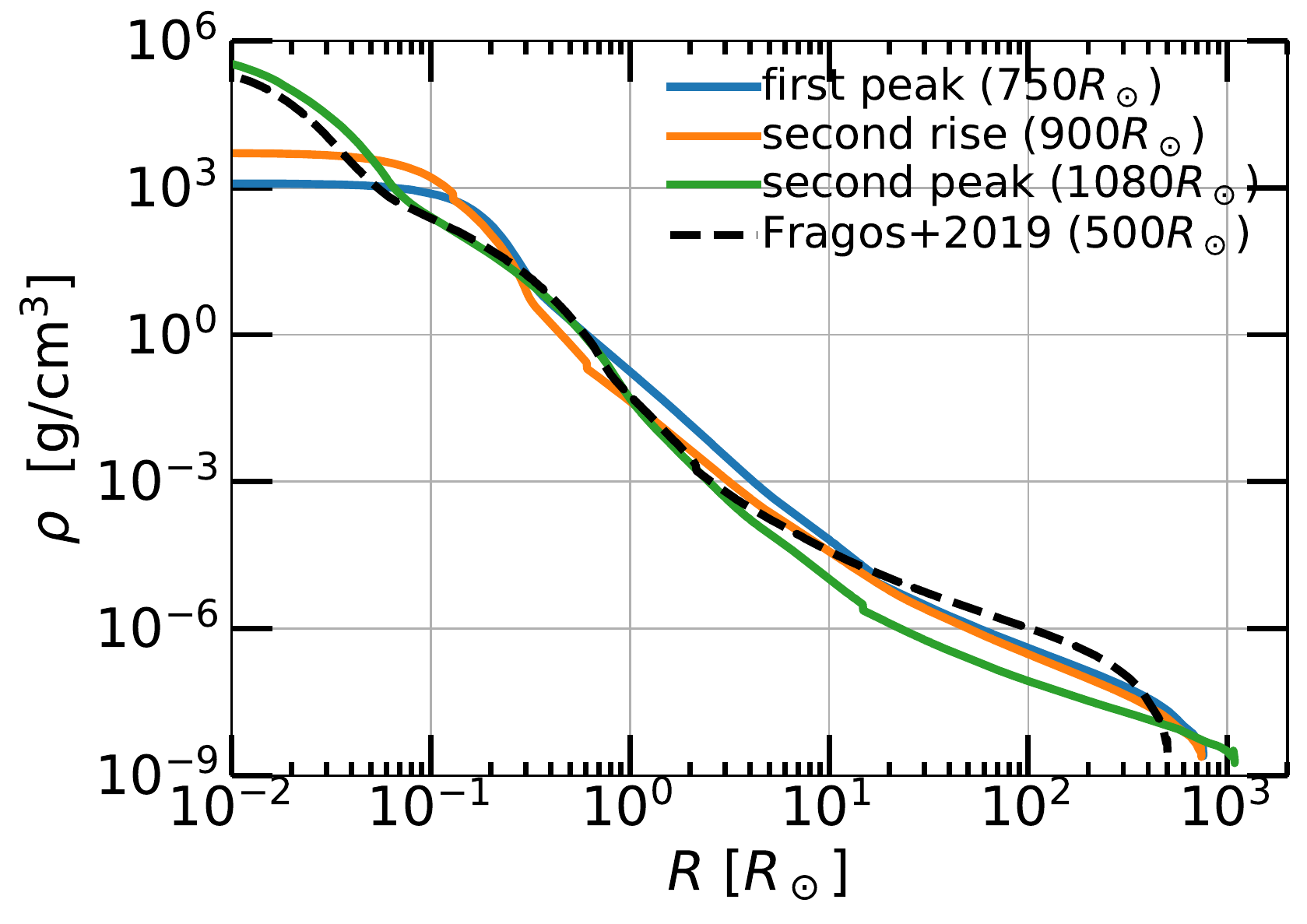}
\epsscale{0.58}
\plotone{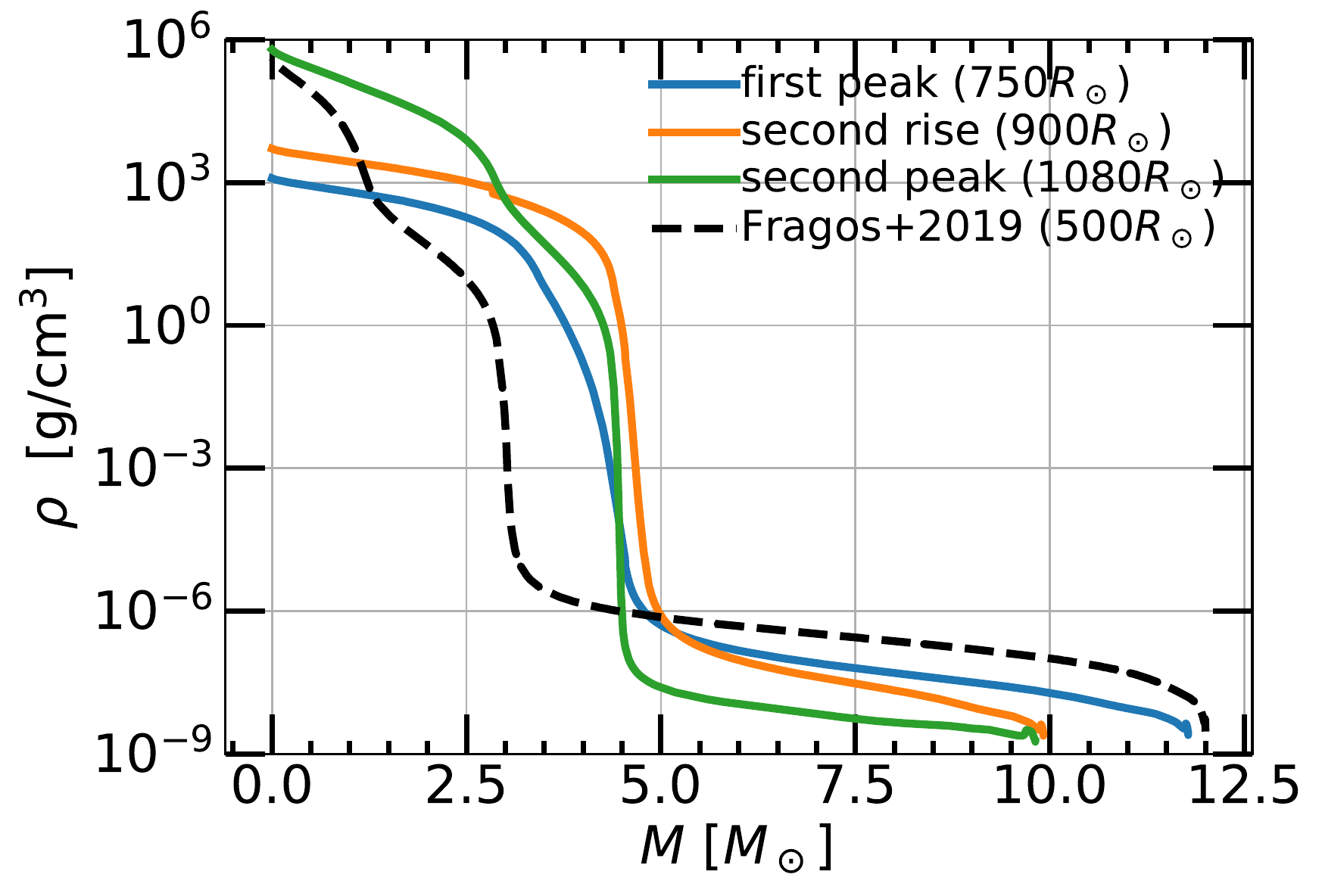}
\plotone{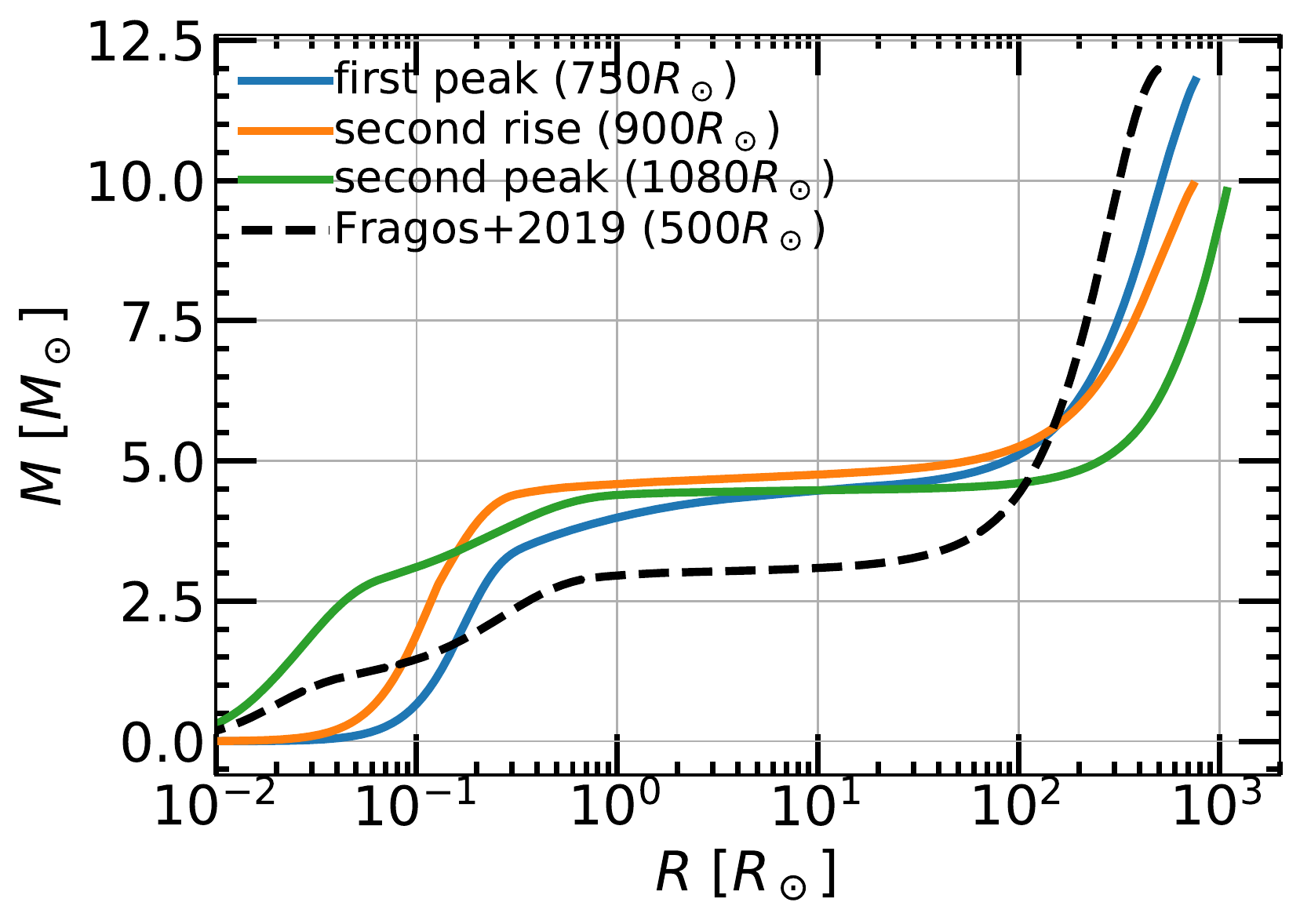}
\epsscale{0.55}
\plotone{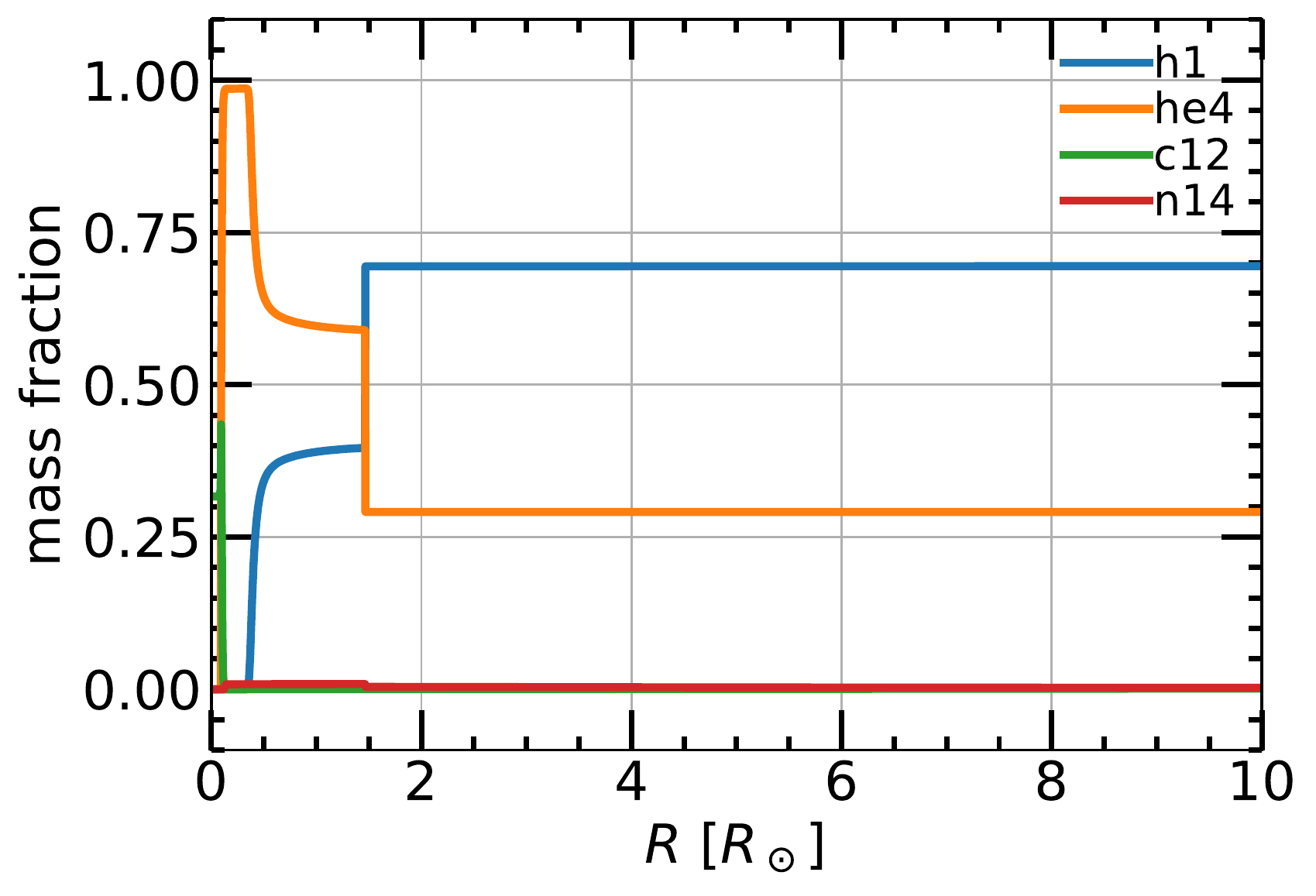}
\caption{
Top row: \texttt{MESA} density profiles vs. radius and mass coordinate for the three models that we simulate in 3D hydrodynamics (see Figure~\ref{fig:MESA}) and for the primary from \citet{2019ApJ...883L..45F}.
Bottom left: enclosed mass vs. radius.
Bottom right: initial 1D composition profiles of hydrogen, helium, carbon, and nitrogen for the $900 R_\odot$ star.
\label{fig:MESA_profiles}
}
\end{figure*}

Here we provide more detail on the 1D stellar models (the initial conditions for the 3D hydrodynamics) built in \texttt{MESA}. 
Our primary is constructed using the setup of \citet{2018A&A...615A..78G}, but for a single star.
The top row of Figure~\ref{fig:MESA_profiles} shows density profiles (vs. radius and mass coordinate) for the three models we simulate in 3D hydrodynamics.
The bottom left panel shows the mass enclosed vs. radius.
We also compare to the primary from the 1D \texttt{MESA} study of CE ejection of \citet{2019ApJ...883L..45F}, which was 12$M_\odot$ and $\approx$500$R_\odot$.
The density profiles are all very similar, being highly centrally concentrated with a core of $\approx 5 M_\odot$ sequestered at $\lesssim 1 R_\odot$.
The greatest difference is in the inner $0.1R_\odot$, where the least centrally concentrated model ($750 R_\sun$) has a central density of $\rho_c \approx 10^3$ g/cm$^3$ and the most centrally concentrated model ($1080 R_\odot$) has a central density of $\rho_c \approx 10^6$ g/cm$^3$ (the reason for this is that the $750 R_\sun$ model has not yet gone through central helium burning and therefore does not have the dense C/O core that the $1080 R_\odot$ model has).
The density drops from a central value of $\rho_c \approx 10^3$--$10^6$ g/cm$^3$ to $\rho \lesssim 10^{-5}$ g/cm $^3$ by $R=10 R_\odot$.  

The bottom right panel of Figure~\ref{fig:MESA_profiles} shows the 1D composition profiles for hydrogen, helium, carbon, and nitrogen at the beginning of the simulation (because the \texttt{MESA} model is mapped exactly into \texttt{FLASH}, these profiles are identical to the \texttt{MESA} composition profiles) for the $900 R_\odot$ star. 
See Figure~\ref{fig:composition_3d} for 3D renderings of the chemical abundance of the system as a function of time.
Note that the sharp composition gradients are a result of the well-defined compositional layering from the \texttt{MESA} model, due to the current or previous convective regions that efficiently mix the material.

\section{Adjusted 1D energy formalism}\label{sec:1D_energy_plots}

\begin{figure*}[htp!]
\epsscale{0.45}
\plotone{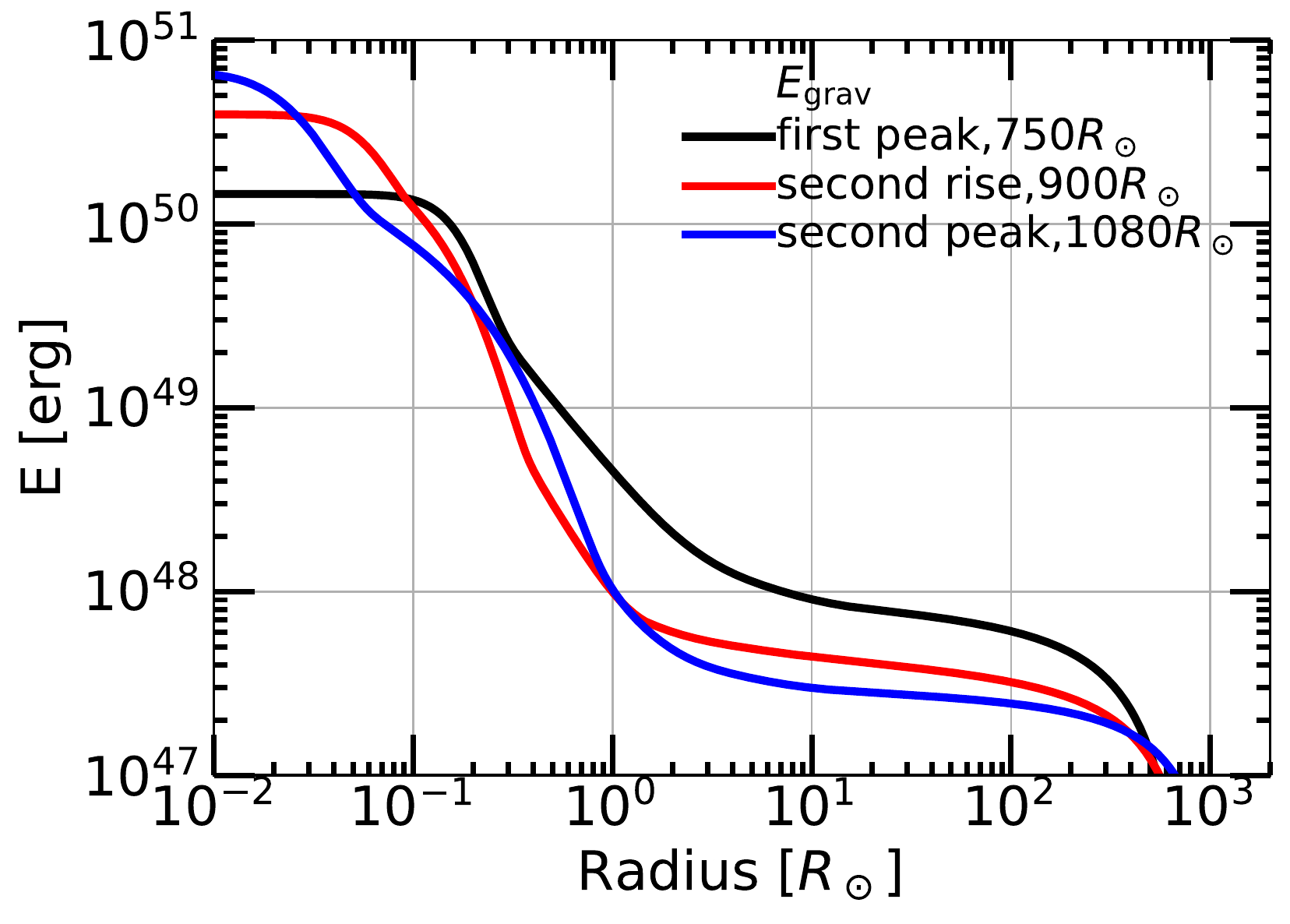}
\plotone{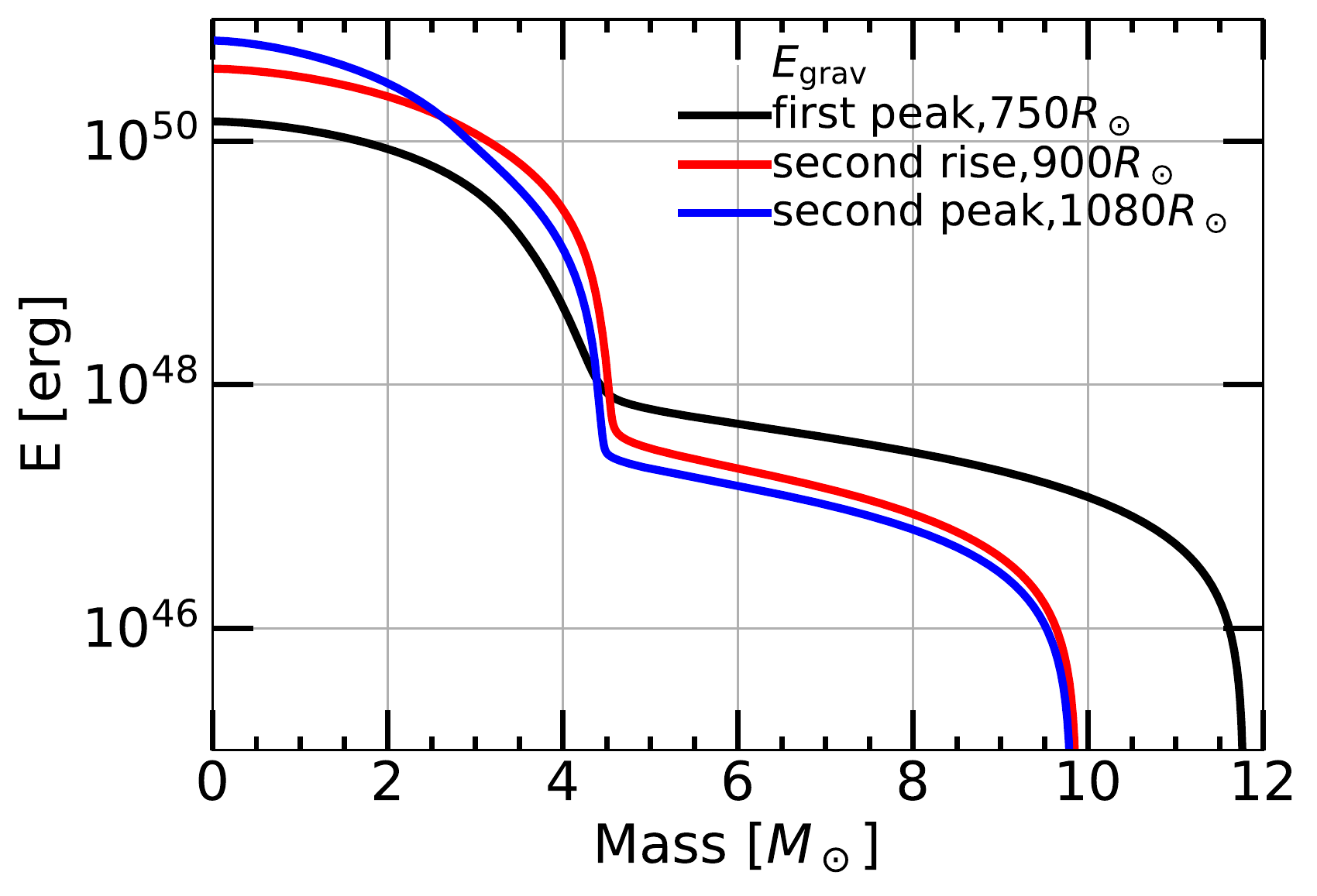}
\plotone{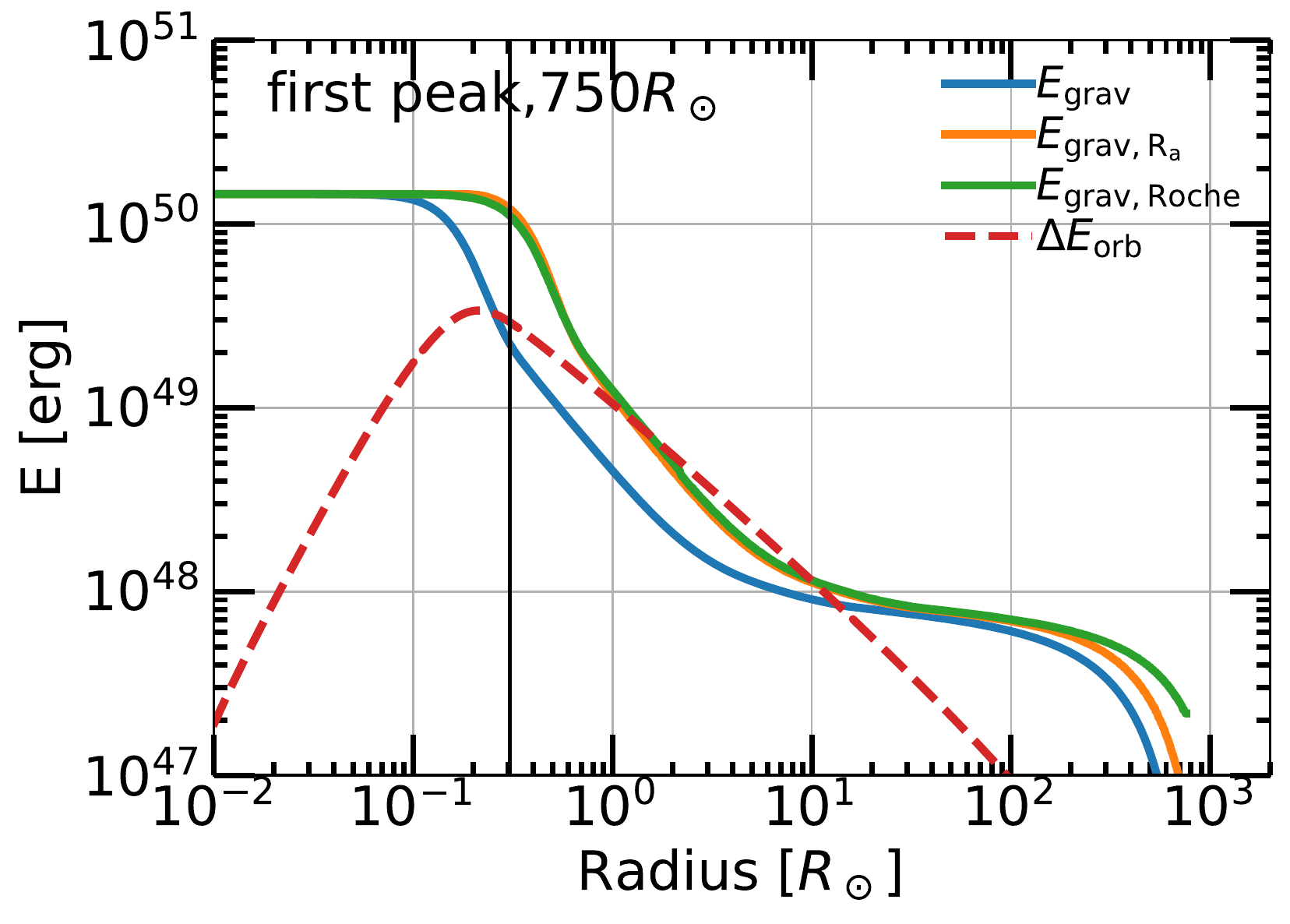}
\plotone{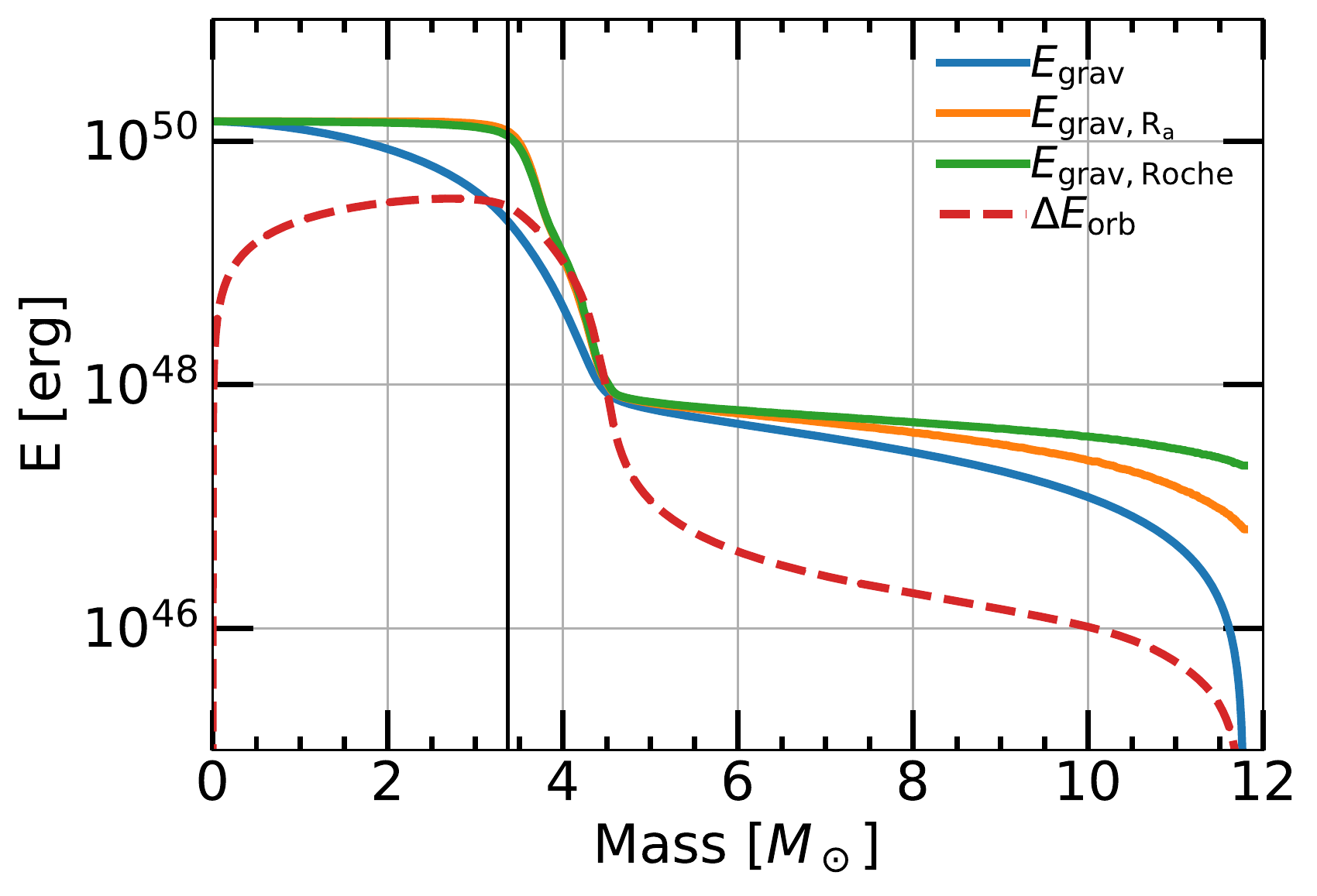}
\plotone{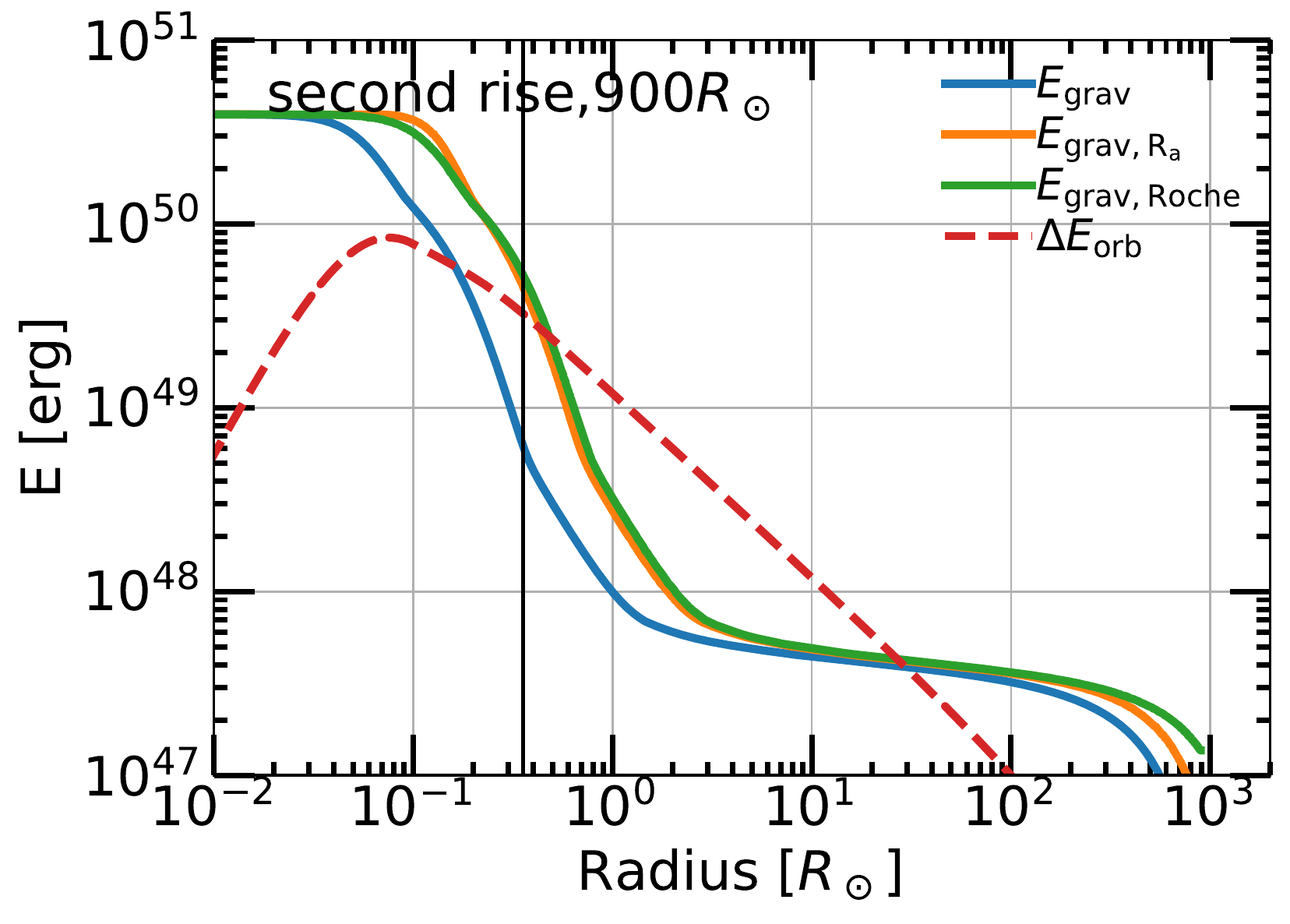}
\plotone{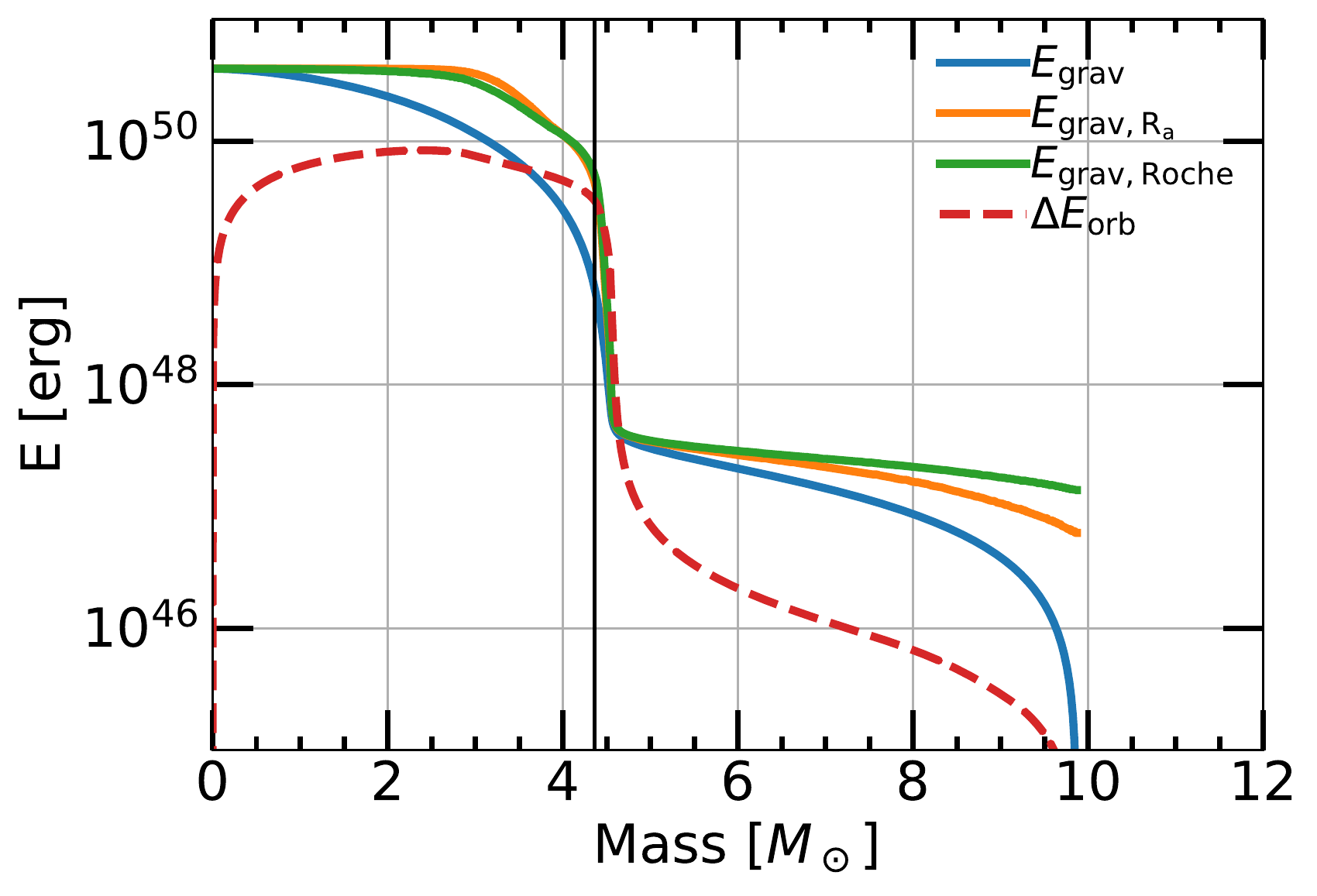}
\plotone{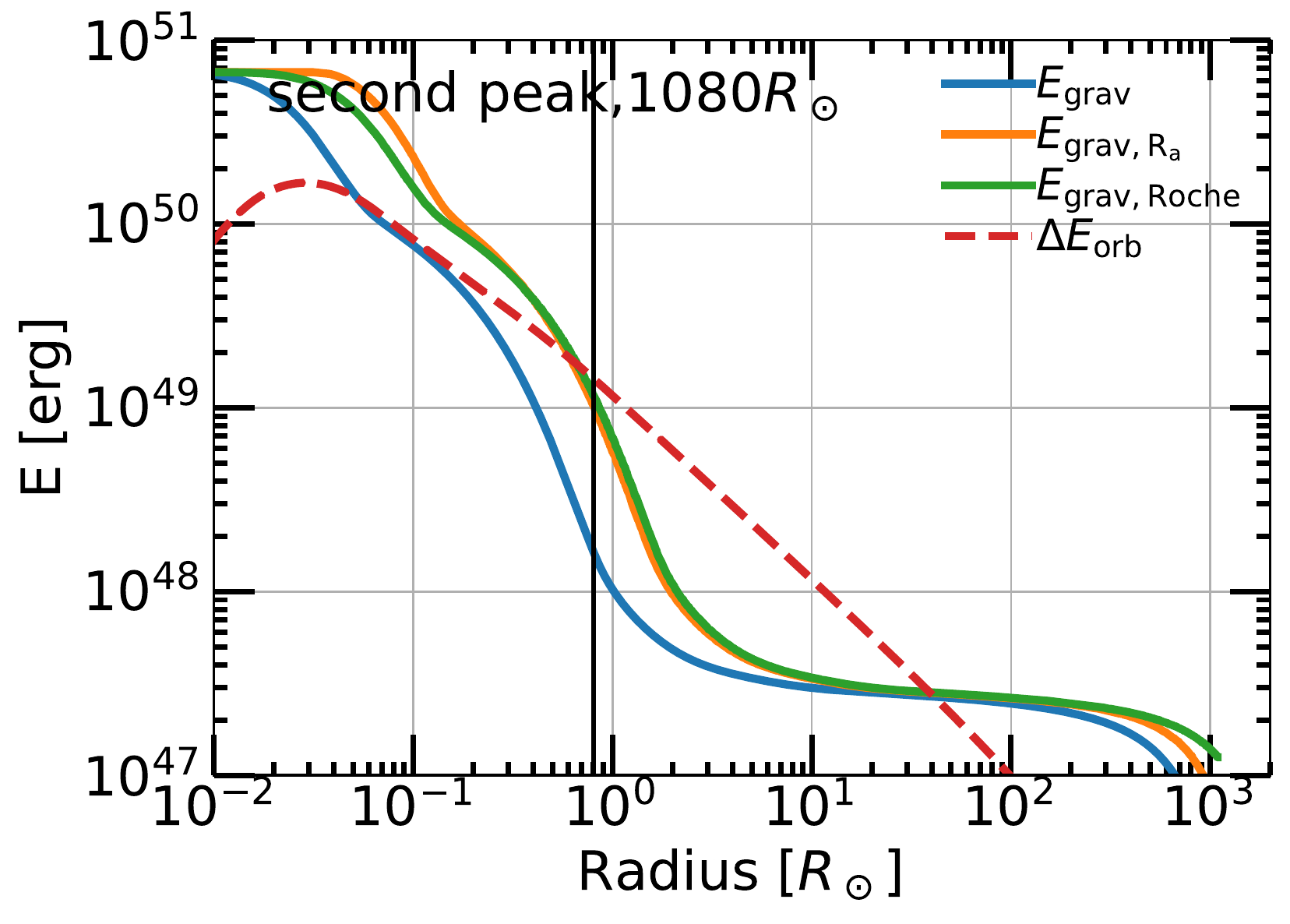}
\plotone{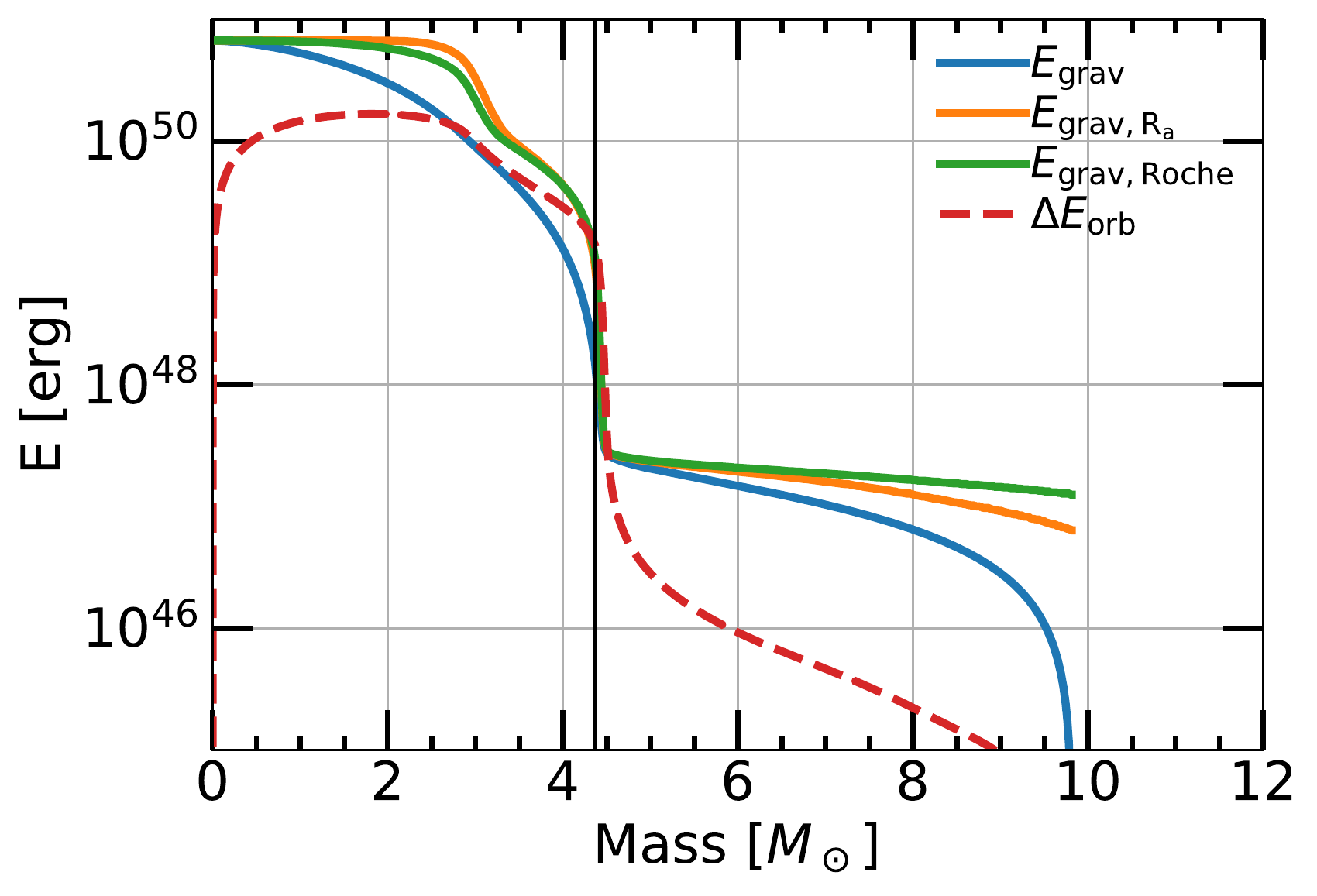}
\caption{
Absolute value of binding and orbital energies vs. radius and mass for the three models we simulate in 3D hydrodynamics (see Figure~\ref{fig:MESA}).
Top row: comparison of the binding energy $E_{\rm grav}$ between all three models.
Other rows: detailed results for each model, including binding energies from the standard $\alpha$ formalism ($E_{\rm grav}$),  the Bondi radius adjusted formalism ($E_{\rm grav, R_a}$),  the Roche radius adjusted formalism ($E_{\rm grav, Roche}$), and the change in orbital energy ($\Delta E_{\rm orb}$).
Vertical line indicates radius of the core (defined by the \texttt{he\_core\_mass} attribute in \texttt{MESA}).
\label{fig:Rosa_profiles}
}
\end{figure*}

Here we provide more detailed results of our 1D energy formalism (method discussed in \S\ref{subsec:1D_energy}).
Figure~\ref{fig:Rosa_profiles} shows binding and orbital energies vs. radius and mass for the three models from that we simulate in 3D hydrodynamics (see Figures~\ref{fig:MESA}, \ref{fig:MESA_profiles}).
In the 1st row we compare gravitational binding energy $E_{\rm grav}$ between all three models.
In other rows we show detailed results for each model including binding energy from the standard $\alpha$ formalism ($E_{\rm grav}$),  the Bondi radius adjusted formalism ($E_{\rm grav, R_a}$),  the Roche radius adjusted formalism ($E_{\rm grav, Roche}$), and the change in orbital energy ($\Delta E_{\rm orb}$).
In general, we see that the binding energy profiles, similar to the density profiles (Figure~\ref{fig:MESA_profiles}), are also highly centrally concentrated and that $<0.1$\% of the binding energy is at radii larger than $10 R_\odot$.
The different calculated energies (for the standard $\alpha$ formalism and for the $r-R_{\rm a}$ and $r-R_{\rm Roche}$ adjusted formalisms) are used to determine the predicted envelope ejection ranges in the 1D energy formalism (see Section~\ref{subsec:1D_energy}).

\section{Forbidden donor radii}\label{sec:forbidden_radii}
The star cannot fill its Roche lobe at an arbitrary moment in its evolution; it needs to have a size large enough such that it would not have filled its Roche lobe before. 
Simply including stellar ages where the star’s radius exceeds any earlier radius it had is not sufficient, as the orbit is changing as well due to wind mass loss and possibly tidal interactions.  

A standard assumption is to think about the orbital changes in the Jeans mode approximation, where the orbital change is a very simple function of the mass loss. It relies on the assumption that (i) mass loss is steady (i.e., in a smooth wind, not a sudden supernova explosion) and (ii) it is lost with a velocity that is high compared to the orbital velocities (such that, e.g.,  it cannot have any tidal interaction with the system) and (iii) it is lost from the vicinity of the mass-losing star in a spherically symmetric fashion in the reference frame of the mass-losing star.   

This gives the following simple analytical result that $a \times (M_1 + M_2) = {\rm constant}$. In this work, this means that any time $t$ the separation $a(t)$ is the following function of the masses and initial parameters:  

\begin{equation}
a (t)  =  a (t=0) \times \frac{M_d(t=0) + M_{NS}}{M_d(t) + M_{NS} }
\end{equation}

We calculate the size of the Roche radius of a system with an initial separation of $a=1301 R_\odot$---this is the initial separation of the widest system to fill its Roche lobe on the first ascent. The system widens with time due to the Jeans mode mass loss. A system with an initial separation slightly larger than $a=1301 R_\odot$ would fill its Roche lobe on the second ascent.
But because of mass loss, the system will have widened in the meantime and the star needs to be $R_\star=900 R_\odot$ or larger. The star can thus not fill its Roche lobe for ages between $t(R=757R_\odot)$ and $t(R=900R_\odot)$, between the first peak and the second rise (see Figure~\ref{fig:MESA}).

In practice this means that the stellar models available to us in this work are:
(a) stars that fill their Roche lobe on the first ascent, that is with radius smaller than $757 R_\odot$, and
(b) stars that fill their Roche lobe on the second ascent, provided their radius is larger than $900 R_\odot$.
In other words, we avoid using models with ``forbidden radii'' (radii between $757$--$900 R_\odot$).

\section{Merger time distribution}\label{sec:merger_time_distribution}

\begin{figure*}[tbp!]
\epsscale{0.5}
\plotone{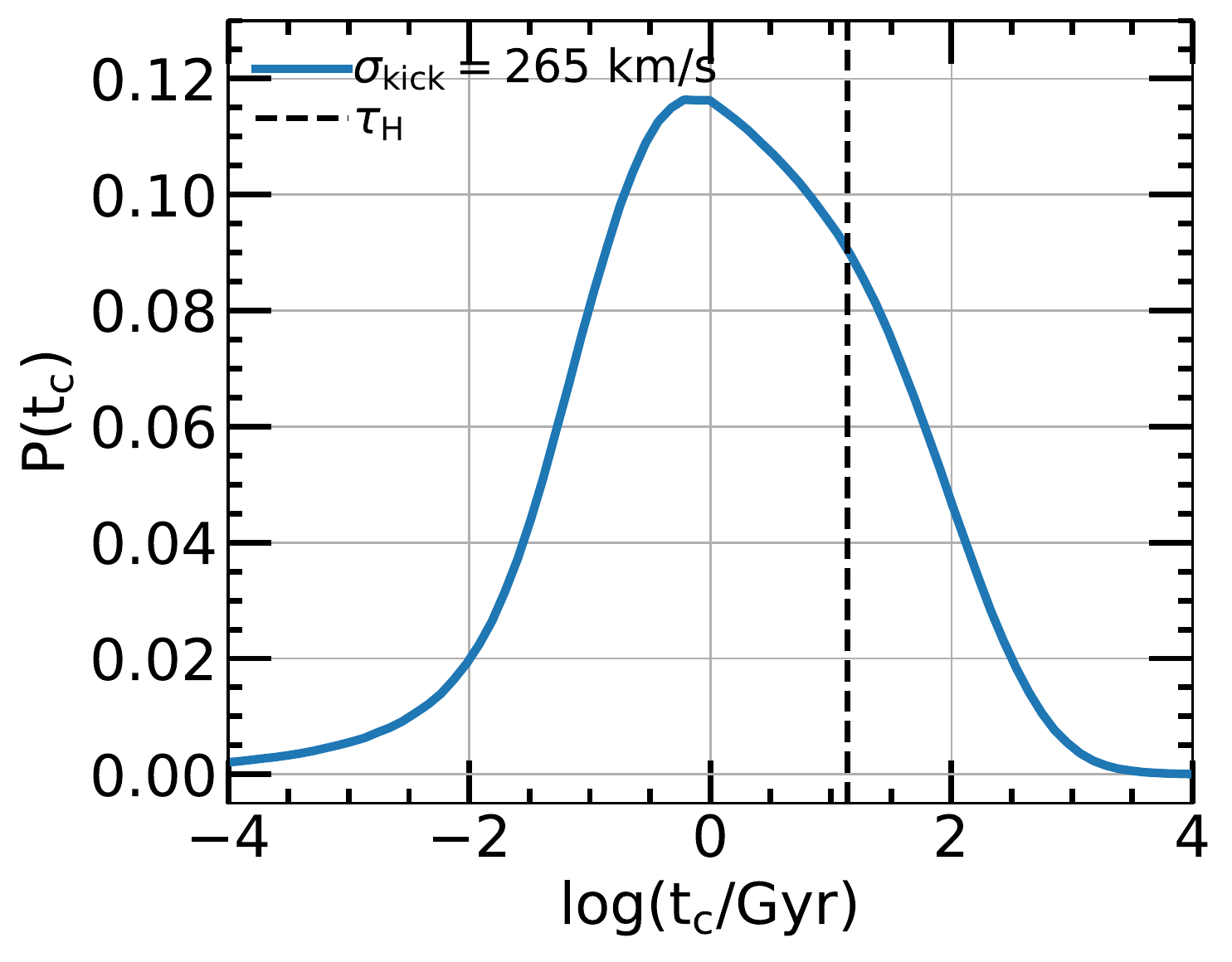}
\caption{
Merger time distribution of the two resulting neutron stars for supernova kick velocities drawn from a Maxwellian distribution with a 1D RMS $\sigma=265$ km/s \citep[following][]{2005MNRAS.360..974H}, for $a_{\rm f}^\ast = 1.3$--$2.8 R_\odot$.
Dashed line indicates the age of the Universe.
\label{fig:merger_time_distribution}
}
\end{figure*}

In order to estimate the merger time distributions of the resulting binary neutron star systems, we take a linear distribution in separations before the supernova (SN) from $a_{\rm f}^\ast = 1.3$--$2.8 R_\odot$ and (separately) $a_{\rm f}^{\ast\ast} = 2.5$--$5.1 R_\odot$.
We then take each separation and run 5000 randomly oriented kicks. 
We sample kick magnitudes from a Maxwellian distribution with a 1D RMS $\sigma=265$ km/s \citep[following][]{2005MNRAS.360..974H}\footnote{This is likely an overestimate, as in a study of Galactic BNSs, \citet{2019MNRAS.487.4847B} find that a majority had a kick velocity much lower than those of standard pulsars \citep[i.e., the distribution of][]{2005MNRAS.360..974H}, with $v_{\rm kick} \lesssim 30$ km/s.} and the final mass of the new neutron star after the SN is $1.4 M_\odot$. To calculate the post-SN orbit we use Eqns. (7) and (8) of \citet{2019MNRAS.486.3213A} and to calculate the merger times of these post-SN orbits we use \citet{1964PhRv..136.1224P}.
Figure~\ref{fig:merger_time_distribution} shows the merger time distribution of the two resulting neutron stars using this procedure for $a_{\rm f}^\ast = 1.3$--$2.8 R_\odot$.

The fraction of bound binaries after the SN is $\approx$37\% (for $a_{\rm f}^\ast$) and $\approx$31\% (for $a_{\rm f}^{\ast\ast}$).
Of the binaries that remain bound after the SN, $\approx$78\% (for $a_{\rm f}^\ast$) and $\approx$57\% (for $a_{\rm f}^{\ast\ast}$) will merge within a Hubble time.\footnote{For comparison, in a delay time distribution study of Galactic BNSs, \citet{2019MNRAS.487.4847B} find that $\gtrsim 40\%$ of BNSs have merger times less than 1 Gyr.}
We caution the reader that after envelope ejection we expect a stable mass transfer phase (case BB) to occur that will likely tighten the binary \citep[see, e.g.,][]{2002MNRAS.331.1027D,2017ApJ...846..170T,2020PASA...37...38V}. As such, this calculation should be taken as an upper limit for the merger timescale. 

\section{Numerical convergence}\label{sec:numerical_convergence}

\begin{figure*}[htp!]
\epsscale{0.53}
\plotone{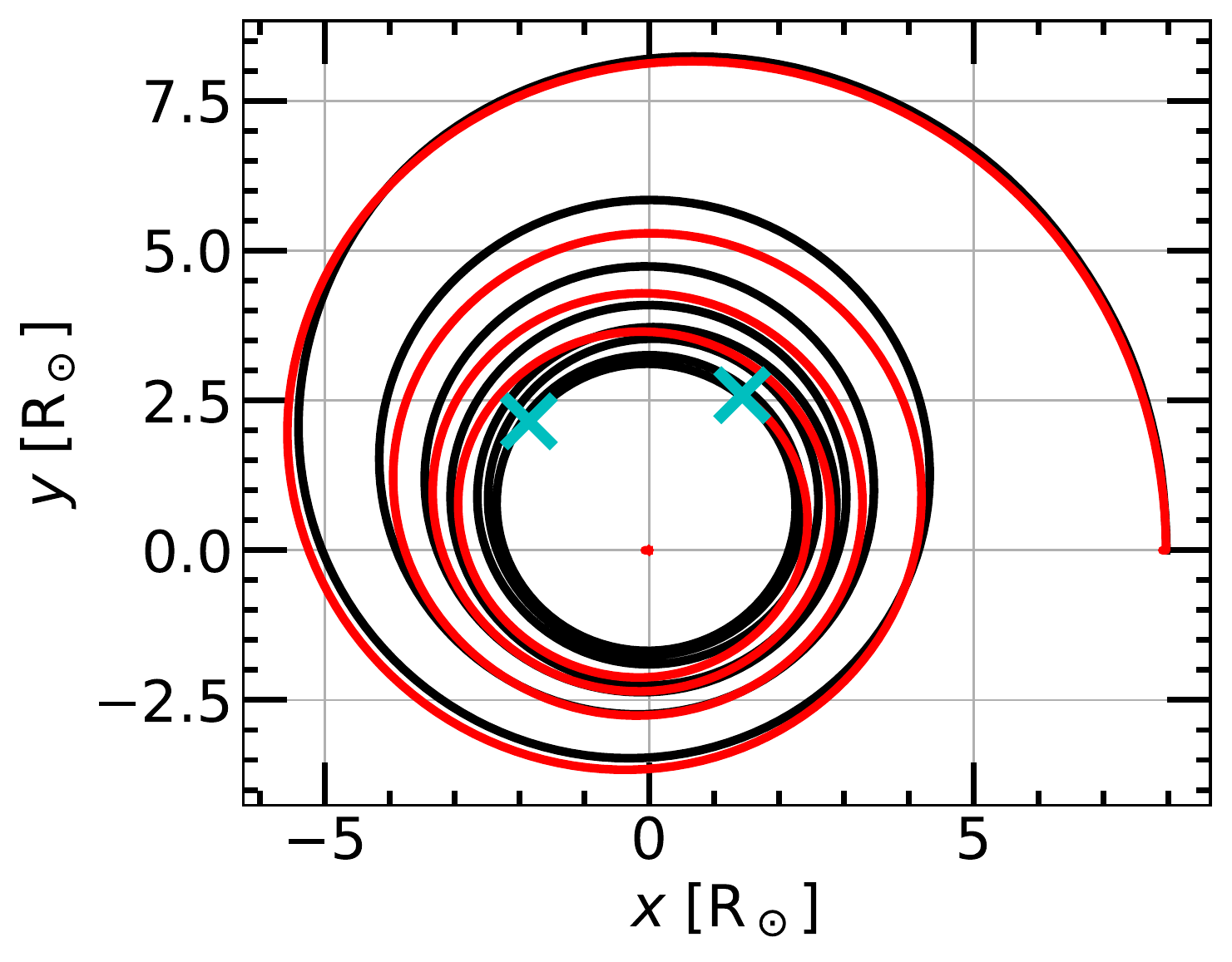}
\plotone{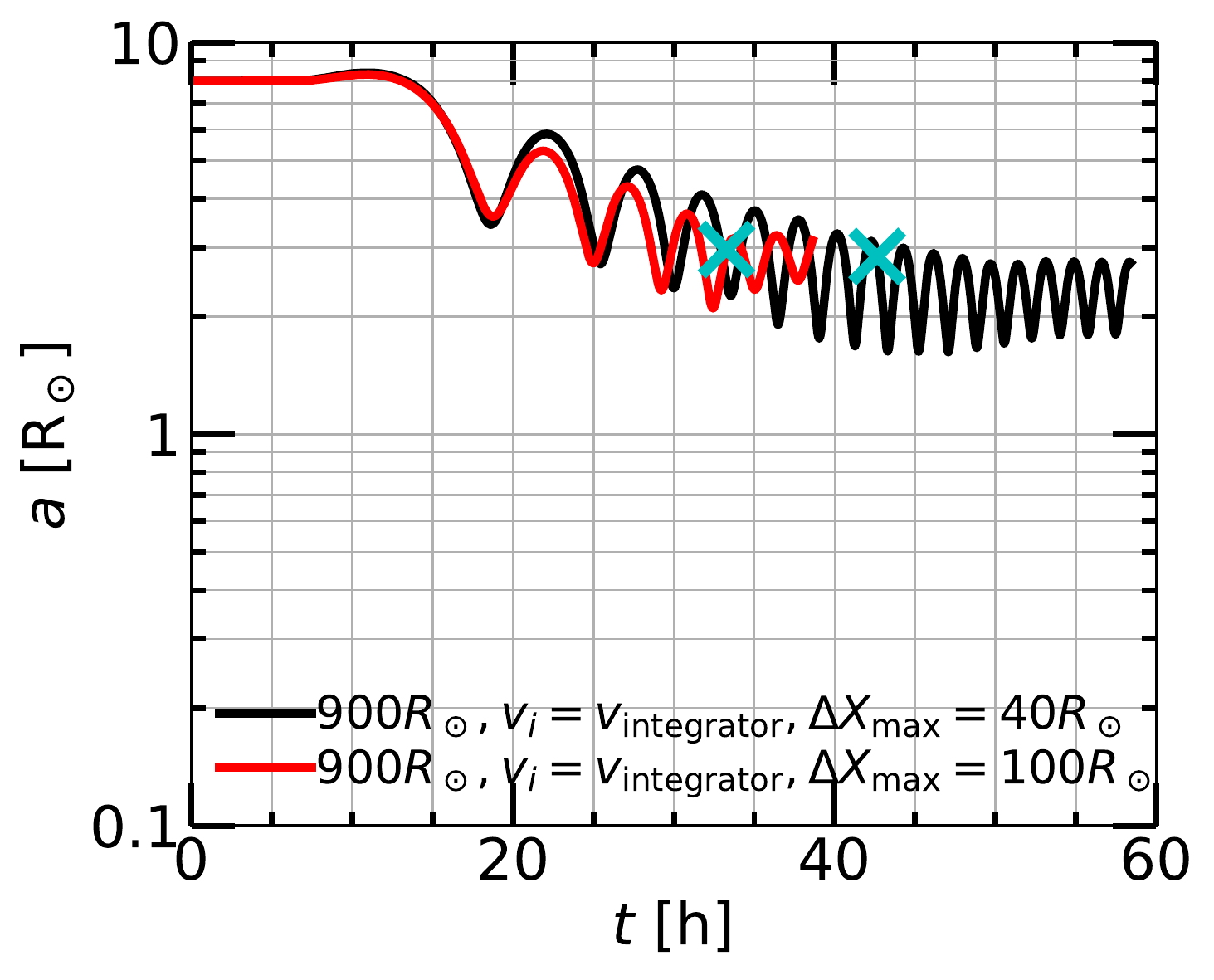}\\
\epsscale{0.53}
\plotone{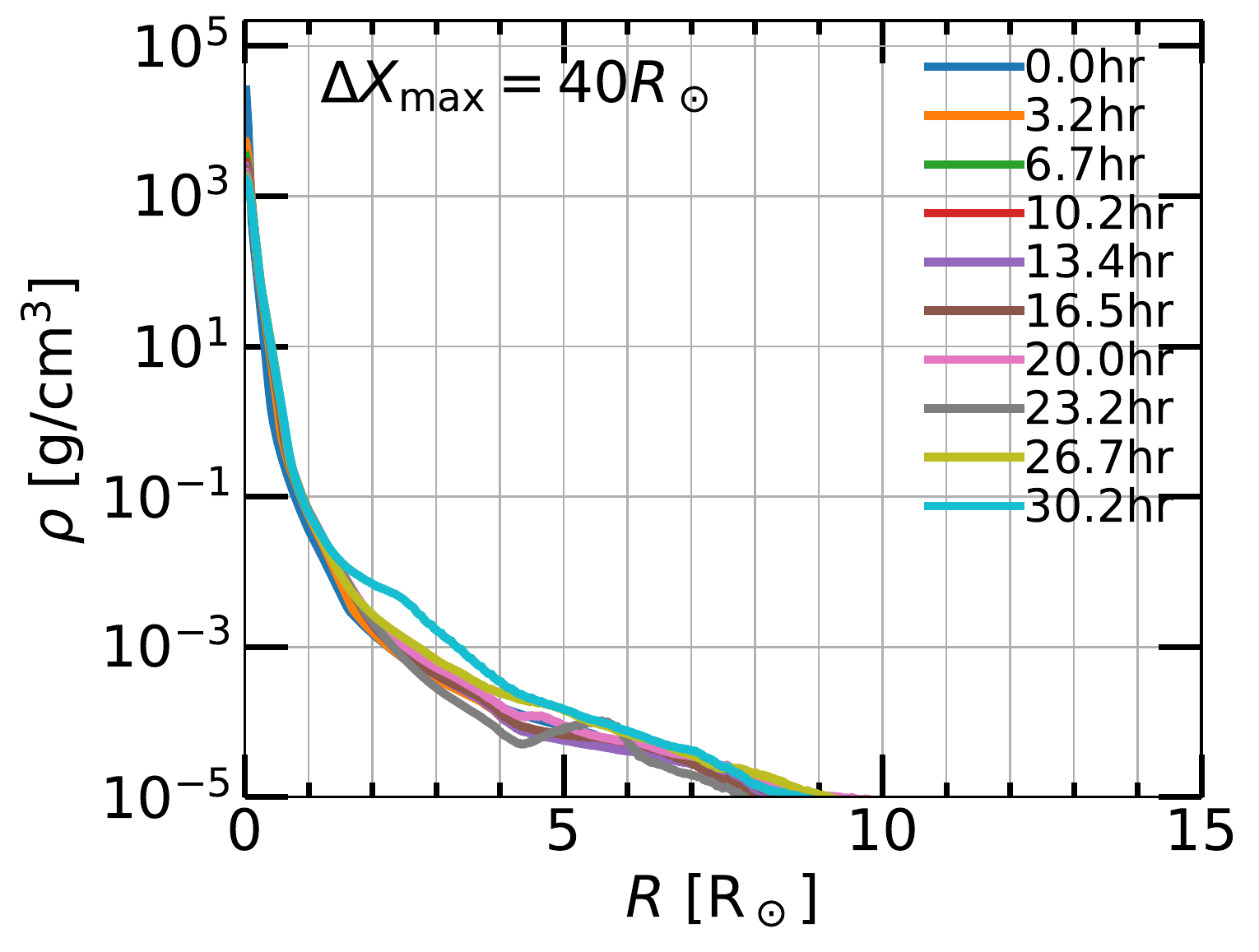}
\plotone{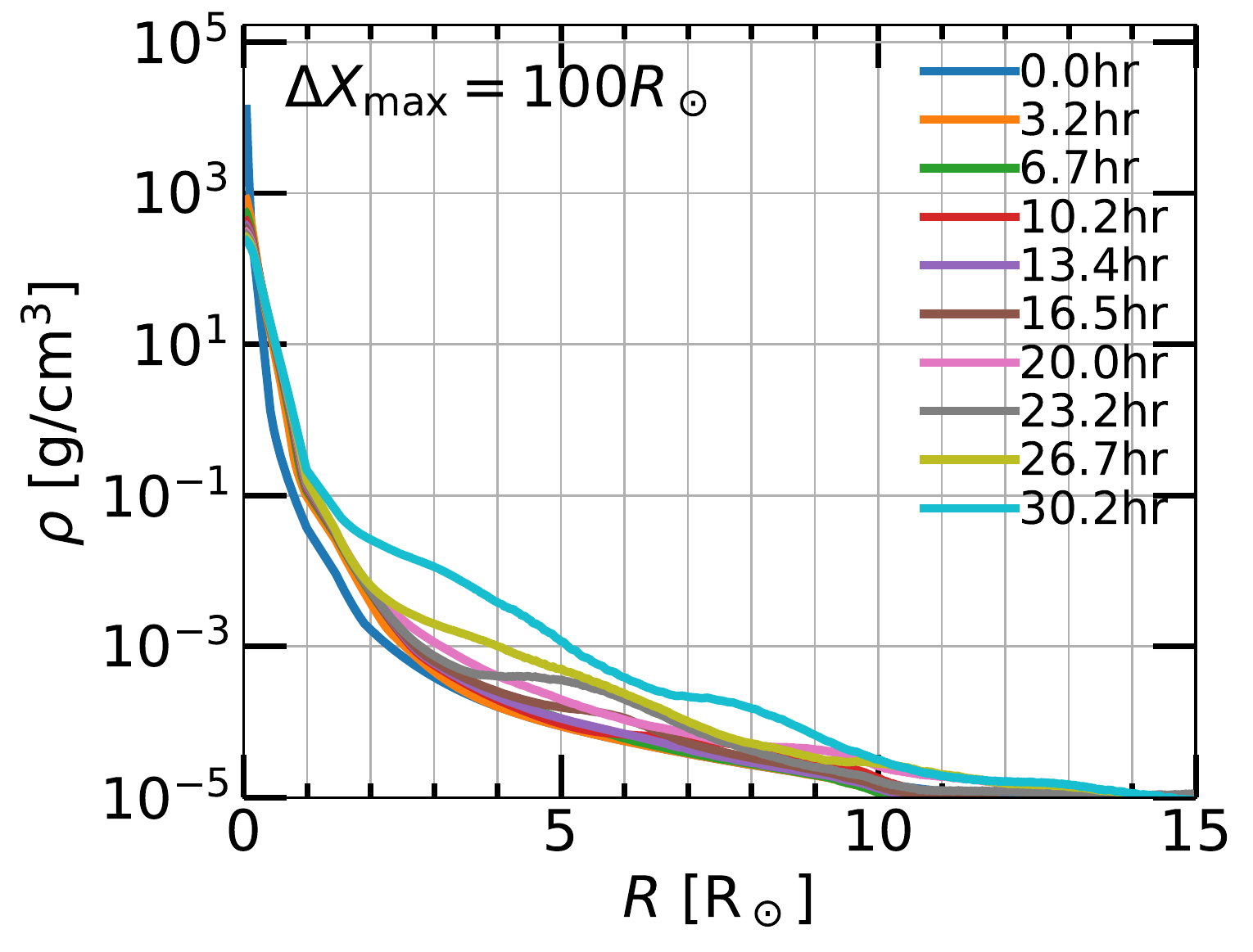}
\epsscale{0.5}
\plotone{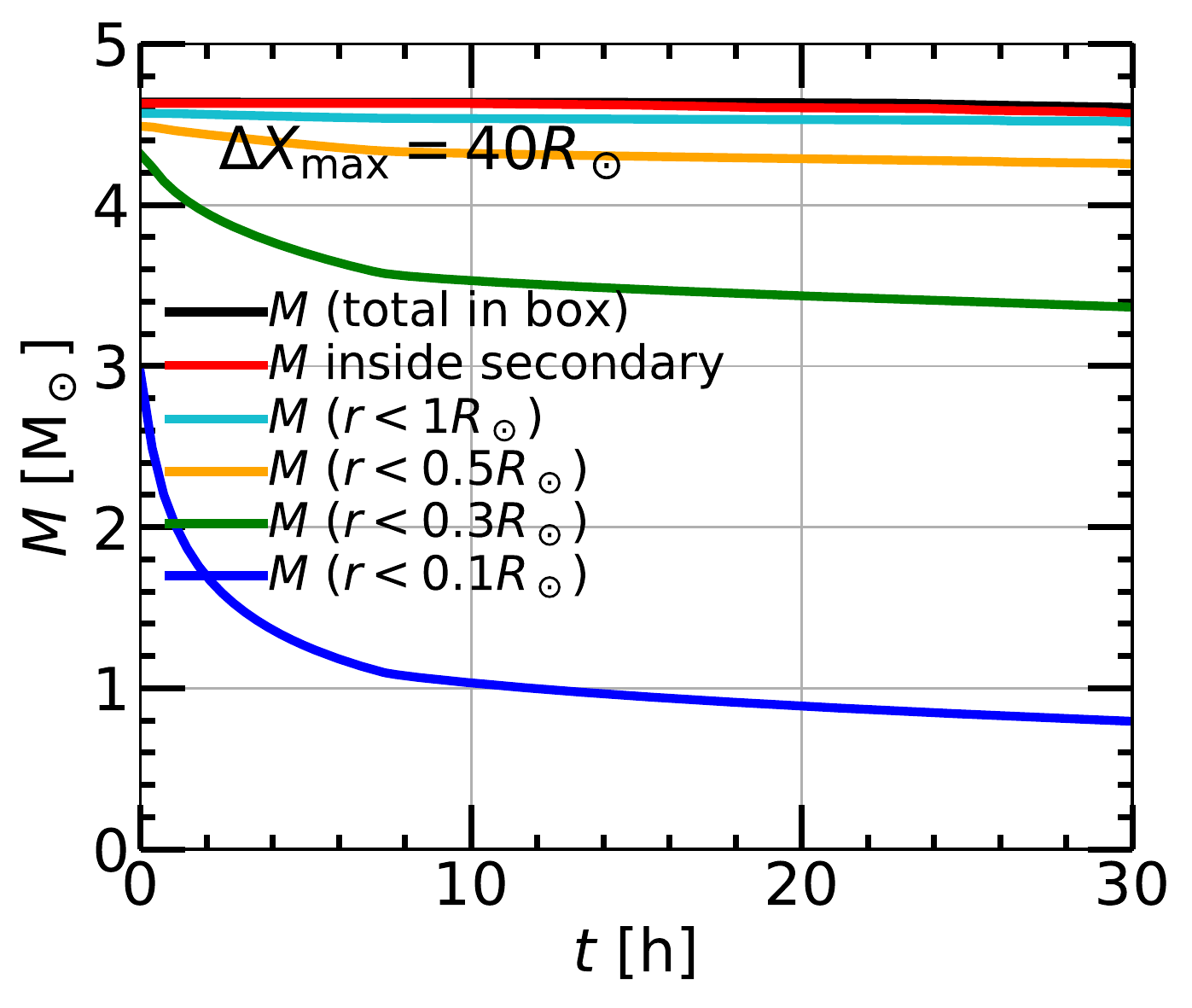}
\plotone{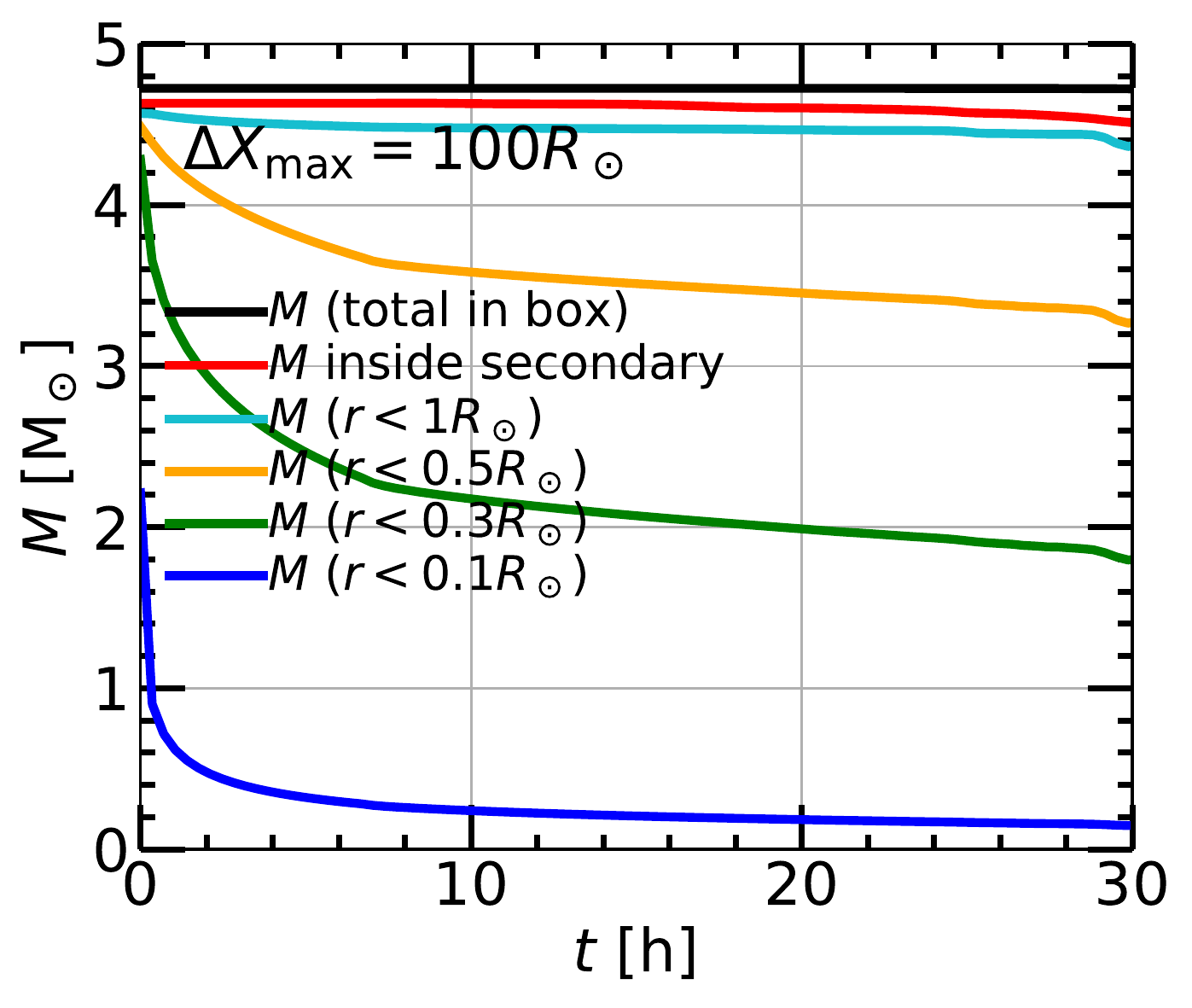}
\caption{
Resolution study.
1st row: orbital separation vs. time for two ($900 R_\odot, v_i=v_{\rm integrator}$) simulations, with box size $\Delta X_{\rm max}=40 R_\odot$ (black) and $\Delta X_{\rm max}=100 R_\odot$ (red, 2.5X lower linear resolution) on a side.
Cyan `X's mark the time at which the envelope is completely ejected (see Figure~\ref{fig:f_ej}).
In the left panel we do not plot the orbital evolution past the point of complete envelope ejection to avoid many overlapping lines.
In the 2nd and 3rd rows, the left panel is the $\Delta X_{\rm max}=40 R_\odot$ run and right panel is the $\Delta X_{\rm max}=100 R_\odot$ run.
2nd row: density profile along one direction in the orbital plane as a function of time.
3rd row: mass enclosed vs. time at several radii. 
\label{fig:numerical_convergence}
}
\end{figure*}

Here we present a brief numerical convergence study.
First, we study the effect of numerical diffusion on the simulation results.
The 1st row of Figure~\ref{fig:numerical_convergence} shows the trajectory and orbital separation vs. time for two ($900 R_\odot, v_i=v_{\rm integrator}$) simulations with different resolutions.
We use the same refinement criteria but two different box sizes: $\Delta X_{\rm max}=(40 R_\odot, 100 R_\odot)$, translating to a factor of 2.5X decrease in linear resolution between the two simulations, which we will denote `high' and `low' resolution in the below.
Note that $\Delta X_{\rm max}=40 R_\odot$ is the nominal resolution used for the simulations in this paper; due to computational constraints we test numerical convergence by decreasing resolution instead of increasing resolution.
The overall orbital evolution and final orbital separations are similar for the two runs. Thus, there is adequate numerical convergence for the purposes of this paper.

The main difference is that the low resolution run ejects the envelope earlier ($t\approx33$ h) than the high resolution run ($t\approx43$ h).
This results in a slightly larger final orbital separation for the low resolution run ($a_{\rm f}^\ast \approx 2.8 R_\odot$) than the high resolution run ($a_{\rm f}^\ast \approx 2.4 R_\odot$).
There are several effects at play here.
The low resolution run has more mass leakage at larger radii (see below), which may aid in ejecting material.
This mass leakage also leads to a higher envelope density (see below) and thus a higher drag force (as $F_{\rm drag} \propto \rho$), which leads to a deeper inspiral and less time to eject the envelope.
In tests at even lower resolution, the secondary experiences so much increased drag that it merges with the helium core before it can eject the envelope.
Nonetheless, a deeper inspiral allows the secondary to deposit orbital energy more rapidly.
In this case, this results in an earlier envelope ejection.

The 2nd row of Figure~\ref{fig:numerical_convergence} shows the density profile along one direction in the orbital plane (other directions are similar) at a few different times throughout the simulation for the same two runs.
For reference, the relaxation process is $t_{\rm relax} \approx 100 t_{\rm dyn, core} \approx 6$ hr.
After relaxation onto the grid, the central density decreases by a factor of $\approx 8$ in the high resolution run and a factor of $\approx 50$ in the low resolution run.
The low resolution run shows lower densities at radii of $r<1 R_\odot$ and higher densities at radii of $r>1R_\odot$, especially at later times, whereas the high resolution run conserves its density profile to later times. This mass leakage from the central regions of the star in the low resolution run leads to a higher envelope density and thus a higher drag force. 
The 3rd row of Figure~\ref{fig:numerical_convergence} shows mass enclosed as a function of time at several radii for the same two runs.
The high resolution run conserves the inner mass shells considerably better than the low resolution run.
However, for both runs, while the core expands and the mass spreads to somewhat larger radii, the mass is primarily redistributed within $r \lesssim 0.5 R_\odot$.

\begin{figure*}[tp!]
\epsscale{0.9}
\plotone{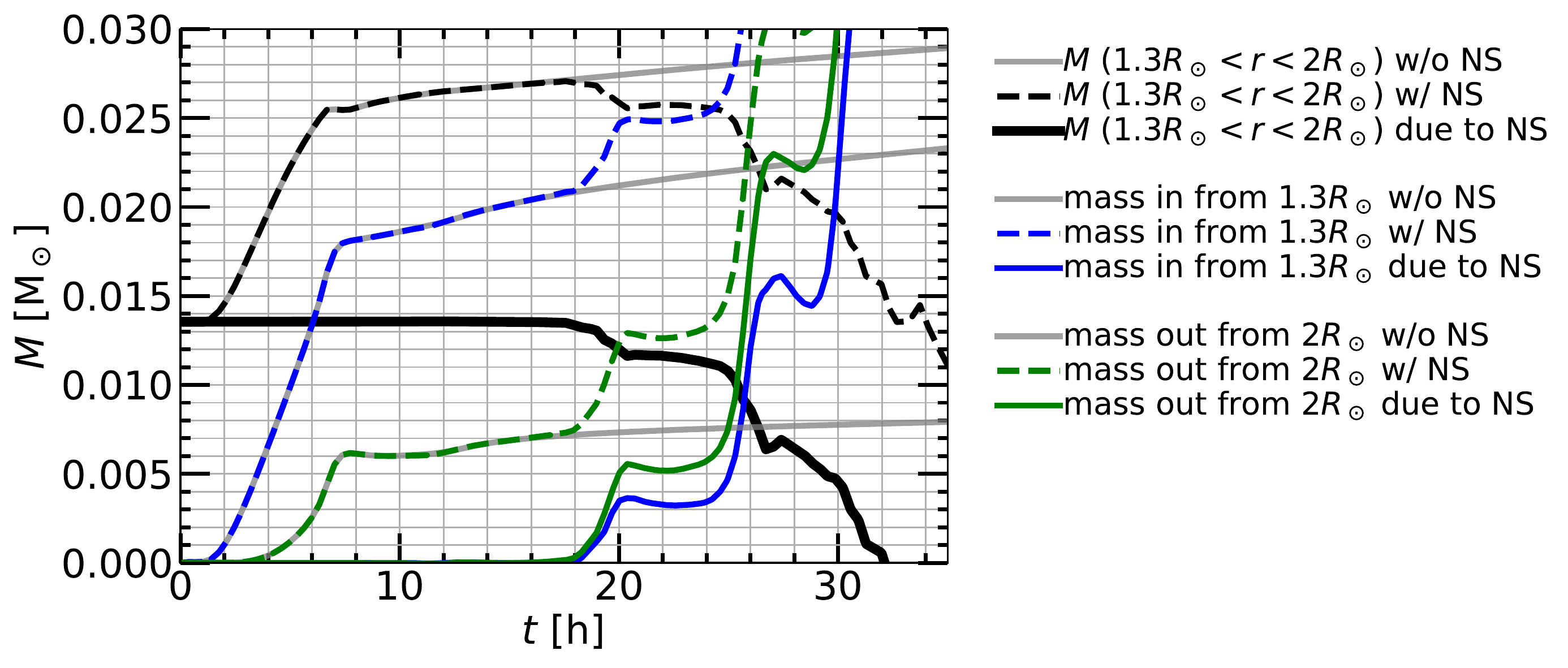}
\caption{
Clearing of material in a representative annulus from $1.3 R_\odot$ to $2 R_\odot$. 
Black lines show mass enclosed in the annulus, blue lines show mass entering the annulus from $1.3 R_\odot$, and green lines show mass exiting the annulus from $2 R_\odot$.
We show data from the ($900 R_\odot, v_i=v_{\rm integrator}$) simulation (dashed lines) and a simulation without the NS (gray lines). 
Solid black, blue, and green lines show the difference between the two simulations---i.e., the effect of the NS.
}
\label{fig:mass_enclosed}
\end{figure*}

Second, while we have shown that the NS ejects the entire envelope outside the helium core (Figure~\ref{fig:f_ej}), we verify that the NS clears the material interior to its orbit and exterior to the Roche radius of the core, which is necessary to successfully stall at a final orbital separation.
Figure~\ref{fig:mass_enclosed} shows data from the ($900 R_\odot, v_i=v_{\rm integrator}$) simulation; other simulations show similar results.
For this analysis we also run a simulation without the NS secondary, to test which effects are numerical (due to resolution-dependent mass leakage; see above) and which are due to the NS.
We consider a representative annulus from $1.3 R_\odot$ (this is approximately the Roche radius of the core) to $2 R_\odot$ (this is approximately the inner boundary of the NS's orbit).
We show the mass enclosed in the annulus as well as the mass entering and exiting it.
The mass enclosed in the annulus rises to double its initial value by $t \approx 18$ h due to mass leakage from $1.3 R_\odot$. It then decreases to below its initial value due to the effect of the NS (namely, the shocks produced by its orbit; see discussion in \S\ref{sec:results} and \S\ref{sec:detailed_time_evolution}).
We subtract the results of the simulation without the NS from the simulation with the NS to isolate the effect of the NS.
This subtracts the mass that is added to the annulus due to numerical leakage alone.
We find that the NS is responsible for clearing the remaining material in the annulus (the solid black line goes to zero).
This analysis also shows that the NS disturbs some material interior to $1.3 R_\odot$ and causes it to enter the annulus.
However (to reiterate), the end result is that after subtracting the mass added to the annulus due to numerical leakage, the NS clears this material as well as the material that was initially in the annulus.

\begin{figure*}[tp!]
\epsscale{0.5}
\plotone{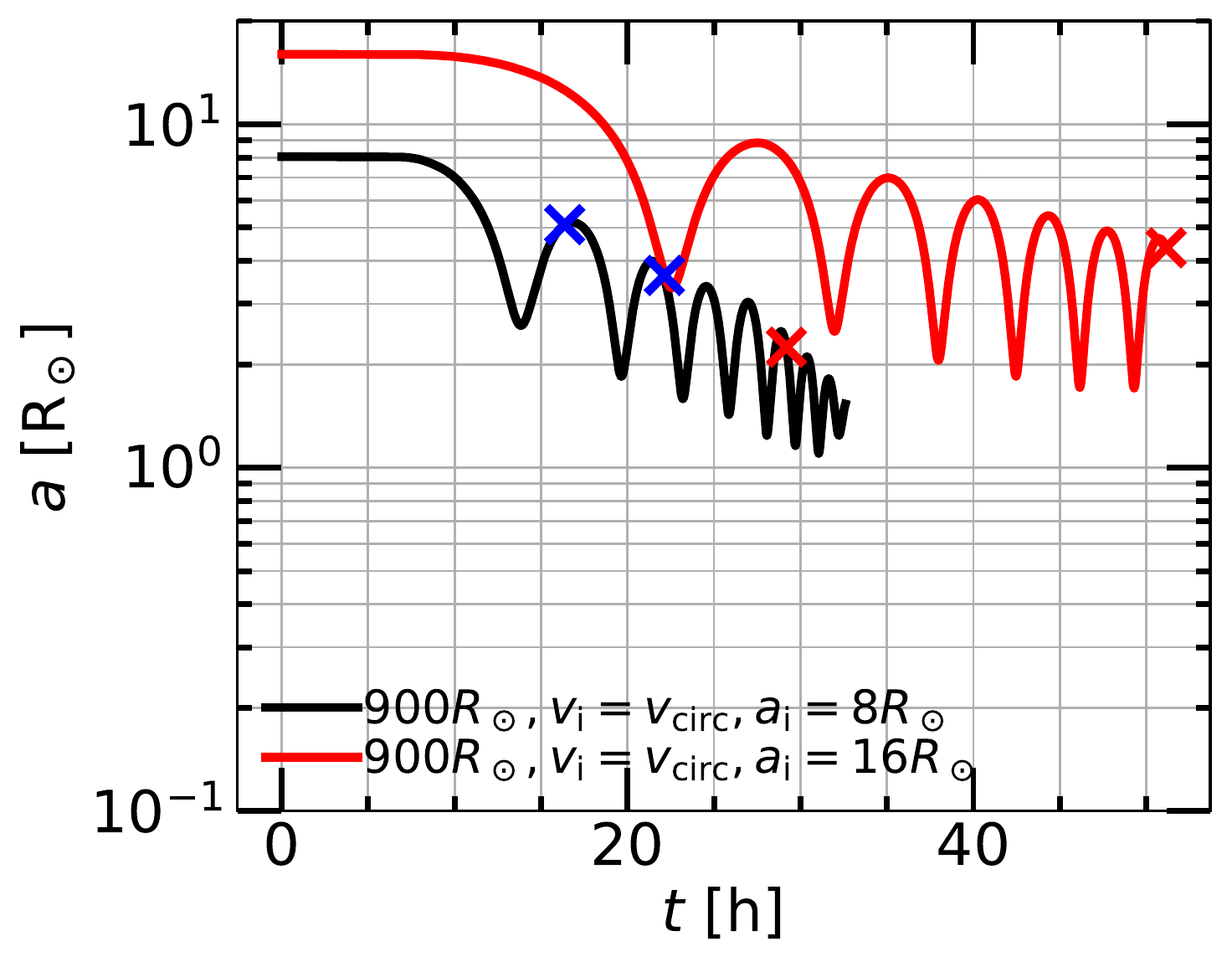}
\caption{
Orbital separation $a(t)$ vs. time for two ($900 R_\odot, v_{\rm i}=v_{\rm circ}$) simulations with different initial orbital separations for the NS secondary. Black line is with $a_{\rm i}=8 R_\odot$ and red line is with $a_{\rm i}=16 R_\odot$.
Blue `X's mark the time at which the envelope outside the NS is unbound (the `X' at $t\approx22$h belongs to the red line) and red `X's mark the time at which the envelope outside the helium core is completely ejected.
}
\label{fig:2xinitialradius}
\end{figure*}

Third, we run an additional simulation where the NS is initialized at $a_{\rm i}=16 R_\odot$, the star is trimmed to $R=20 R_\odot$, and $\Delta X_{\rm max}=80 R_\odot$, twice the values in our nominal study, but with the same linear resolution as in our nominal study (thus, this simulation is $2^3=8$ times more computationally expensive), in order to verify that the NS does not stall exterior to our initial conditions if we start our simulation at this larger radius.
Figure~\ref{fig:2xinitialradius} shows the orbital separation $a(t)$ vs. time for this simulation in comparison with the ($900 R_\odot, v_{\rm i}=v_{\rm circ}, a_{\rm i}=8 R_\odot$) simulation.
The NS in each simulation is initialized with a circular velocity vector.
While the locations of the first few minima of $a(t)$ are different, we verify that the NS reaches a radius $r<a_{\rm i}=8 R_\odot$ at which we start our nominal simulations.
We run this simulation until the point of complete ejection of the envelope outside the helium core (see \S\ref{sec:results}), at $t\approx 51$h.
The NS unbinds the material outside its orbit (see \S\ref{sec:old_fig4}) relatively early in its evolution, at $t\approx 22$h.
The final orbital separations calculated using the two criteria are $a_{\rm f}^\ast \approx 3.2 R_\odot$ and $a_{\rm f}^{\ast\ast} \approx 3.6 R_\odot$.
For the nominal simulation,
$a_{\rm f}^\ast \approx 1.9 R_\odot$ and $a_{\rm f}^{\ast\ast} \approx 5.1 R_\odot$.
The difference in $a_{\rm f}^{\ast\ast}$ is perhaps not as meaningful, due to the method by which we calculate it, which does not take an average orbital separation.
The factor of $\approx 1.7$ difference in $a_{\rm f}^\ast$, however, where we do take an average orbital separation, indicates that our simulations have not completely converged in this quantity.

\section{Alternate criterion for envelope ejection}\label{sec:old_fig4}

Here we discuss an alternate criterion for envelope ejection.
Rather than consider the entire mass of the envelope outside the helium core (which only extends to $R_{\rm core}=0.31 R_\odot$, $0.36 R_\odot$, and $0.8 R_\odot$ for the $750 R_\odot$, $900 R_\odot$, and $1080 R_\odot$ models respectively), as in Figure~\ref{fig:f_ej}, here we consider only the material outside the current orbit of the secondary.
This is a natural criterion in the sense of the $\alpha$-formalism (see e.g. Figures~\ref{fig:MESA} and \ref{fig:Rosa_profiles}), as we calculate the total energy deposited in the envelope exterior to the secondary.
We also note that this criterion is functionally similar to what \citet{2022MNRAS.512.5462L} and \citet{2021arXiv211112112M} do, as the final separations of their NSs are within their $R_{\rm cut}$'s. So \citet{2022MNRAS.512.5462L} consider the mass of the envelope at $r>R_{\rm cut}=18.5 R_\odot$ in determining ejection and \citet{2021arXiv211112112M} at $r>R_{\rm cut}=20 R_\odot$ (note they also perform a resolution convergence study with $R_{\rm cut}=10 R_\odot$). 
However, we also note that these studies both simulate the entire envelope exterior to this point and do not excise it exterior to $10 R_\odot$ as we do in this work, so the total envelope mass they consider is much higher than we do.

\begin{figure*}[tp!]
\epsscale{1.17}
\plotone{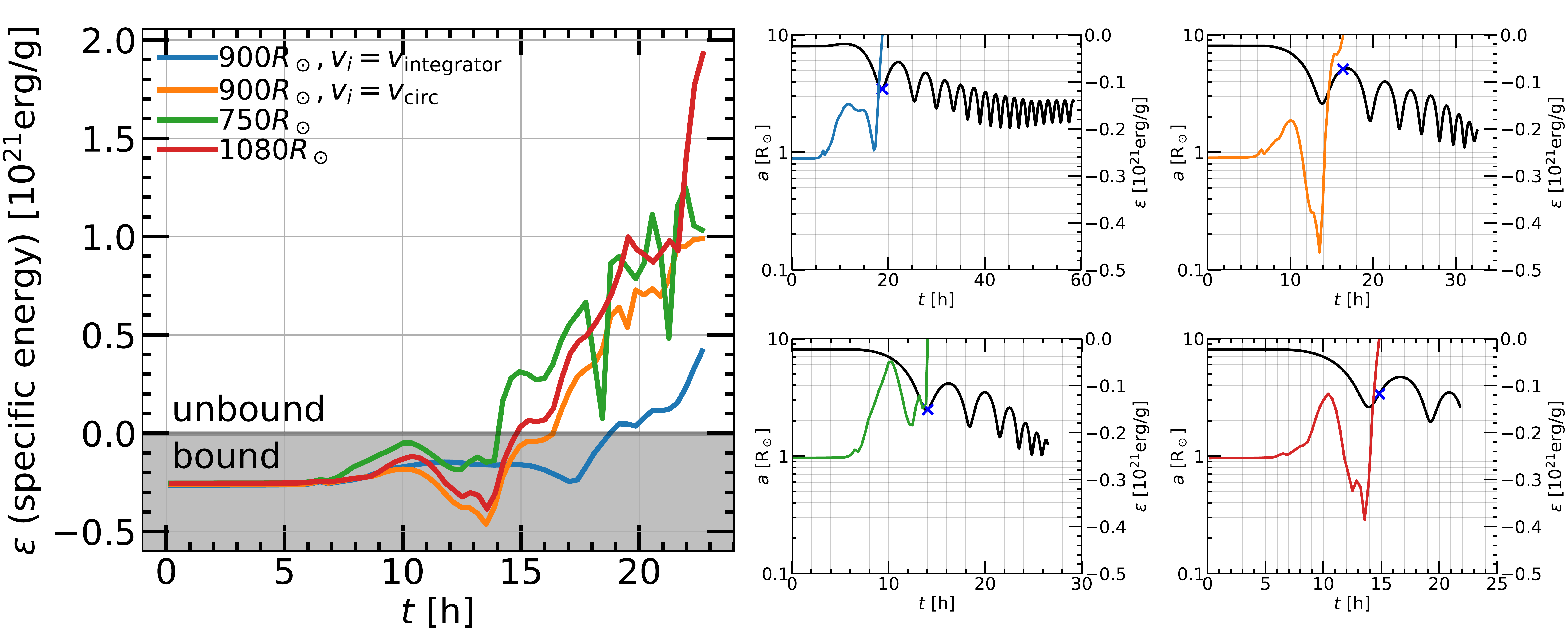}
\caption{
Sum of specific kinetic and potential energy ($\varepsilon = \varepsilon_{\rm kin} + \varepsilon_{\rm grav}$) vs. time for material outside of the current orbit of the secondary for the four main simulations (see \S\ref{sec:numerical_convergence} for additional numerical convergence simulations).
Left panel: all the $\varepsilon(t)$ on one plot.
Envelope outside the secondary is bound for $\varepsilon<0$ (gray region) and is unbound (ejected) for $\varepsilon>0$. 
Right panels: orbital separation $a(t)$ vs. time (black lines) for each simulation with $\varepsilon(t)$ (colored lines; colors match left panel) overlaid.
Blue `X's mark the time at which the envelope outside the secondary is unbound.
\label{fig:E_bind_vs_time}
}
\end{figure*}

Figure~\ref{fig:E_bind_vs_time} shows the total sum of the specific kinetic and gravitational potential energy ($\varepsilon = \varepsilon_{\rm kin} + \varepsilon_{\rm grav}$) of each cell outside of the current orbit of the secondary for the four main 3D hydrodynamics simulations in this paper.
We consider this quantity, rather than the fraction of envelope mass unbound (as in Figure~\ref{fig:f_ej}), as this shows the total energy deposited in the envelope as a result of the change in orbital energy of the secondary.
The plot for the fraction of envelope mass outside the secondary that is ejected, $f_{\rm ej}^{\ast\ast}$, draws the same conclusion.
We note that the recombination energy, while not included in our simulations, is small compared to the envelope energy in our simulations (i.e., for the models we study and at the radii we simulate); e.g., for hydrogen, $\varepsilon_{\rm recomb} \approx 13.6~{\rm eV}/m_{\rm p} \approx 10^{13}~{\rm erg/g}$, whereas the envelope energy is $\varepsilon \approx 10^{21}~{\rm erg/g}$ (Figure~\ref{fig:E_bind_vs_time}).

The left panel shows $\varepsilon(t)$ for all of the simulations and the right panels show the orbital separation $a(t)$ vs. time for each simulation with $\varepsilon(t)$ overlaid.
The total energy of the material outside the secondary increases with time, transitioning from negative (bound) to positive (unbound) at $t \approx 15$--$20$ h for all models.
Small-scale variations correspond to the oscillations in orbital separation as a function of time with each successive orbit of the secondary.
The final orbital separations at the point when the envelope outside the secondary is unbound range from $a_{\rm f}^{\ast\ast} \approx 2.5$--$5.1 R_\odot$.
Note that here we do not quote the average of the local maximum and minimum orbital separation in the neighborhood of this point, as we did for $a_{\rm f}^\ast$ and Figure~\ref{fig:f_ej}, because this point occurs early enough in the evolution that there is not as well-defined an average orbital separation.
The values are listed in Table~\ref{tab:table1}, as well as the associated $\alpha_{\rm CE}$-equivalent efficiencies, $\alpha_{\rm CE}^{\ast\ast}$.


\bibliography{ms}
\bibliographystyle{aasjournal}

\end{document}